\documentclass[ALICE,manyauthors]{cernphprep}
\usepackage[comma,square,numbers,sort&compress]{natbib}
\usepackage{hyperref}
\usepackage{lineno}
\usepackage{xspace}
\usepackage{tabularx,booktabs}
\usepackage{multirow}
\usepackage{makecell}
\usepackage[binary-units,range-phrase=--,range-units=single,exponent-product=\cdot]{siunitx}
\usepackage[T1]{fontenc}
\usepackage{orcidlink} 
\usepackage{xspace}
\usepackage{color}
\usepackage{xcolor}
\usepackage{rotating}
\definecolor{light-gray}{gray}{0.8}

\newcommand{\pp}{\ensuremath{\rm pp}\xspace}
\newcommand{\pPb}{p--Pb\xspace}

\newcommand{\GeVc}{\ensuremath{{\rm GeV/}c}\xspace}

\newcommand{\TeV}{\ensuremath{\rm TeV}\xspace}

\renewcommand{\d}{\ensuremath{\rm d}\xspace}

\newcommand{\sqrts}{\ensuremath{\sqrt{s}}\xspace}

\newcommand{\sqrtsNN}{\ensuremath{\sqrt{s_{\rm NN}}}\xspace}
\newcommand{\pt}{\ensuremath{{\it p}_{\rm T}}\xspace}

\newcommand{\ntracklets}{\ensuremath{N_{\rm tracklets}}\xspace}

\newcommand{\dnchdeta}{\ensuremath{{\rm d}N_{\rm ch}/{\rm d}\eta}\xspace}

\newcommand{\etaless}[1]{\ensuremath{\left|\eta\right| < #1}\xspace}

\newcommand{\zvertex}{\ensuremath{z_{\rm vtx}}\xspace}

\newcommand{\Dzero}{\ensuremath{{\rm D}^{0}}\xspace}

\newcommand{\cmnt}[1]{}

\begin{document}

\begin{titlepage}
\PHyear{2023}       
\PHnumber{041}      
\PHdate{17 March}  

\title{Inclusive and multiplicity dependent production of electrons from heavy-flavour hadron decays in pp and \pPb collisions}
\ShortTitle{Electrons from heavy-flavour hadron decays in pp and \pPb collisions}   

\Collaboration{ALICE Collaboration\thanks{See Appendix~\ref{app:collab} for the list of collaboration members}}
\ShortAuthor{ALICE Collaboration} 

\begin{abstract}
Measurements of the production of electrons from heavy-flavour hadron decays in pp collisions at $\sqrt{s} = {13~\rm TeV}$ at midrapidity with the ALICE detector are presented down to a transverse momentum (\pt) of 0.2 GeV$/c$ and up to $p_{\rm T} = 35~{\rm GeV/}c$, which is the largest momentum range probed for inclusive electron measurements in ALICE.
In \pPb collisions, the production cross section and the nuclear modification factor of electrons from heavy-flavour hadron decays are measured in the \pt range 0.5 $< p_{\rm T} <$ 26 GeV$/c$ at $\sqrt{s_{\textrm{NN}}}=8.16~{\rm TeV}$. The nuclear modification factor is found to be consistent with unity within the statistical and systematic uncertainties. 
In both collision systems, first measurements of the yields of electrons from heavy-flavour hadron decays in different multiplicity intervals normalised to the multiplicity-integrated  yield (self-normalised yield) at midrapidity are reported as a function of the self-normalised charged-particle multiplicity estimated at midrapidity.
The self-normalised yields in pp and \pPb collisions grow faster than linear with the self-normalised multiplicity. A strong \pt~dependence is observed in pp collisions, where the yield of high-\pt electrons increases faster as a function of multiplicity than the one of low-\pt electrons. The measurement in \pPb collisions shows no \pt~dependence within uncertainties.
The self-normalised yields in pp and \pPb collisions are compared with measurements of other heavy-flavour, light-flavour, and strange particles, and with Monte Carlo simulations.


\end{abstract}
\end{titlepage}

\setcounter{page}{2} 


\section{Introduction}\label{section:introduction}
In high-energy hadronic collisions, heavy quarks are mainly produced in hard parton scattering processes. Due to their large masses, their production cross sections can be calculated in the framework of perturbative quantum chromodynamics (pQCD) down to low transverse momenta~\cite{Kniehl:2008zza,Cacciari:2003uh,Kniehl:2005mk,Cacciari:2003zu}. 
Measurements of production cross sections of open heavy-flavour hadrons and their decay products in pp and Pb--Pb collisions were performed by the ALICE, CMS, ATLAS and LHCb Collaborations at the LHC at both mid and forward rapidity~\cite{ATLAS:2015igt, ALICE:2019nxm,ALICE:2012msv,Abelev:2012vra,ALICE:2012mzy,Aaij:2016jht,CMS:2017qjw,LHCb:2016ikn,LHCb:2015swx,ALICE:2021npz,ALICE:2012acz,ALICE:2019nuy,ATLAS:2021xtw,Abelev:2012gx,Adam:2016wyz,ATLAS:2013cia,Khachatryan:2010yr,Aaij:2010gn,ALICE:2021mgk,ALICE:2020mfy,CMS:2016plw,LHCb:2016qpe,LHCb:2015foc,Abelev:2012sca}. These measurements are described by theoretical predictions based on pQCD calculations with the collinear factorisation approach at next-to-leading order with next-to-leading log resummation e.g.~in the GM-VFNS (general-mass variable-flavour-number scheme)~\cite{Kniehl:2008eu,Kniehl:2011bk,Kniehl:2005ej,Bolzoni:2012kx,Bolzoni_2014}) or the FONLL (fixed order with next-to-leading-log resummation)~\cite{Cacciari:2012ny} frameworks, within theoretical uncertainties.
 Measurements of charm-baryon production at midrapidity in pp collisions show an enhancement of the $\Lambda_{\rm{c}}^{+}/\rm{D}^0$~\cite{ALICE:2021npz,Acharya:2017kfy,ALICE:2020wfu,ALICE:2020wla,Sirunyan:2019fnc}, $\Xi_{\rm{c}}^{+,0}/\Dzero$~\cite{Acharya:2017lwf,ALICE:2021psx,ALICE:2021bli}, $\Sigma_{\rm{c}}^{+,0}/\Dzero$~\cite{ALICE:2021rzj}, and $\Omega_{\rm{c}}^{+,0}/\Dzero$~\cite{ALICE:2022cop} ratios with respect to those measured in ${\rm e^{+}e^{-}}$ and ep collisions~\cite{Gladilin:2014tba}. A multiplicity dependence measurement of the $\Lambda_{\rm{c}}^{+}/\rm{D}^0$ ratio~\cite{ALICE:2021npz} has revealed a significant increase from the lowest to the highest multiplicity. These observations indicate that the hadronisation of charm quarks into charm hadrons is not a universal process among different collision systems. These findings are similar to those obtained in the beauty sector by the CDF Collaboration at the Tevatron~\cite{Aaltonen:2008zd} and by the LHCb Collaboration at the LHC~\cite{Aaij:2011jp,Aaij:2019pqz}.

In proton--nucleus collisions, the so-called cold nuclear matter (CNM) effects occur due to the presence of a nucleus in the colliding system and to the large density of produced particles. In particular, the parton distribution functions (PDFs) of nucleons bound in nuclei are modified with respect to those of free nucleons, which can be described by phenomenological parameterisations referred to as nuclear PDFs (nPDFs)~\cite{Eskola:2009uj,deFlorian:2003qf,Hirai:2007sx,Eskola:2016oht}. When the production process is dominated by gluons at low Bjorken-$x$, the nucleus can be described by the Colour Glass Condensate (CGC) effective theory as a coherent and saturated gluonic system~\cite{Fujii:2013yja,Tribedy:2011aa,Albacete:2012xq,Rezaeian:2012ye}. The kinematics of the partons in the initial state can be affected by multiple scatterings
~\cite{Lev:1983hh,Kopeliovich:2002yh} or by gluon radiation (energy loss) before or after the heavy-quark pair is produced~\cite{Vitev:2007ve}. Measurements of heavy-flavour production in \pPb collisions at the LHC will allow a study of the above mentioned effects. Previous measurements of the nuclear modification factor of leptons from heavy-flavour hadron decays in \pPb collisions at $\sqrt{s_{\textrm{NN}}}=5.02$ TeV by the ALICE Collaboration indicate no significant modification of their yields due to CNM effects in the measured  transverse momentum (\pt) region within uncertainties~\cite{Adam:2015qda,Adam:2016wyz}. For a nucleus--nucleus collisions (AA), the nuclear modification factor ($R_{\rm AA}$) is the ratio of the yield in nucleus--nucleus collisions with respect to the yield in proton--proton collisions scaled by the number of binary nucleon--nucleon collision in AA. It quantifies the interaction of a particle and its energy loss while traversing through a medium formed in AA collisions with respect to pp collisions.  Measurements of the nuclear modification factor of open heavy flavour and quarkonia  at mid, forward, and backward rapidity in p–Pb collisions were performed by the ALICE~\cite{ALICE:2019fhe,ALICE:2021lmn,ALICE:2022zig, ALICE:2018mml,Abelev:2014hha,ALICE:2022ych}, ATLAS~\cite{ATLAS:2017prf}, CMS~\cite{CMS:2017exb,CMS:2015sfx}, and LHCb~\cite{LHCb:2013gmv,LHCb:2017ygo,LHCb:2019avm,LHCb:2022rlh} collaborations. The results can be described qualitatively by the various theoretical calculations  mentioned above.

Recent measurements of light-flavour~\cite{CMS:2010ifv,ATLAS:2015hzw,CMS:2015fgy,Abelev:2012ola,Aaboud:2016yar,Chatrchyan:2013nka,ABELEV:2013wsa,Khachatryan:2014jra,Adare:2013piz,Adamczyk:2015xjc,Adare:2015ctn,Khachatryan:2010gv,CMS:2012qk,ALICE:2012eyl,ATLAS:2012cix,LHCb:2015coe} and heavy-flavour hadrons~\cite{Acharya:2018dxy,Sirunyan:2018toe,Sirunyan:2018kiz,ALICE:2017smo,CMS:2020qul} in high-multiplicity pp, p--A, and d--A collisions at different energies have revealed strong flow-like effects in these small systems~\cite{ALICE:2022wpn}. The origin of this phenomenon is debated. Models that incorporate hydrodynamical evolution of the system~\cite{Werner:2010ss,Deng:2011at,Werner:2013ipa,Schenke:2019pmk}, overlapping strings~\cite{Bierlich:2014xba}, string percolation~\cite{Bautista:2015kwa}, or multiple-parton
interactions together with colour reconnection~\cite{Sjostrand:2014zea,Ortiz:2013yxa} can describe
qualitatively the observed features in high-multiplicity events. A multiphase transport model~\cite{Koop:2015wea}, as well as calculations based on the fragmentation of saturated gluon states~\cite{Schlichting:2016sqo,Schenke:2016lrs}, are also able to describe some features of the data.
The measurement of heavy-flavour production in small systems as a function of the charged-particle multiplicity produced in the collision could thus provide further insight into the processes occurring in the collision at the partonic level and the interplay between the hard and soft mechanisms in particle production in pp and \pPb collisions. 

Measurements of charm and beauty productions~\cite{Adam:2015ota,ALICE:2020msa,Adam:2018jmp,Chatrchyan:2013nza,Acharya:2020giw,Adam:2016mkz} indicate an increase of heavy-flavour production with charged-particle multiplicity measured at midrapidity.
The D meson~\cite{Adam:2015ota} and J$/\psi$~\cite{ALICE:2020msa} productions normalised to their corresponding multiplicity-integrated yields in minimum bias events (self-normalised yields), as a function of self-normalised event multiplicities, (i.e., normalised to the average multiplicity in minimum bias collisions)  are measured in pp collisions at $\sqrt{s} = 7$ TeV and $\mbox{$\sqrt{s} = 13$ TeV}$ by the ALICE Collaboration at the LHC, and at $\sqrt{s} = 0.2$ TeV by the STAR Collaboration at RHIC~\cite{Adam:2018jmp}. These measurements show a stronger than linear increase of self-normalised yields as a function of self-normalised multiplicity.  Measurements of the $\Upsilon$(nS) production in pp collisions at $\sqrt{s}$ = 2.76 TeV and $\sqrt{s}$ = 7 TeV by the CMS Collaboration at midrapidity indicate a linear increase with the event activity, when measuring it at forward rapidity, and a stronger than linear increase with the event activity measured at midrapidity~\cite{Chatrchyan:2013nza}. A comprehensive review of the connection between the $\Upsilon$(nS)
production and the underlying event, is presented by the CMS Collaboration in pp collisions at $\sqrt{s}$ = 2.76 TeV~\cite{CMS:2020fae}. Measurements of  multiplicity dependence of $\Upsilon$(nS) production at forward rapidity, is presented by the ALICE Collaboration in pp collisions at $\sqrt{s}$ = 13 TeV~\cite{ALICE:2022yzs}. In \pPb collisions, the self-normalised D meson yield at midrapidity increases with a faster than linear trend as a function of the self-normalised charged-particle multiplicity at midrapidity and is consistent with a linear growth for multiplicity measured at large rapidities~\cite{Adam:2016mkz}. The self-normalised J/$\psi$ yield at larger rapidities also exhibits an increase with increasing normalised charged-particle pseudorapidity density, where the yield at backward rapidity grows faster than the forward rapidity one~\cite{Acharya:2020giw}. A possible correlation with the event multiplicity (and event shape) is also observed for the inclusive charged-particle production~\cite{Acharya:2019mzb}, and for identified particles, including multi-strange hyperons~\cite{Acharya:2018orn}. The trends are qualitatively, and for some of the calculations quantitatively, reproduced by QCD-inspired event generators such as PYTHIA 8~\cite{Sjostrand:2006za}, and EPOS LHC and EPOS 3~\cite{Werner:2013tya,Pierog:2013ria}. But a critical evaluation of the similarities and differences between the physics mechanisms at play in various models is yet to be performed. More stringent tests of the models would be important in this direction. A comparison of the multiplicity-dependent measurements for different particle species would also provide insight into the origin of the observed phenomena~\cite{LHCb:2022syj,ALICE:2021npz}.  

In this article, measurements of the production cross section of electrons from heavy-flavour hadron decays at midrapidity in pp collisions at $\sqrt{s}=13$ TeV and \pPb collisions at $\sqrt{s_{\textrm{NN}}}=8.16$ TeV are presented. The cross section of electrons from heavy-flavour hadron decays was measured as a function of transverse momentum down to 0.2 GeV$/c$ and up to 35 GeV$/c$ in pp collisions, which is the lowest and highest $\pt$-reach attained for electrons from heavy-flavour hadron decays with the ALICE detector. 
Results of the nuclear modification factor ($R_{\rm{pPb}}$) of electrons from heavy-flavour hadron decays at midrapidity in \pPb collisions at $\sqrt{s_{\textrm{NN}}}=8.16$ TeV are reported as well. The self-normalised yields of electrons from heavy-flavour hadron decays measured for the first time as a function of charged-particle multiplicity estimated at midrapidity  ($|\eta|$ $<$ 1) in pp and \pPb collisions are also presented. The comparison of the self-normalised yields of electrons from heavy-flavour hadron decays with other particles measured using the ALICE detector and with Monte Carlo (MC) simulations is discussed.  

The article is structured as follows. In Sec.~\ref{sec:datasampleandselection}, the ALICE apparatus, its main detectors and the data samples used for the analysis are reported. The definition of multiplicity and the calculation of the charged-particle pseudorapidity density is addressed in Sec.~\ref{section:multiplicityanalysis}. In Sec.~\ref{section:heavyflavour}, the procedure employed to obtain the production cross sections of electrons from heavy-flavour hadron decays is explained. Section~\ref{sec:systematics} describes the systematic uncertainties associated with the measurements. The results of the analysis are presented and discussed in Sec.~\ref{section:results}. Finally, the article is summarised in Sec.~\ref{section:summary}.

\section{Experimental apparatus and data sample}\label{sec:datasampleandselection}

In LHC Run 2, the ALICE apparatus consisted of a central barrel, covering the pseudorapidity region $|\eta| < 0.9$, a muon
spectrometer with $-4 < \eta < -2.5$ coverage, and forward- and backward-pseudorapidity detectors employed for triggering, background rejection, and event characterisation. A complete description of the
detector and an overview of its performance are presented in Refs.~\cite{ALICE:2022wpn,Aamodt:2008zz, Abelev:2014ffa}. 

The central barrel detectors used in the analysis are the Inner Tracking System (ITS)~\cite{CERN-LHCC-99-012}, the Time Projection Chamber (TPC)~\cite{Dellacasa:451098}, the Time-Of-Flight detector (TOF)~\cite{CERN-LHCC-2000-012,Cortese:545834}, and the Electromagnetic Calorimeters (EMCal and DCal)~\cite{Cortese:1121574, Allen:2010stl}. They are embedded in a large solenoidal magnet that provides a magnetic field parallel to the beams axis. The ITS consists of six layers of silicon detectors, with the innermost two composed of Silicon Pixel Detectors (SPD). The ITS is used to reconstruct the primary vertex and to track charged particles. 
The TPC is the main tracking detector of the central barrel. It is a gas detector  placed co-axially with the beam axis next to the ITS radially. It also enables charged-particle identification via the measurement of the particle specific energy loss (d$E$$/$d$x$) in the detector gas. The particle identification capabilities of the TPC are supplemented with the TOF detector, which provides a measurement of the time-of-flight of charged particles. The TOF is a gas detector which uses
Multigap Resistive Plate Chamber (MRPC)~\cite{CerronZeballos:1995iy} as its basic detecting element. The TOF detector has the capability to distinguish the electrons from pions, kaons, and protons up to $p_{\rm T}$ $ \approx$ 1 GeV$/c$, $p_{\rm T}$ $ \approx$ 2.5 GeV$/c$, and $p_{\rm T}$ $ \approx$ 4 GeV$/c$, respectively. The EMCal and DCal detectors are shashlik-type sampling calorimeters consisting of alternate layers of lead absorber and scintillator material. The EMCal covers $|\eta| < 0.7$ in pseudorapidity and $\Delta\varphi = 107^{\circ}$ in azimuth. The DCal is located azimuthally opposite to the EMCal covering $0.22 < |\eta| < 0.7$ and $\Delta\varphi = 60^{\circ}$ plus $|\eta| < 0.7$ and $\Delta\varphi = 7^{\circ}$. In the following, EMCal and DCal will be together referred to as EMCal, as they are part of the same detector system. The smallest segmentation of the EMCal is a cell, which has a dimension of $6 \times 6$~cm$^{\rm 2}$ ($0.0143$~rad $\times$~ 0.0143~rad) in its base placed in the $\eta \times \varphi$ direction. 
The electromagnetic calorimeters were used for electron identification and for triggering on rare events with high momentum particles in their acceptance. 

The detectors at forward rapidity used in the analysis are the V0~\cite{Cortese:781854} and T0~\cite{Cortese:781854} detectors. The V0 detector, composed of two scintillator arrays placed on either side of the interaction point along the beam axis (with pseudorapidity coverage $ 2.8 < \eta < 5.1$ and $-3.7 < \eta < -1.7$), was utilised for triggering and for offline rejection of beam-induced background events. In \pPb collisions, the contamination from beam--background interactions and electromagnetic interactions was further removed using the information from the Zero Degree Calorimeters (ZDC)~\cite{Gallio:381433} located at 112.5 m on both sides of the interaction point along the beam axis. The T0 detector, composed of two arrays of quartz Cherenkov counters, covers an acceptance of $4.6 < \eta < 4.9$ and $-3.3 < \eta < -3.0$, and is used to provide the start time for the TOF detector. The V0 and T0 detectors were also employed to determine the integrated luminosity.

The results presented in this article were obtained using data recorded by ALICE during the LHC Run 2 data taking periods between the years 2016 and 2018 
for pp collisions at $\sqrt{s}=13$ TeV, and in 2016 
for $\mbox{\pPb}$ collisions at $\sqrt{s_{\rm{NN}}}=8.16$ TeV. While the nominal magnetic field used during the data taking is 0.5~T, for a subset of periods in pp collisions the magnetic field was reduced to 0.2~T
(will be referred to as low-$B$ field data set in the following sections), allowing for the measurement of electrons down to a \pt of 0.2 \GeVc.
In p--Pb collisions, a centre-of-mass energy per nucleon--nucleon collision  of $\mbox{ $\sqrt{s_{\rm{NN}}}=8.16$ TeV}$ was obtained by delivering proton and lead beams with energies of 6.5 TeV and 2.56 TeV per nucleon, respectively. Due to this asymmetry of the beam energy per nucleon, the proton–nucleon centre-of-mass rapidity frame is shifted by $\Delta y = 0.465$ in the direction of the proton beam. 

Events used in the analyses were obtained using the minimum bias (MB) trigger provided by the V0 detector, and two single shower triggers 
based on the energy deposited in the EMCal~\cite{Abelev:2014ffa, ALICE:2022qhn}. The MB trigger condition requires coincident signals in both scintillator arrays of the V0 detector. The EMCal trigger is based on the sum of energy in a sliding window of $4\times4$ cells above a given threshold. The energy thresholds of the two EMCal triggers were set to 4~GeV~(EG2) and 9~GeV~(EG1) for the pp data sets, and 5.5~GeV~(EG2) and 8~GeV~(EG1) for the \pPb data sets.

In order to obtain a uniform acceptance of the detectors, only events with a reconstructed primary vertex within $\pm10$ cm from the centre of the detector along the beam line (\zvertex) were considered for both pp and p--Pb collisions. The number of selected events in pp and p--Pb collisions for different triggers, and the corresponding integrated luminosities~\cite{ALICE-PUBLIC-2016-002,ALICE-PUBLIC-2018-002} are listed in Table~\ref{table:EventStat}. 
In-bunch pileup events, where more than one collision occurs in the same bunch crossing  and are recorded as a single event, were rejected using an algorithm based on track segments reconstructed with the SPD to detect multiple primary vertices. Out-of-bunch pileup events, where one or more collisions occur in bunch crossings different from the one that triggered the data acquisition, were then rejected based on the timing information provided by the V0 detector. 

\begin{table}[!ht]
\caption{Number of selected events in pp and \pPb collisions for different triggers, and the corresponding integrated luminosities and their uncertainties.}
\small
 \label{table:EventStat}
 	\renewcommand{\arraystretch}{1.4}
  \begin{tabular*}{\textwidth}{@{\extracolsep{\fill}} |c|cccc|ccc}
    \toprule
     \multicolumn{1}{c|}{}&
       \multicolumn{4}{c|}{pp $\sqrt{s} = 13$ TeV}&
       \multicolumn{3}{c}{\pPb $\sqrt{s_{\rm NN}} = 8.16$ TeV}\\
 \hline 
     \multicolumn{1}{c|}{Magnetic field (T)}&
       \multicolumn{1}{c|}{0.2}&
       \multicolumn{3}{c|}{0.5}&
       \multicolumn{3}{c}{0.5}\\
\hline 
     \multicolumn{1}{c|}{Trigger}&
       \multicolumn{1}{c|}{MB}&
       \multicolumn{1}{c}{MB}&
       \multicolumn{1}{c}{EG2}&
       \multicolumn{1}{c|}{EG1}&
       \multicolumn{1}{c}{MB}&
       \multicolumn{1}{c}{EG2}&
       \multicolumn{1}{c}{EG1}\\
       \hline
       
 \multicolumn{1}{c|}{Number of events ($ 10^{6}$)}&
       \multicolumn{1}{c|}{438}&
       \multicolumn{1}{c}{1755}&
       \multicolumn{1}{c}{116}&
    \multicolumn{1}{c|}{96}&
    \multicolumn{1}{c}{39}&
       \multicolumn{1}{c}{0.6}&
       \multicolumn{1}{c}{3.4}\\
\hline

 \multicolumn{1}{c|}{Luminosity (nb$^{-1}$)}&
       \multicolumn{1}{c|}{{7.6}}&
       \multicolumn{1}{c}{{30.3}}&
       \multicolumn{1}{c}{{811.3}}&
       \multicolumn{1}{c|}{{8214.5}}&
       \multicolumn{1}{c}{{0.0190}}&
       \multicolumn{1}{c}{{0.0860}}&
       \multicolumn{1}{c}{{1.65}}\\
  \multicolumn{1}{c|}{}&
       \multicolumn{1}{c|}{{$\pm$0.2}}&
       \multicolumn{1}{c}{{$\pm$0.7}}&
       \multicolumn{1}{c}{{$\pm$30.7}}&
       \multicolumn{1}{c|}{{$\pm$378.7}}&
       \multicolumn{1}{c}{{$\pm$0.0005}}&
       \multicolumn{1}{c}{{$\pm$0.0025}}&
       \multicolumn{1}{c}{{$\pm$0.05}}\\

\hline

  \end{tabular*}

\end{table}

\section{Multiplicity definition and corrections}\label{section:multiplicityanalysis}

The production of electrons from heavy-flavour hadron decays was investigated as a function of charged-particle pseudorapidity density (\dnchdeta) in pp and \pPb collisions.
The \dnchdeta was measured in the pseudorapidity range \etaless 1. It was evaluated using the number of tracklets (\ntracklets) in the SPD~\cite{ALICE:2012xs,Acharya:2018egz}, defined as track segments pointing to the primary vertex and formed by joining pairs of hits in the two SPD layers. 

The number of raw tracklets (\ntracklets) in an event were corrected ($N^{\rm{corr}}_{\rm{tracklets}}$) for the variation of the detector conditions with time (fraction of active SPD channels) and its limited acceptance as a function of \zvertex using a data-driven event-by-event correction, following the procedure discussed in Refs.~\cite{Abelev:2012rz}.
The corrections were done by applying a \zvertex and time-dependent correction factor such that the measured average multiplicity is equalised to a reference value, which was chosen to be the largest mean SPD tracklet multiplicity observed over time. The correction factor for each event was randomly smeared using a Poisson distribution to take into account event-by-event fluctuations. The number of events, sliced in $N^{\rm{corr}}_{\rm{tracklets}}$ intervals, were corrected for the trigger and primary vertex finding efficiencies, following the procedure discussed in~\cite{ALICE:2020msa}.
The former was estimated from MC simulations and the latter with a data-driven approach. 
The efficiencies were close to unity for all multiplicity classes except for the lowest multiplicity class interval, where the efficiency was close to 90\%.

Detector inefficiencies, production of secondary particles due to interactions with the detector material, and particle decays give a different number of reconstructed tracklets compared to the
true primary charged-particle multiplicity value $N_{\rm{ch}}$~\cite{Adam:2015pza}. MC simulations using the PYTHIA 8.2~\cite{Sjostrand:2006za} and the DPMJET~\cite{Roesler:2000he} event generators, for pp and \pPb collisions respectively, and the GEANT 3~\cite{Brun:1073159} transport code were used to estimate $N_{\rm{ch}}$ from $N^{\rm{corr}}_{\rm{tracklets}}$ and later \dnchdeta. A second-order polynomial correlation was assumed between the two quantities, $N_{\rm{ch}}$ and $N^{\rm{corr}}_{\rm{tracklets}}$, for the full $N^{\rm{corr}}_{\rm{tracklets}}$ range. 
To estimate the systematic uncertainties on \dnchdeta, possible deviations from the second-order polynomial correlation between $N_{\rm{ch}}$ and $N^{\rm{corr}}_{\rm{tracklets}}$ were estimated using a linear function.
The systematic uncertainty on the residual $z_{\rm vtx}$ dependence due to differences between data and MC amounts to about 1\% in pp collisions and is negligible in \pPb collisions. The total systematic uncertainty on $\dnchdeta$ is about $5\%$ in all multiplicity intervals for both pp and \pPb collisions.

The average charged-particle pseudorapidity density was normalised to its average value in $\rm{INEL}>0$ events in pp and \pPb collisions. The $\rm{INEL}>0$ event class contains all events with at least one charged particle within \etaless 1. The average charged-particle pseudorapidity densities $\left(\left<\dnchdeta\right>\right)$ for $\rm{INEL}>0$ were found to be
in agreement with the previous published ALICE measurements~\cite{2021,ALICE:2020msa,Acharya:2020giw}. The resulting values of the self-normalised charged-particle pseudorapidity density $\left( \dnchdeta/\left<\dnchdeta\right> \right)$ for the event classes considered in the analyses presented here are summarised in Table~\ref{Table:dndchValues}.

\begin{table}[!ht]
\caption{Average self-normalised charged-particle pseudorapidity density $\left( \dnchdeta/\left<\dnchdeta\right> \right)$ in \etaless 1.0 for each event class selected in pp and \pPb collisions.}
\label{Table:dndchValues}
	\renewcommand{\arraystretch}{1.4}
  \begin{tabular*}{\textwidth}{@{\extracolsep{\fill}} c|cc|cc}
    \hline
  \multicolumn{1}{c|}{ } &
   
   \multicolumn{2}{c|}{pp $\sqrt{s} = 13$ TeV} & \multicolumn{2}{c}{\pPb $\sqrt{s_{\rm NN}} = 8.16$ TeV} \\
 \hline

 \multicolumn{1}{c|}{Multiplicity class} &
     \multicolumn{1}{c}{$N^{\rm{corr}}_{\rm{tracklets}}$} & \multicolumn{1}{c|}{$\dnchdeta/\left<\dnchdeta\right>$} &  \multicolumn{1}{c}{$N^{\rm{corr}}_{\rm{tracklets}}$} & \multicolumn{1}{c}{$\dnchdeta/\left<\dnchdeta\right>$} \\
  \hline    
     \multicolumn{1}{c|}{I} &
     \multicolumn{1}{c}{1--14} & \multicolumn{1}{c|}{0.48} &  \multicolumn{1}{c}{1--38} & \multicolumn{1}{c}{0.51} \\
       \hline    
     \multicolumn{1}{c|}{II} &
     \multicolumn{1}{c}{15--24} & \multicolumn{1}{c|}{1.63} &  \multicolumn{1}{c}{39--55} & \multicolumn{1}{c}{1.32} \\
       \hline    
     \multicolumn{1}{c|}{III} &
     \multicolumn{1}{c}{25--34} & \multicolumn{1}{c|}{2.50} &  \multicolumn{1}{c}{56--95} & \multicolumn{1}{c}{2.03} \\
       \hline    
     \multicolumn{1}{c|}{IV} &
     \multicolumn{1}{c}{35--44} & \multicolumn{1}{c|}{3.34} &  \multicolumn{1}{c}{96--121} & \multicolumn{1}{c}{3.01} \\
       \hline    
     \multicolumn{1}{c|}{V} &
     \multicolumn{1}{c}{45--54} & \multicolumn{1}{c|}{4.16} &  \multicolumn{1}{c}{122--300} & \multicolumn{1}{c}{3.85} \\
       \hline    
     \multicolumn{1}{c|}{VI} &
     \multicolumn{1}{c}{55--64} & \multicolumn{1}{c|}{4.97} &  \multicolumn{1}{c}{} & \multicolumn{1}{c}{} \\
       \hline    
     \multicolumn{1}{c|}{VII} &
     \multicolumn{1}{c}{65--120} & \multicolumn{1}{c|}{6.05} &  \multicolumn{1}{c}{} & \multicolumn{1}{c}{} \\
   \hline
  \end{tabular*}
\end{table}

\section{Analysis overview}\label{section:heavyflavour}
Measurements of electrons from heavy-flavour hadron decays were  obtained by selecting an inclusive electron sample and subtracting electrons which do not originate from heavy-flavour hadron decays. The measurements were performed by identifying electrons using the TPC and TOF detectors at low \pt ($p_{\rm T} < 4~{\rm GeV}/{c}$) and the TPC and EMCal detectors at higher \pt (\pt $>$ 3 \GeVc) offering the largest \pt reach. In particular, this ensures that the systematic uncertainties and the hadron contamination are small over the whole transverse momentum range. In the interval $3 < p_{\rm{T}} < 4$ \GeVc, where the heavy-flavour decay electron production was measured with both techniques, the TPC--TOF analysis was used for the final results, while the TPC--EMCal analysis was utilised as a consistency check.
This choice was motivated by the precision of the measurements based on the statistical and systematic uncertainties, as will be further discussed in Sec.~\ref{section:results}. Throughout the article, the term ‘electron’ is used for electrons and positrons. 

\subsection{Electron identification}\label{subsec:elecId}
Reconstructed tracks were selected based on the criteria listed in Table~\ref{Table:InclusiveTrackSelection}, which are similar to those used in the analysis described in~\cite{Acharya:2019hao, Acharya:2019mom}. These requirements were applied depending on the data sample as well as the transverse momentum region of the analysis.
The rapidity ranges used in the nominal-$B$ field TPC--TOF analysis and the TPC--EMCal analysis were limited to $|y|<$ 0.8 and $|y|<$ 0.6, respectively, to avoid the edges of the detectors, where the systematic uncertainties related to the particle identification increase. In the low-$B$ field TPC--TOF analyses for pp collisions, the rapidity interval was restricted to $|y|<$ 0.5, to ensure a stable estimation of the photonic electron background (Sec.~\ref{subsection:Nonheavyflavour}), which significantly increases in the low-$B$ field sample for small \pt~and large rapidities, resulting in a small signal over background ratio.
A charged particle passing through the TPC deposits energy inducing signals in the pad rows of the detector. The reconstructed
space points are known as clusters.
The number of crossed rows which is equivalent to the effective cluster track length is used as a criteria for selecting tracks. A threshold of a minimum of 70 out of the total 159 crossed rows of the TPC for track reconstruction and 80 clusters for particle identification were used.
 The $\chi^{2}$ of the Kalman fit of the reconstructed track in the TPC, normalised to the number of TPC clusters ($\chi^{2} / N^{\rm{cls}}_{\rm{TPC}}$), had to be smaller than 4 to select tracks with good quality and reduce the contribution from wrongly attached clusters to the reconstructed track. Only tracks with a distance of closest approach (DCA) to the primary vertex smaller than 1 cm in the transverse plane and 2 cm in the longitudinal direction were selected in order to reject background and non-primary tracks. In the TPC--TOF analyses, all tracks were required to have an associated hit in each of the two innermost layers of the ITS to reduce the background electrons from photon conversions in the material, and to reduce wrong assignations of hits in the first layer of the ITS. For the TPC--EMCal analyses, the tracks were required to have at least one hit in one of the two innermost layers of the ITS. This reduces the impact of the inactive channels in the first ITS layer in the acceptance window of the EMCal. As the photon conversion background decreases with increasing \pt, the relaxed requirement does not affect the signal over background ratio significantly in the \pt range where the TPC--EMCal analyses were performed. Moreover, it is important to note that the track selection criteria on SPD, TOF, and EMCAL detectors sufficiently suppress background tracks originating from out-of-bunch pileup.

\begin{table}[h!]
\caption{Summary of the track selection criteria imposed on the inclusive electron candidates for different data sets and electron identification strategies. Details can be found in Sec.~\ref{subsec:elecId}.}
\begin{center}
\small
	\renewcommand{\arraystretch}{2}
\begin{tabular}{c|ccc|cc}
\hline\hline
 & \multicolumn{3}{c|}{pp $\sqrt{s} = 13$ TeV} & \multicolumn{2}{c}{\pPb $\sqrt{s_{\rm NN}} = 8.16$ TeV}  \\ 
  \hline
\pt interval (\GeVc) & 0.2--4.0 & 0.5--4.0  & 3.0--35.0 & 0.5--4.0 & 3.0--26.0 \\
 
\makecell{Track selection\\ criteria} & \makecell{Low-$B$\\TPC--TOF} & \makecell{Nominal-$B$\\TPC--TOF}  & \makecell{Nominal-$B$\\TPC--EMCal} & \makecell{Nominal-$B$\\TPC--TOF} & \makecell{Nominal-$B$\\TPC--EMCal} \\ \hline\hline
$|y|$ & $<$ 0.5 & $<$ 0.8 & $<$ {0.6} & $<$ 0.8 & $<$ 0.6 \\\hline
\makecell{No. of TPC\\crossed rows} & $\geq$ 70 & $\geq$ 70  &  $\geq$  70   & $\geq$ 70  & $\geq$ 70   \\\hline
\makecell{No. of TPC d$E/$d$x$\\clusters for PID}& $\geq$ 80 & $\geq$ 80 &  $\geq$ {80} & $\geq$ 80 & $\geq$ 80\\\hline
Number of ITS hits & $\geq$ {3} & $\geq$ {3}& $\geq$ {3} & $\geq$ 3& $\geq$ 3\\\hline
$\chi^{2} / N^{\rm{cls}}_{\rm{TPC}}$ & $<$ 4 & $<$ 4 & $<$ 4 &$<$ 4&$<$ 4\\\hline
\makecell{Minimum number of\\hits in the SPD }& 2 & 2 & 1 & 2 & 1\\\hline
$|\rm DCA_{xy}|$ & $<$ 1 cm & $<$ 1 cm & $<$ 1 cm & $<$ 1 cm& $<$ 1 cm\\\hline
$|\rm DCA_{z}|$ & $<$ 2 cm & $<$ 2 cm & $<$ 2 cm & $<$ 2 cm & $<$ 2 cm\\ \hline
\end{tabular}
\label{Table:InclusiveTrackSelection}
\end{center}
\end{table}

\begin{figure}[tb!]
    \includegraphics[scale = 0.4]{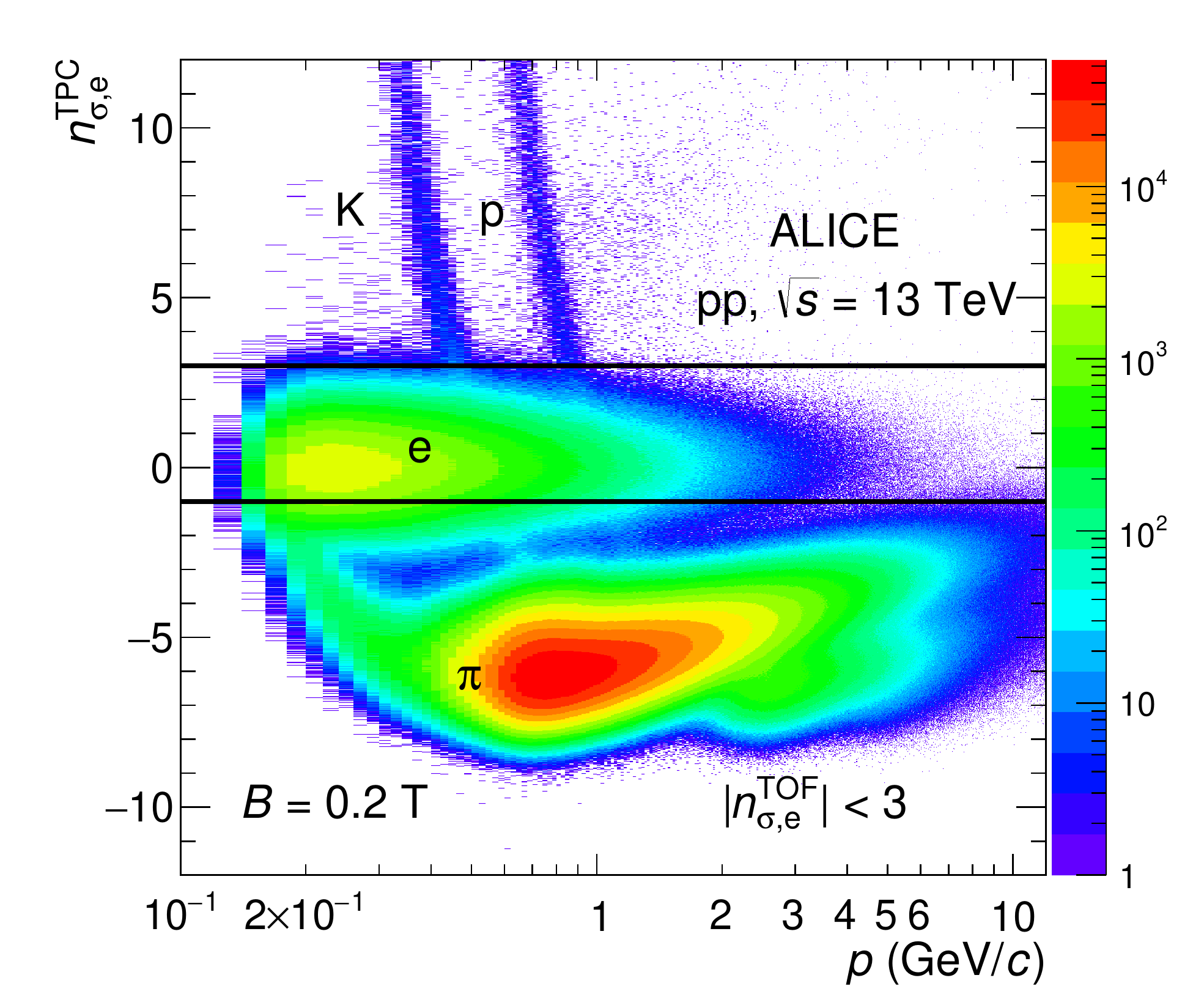}
    \includegraphics[scale = 0.4]{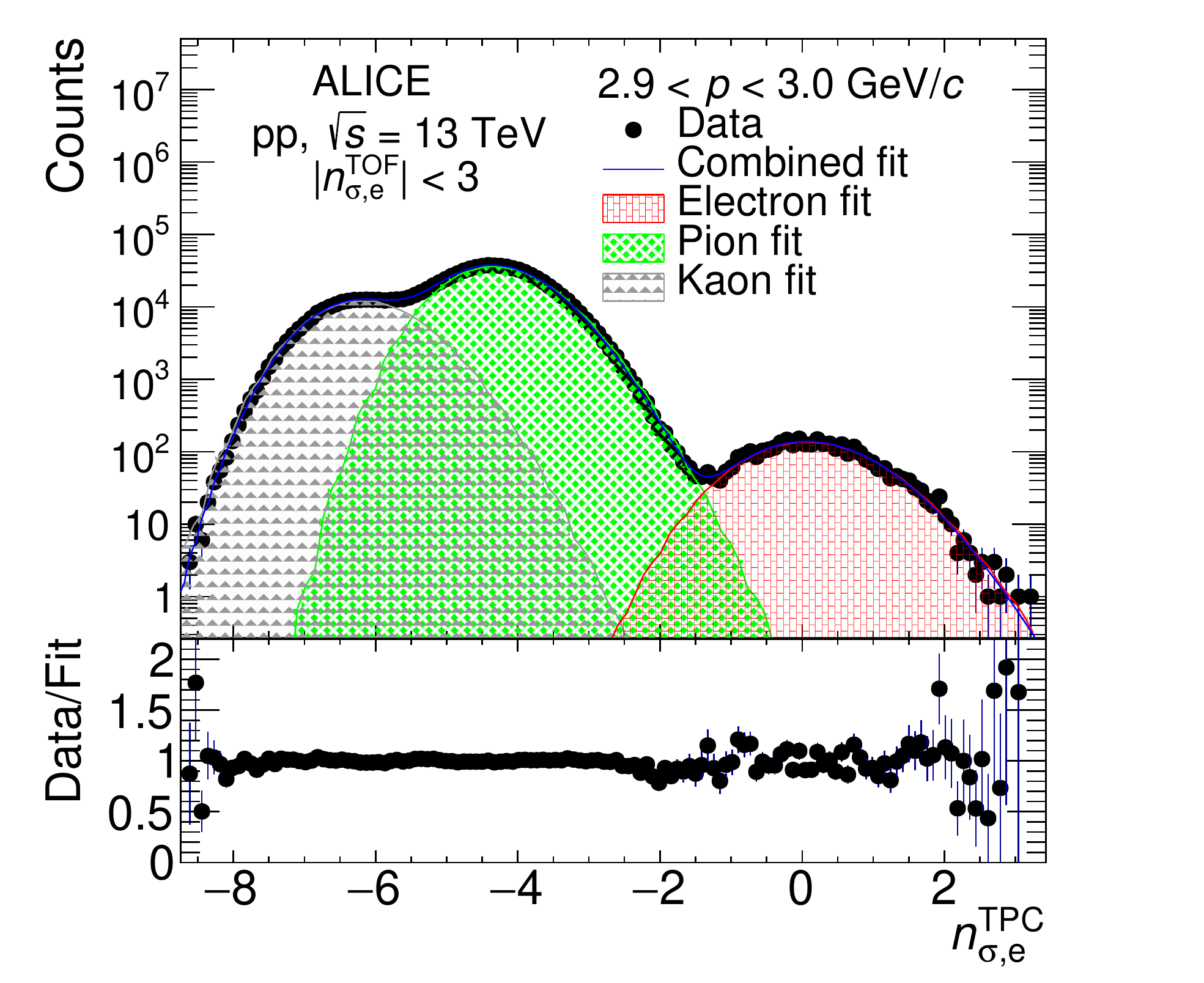}
    \caption{TPC d$E/$d$x$ signal, expressed in terms of deviation from the expected electron energy loss as a function of momentum (left panel) and fit of the measured $n^{\rm{TPC}}_{\sigma,\rm{e}}$ distribution after the $n^{\rm{TOF}}_{\sigma,\rm{e}}$ requirement in the momentum range 2.9 $< p <$ 3.0 \GeVc (right panel) in pp collisions at $\sqrt{s}$ $=$ 13 TeV for the low-$B$ field data set.}
    \label{fig:pp_13_TPCNsigma}
\end{figure}

To identify electrons at low $p_{\rm T}$ ($p_{\rm T}$ $<$ 4 GeV$/c$), the specific energy deposition (d$E/$d$x$) in the TPC and the time-of-flight measurement in the TOF detector were used. The discriminant variable used for the TPC (TOF) detector is the deviation of d$E/$d$x$ (particle time-of-flight) from the parameterised electron Bethe--Bloch (electron time-of-flight) expectation value~\cite{Bethe:1930ku}, expressed in terms of the d$E/$d$x$ (time-of-flight) resolution, $n^{\rm{TPC}}_{\sigma,\rm{e}}$ ($n^{\rm{TOF}}_{\sigma,\rm{e}}$). In the left panel of Fig.~\ref{fig:pp_13_TPCNsigma}, $n^{\rm{TPC}}_{\sigma,\rm{e}}$ is shown as a function of the momentum of the track ($p$) after TOF selection.  
For $0.2 < p_{\rm T} < 4$ GeV$/c$, electron candidates were selected by requiring $|n^{\rm{TOF}}_{\sigma,\rm{e}}|$ $<$ 3 and $-$1 $<$ $n^{\rm{TPC}}_{\sigma,\rm{e}}$ $<$ 3, resulting in a 100\% pure electron sample at \pt$\approx$ 0.2 \GeVc, and a sample with a purity of about 90\% at 4 \GeVc.
The remaining hadron contamination in the sample, after TOF selection, was estimated and subtracted by parameterising the TPC d$E/$d$x$ distribution for each particle species with an analytical function in different momentum regions as shown in the right panel of Fig.~\ref{fig:pp_13_TPCNsigma}, and as performed in previous analyses~\cite{Acharya:2019hao, Acharya:2019mom}. 

\begin{figure}[ht!]
    \centering
\includegraphics[height=7.5cm]{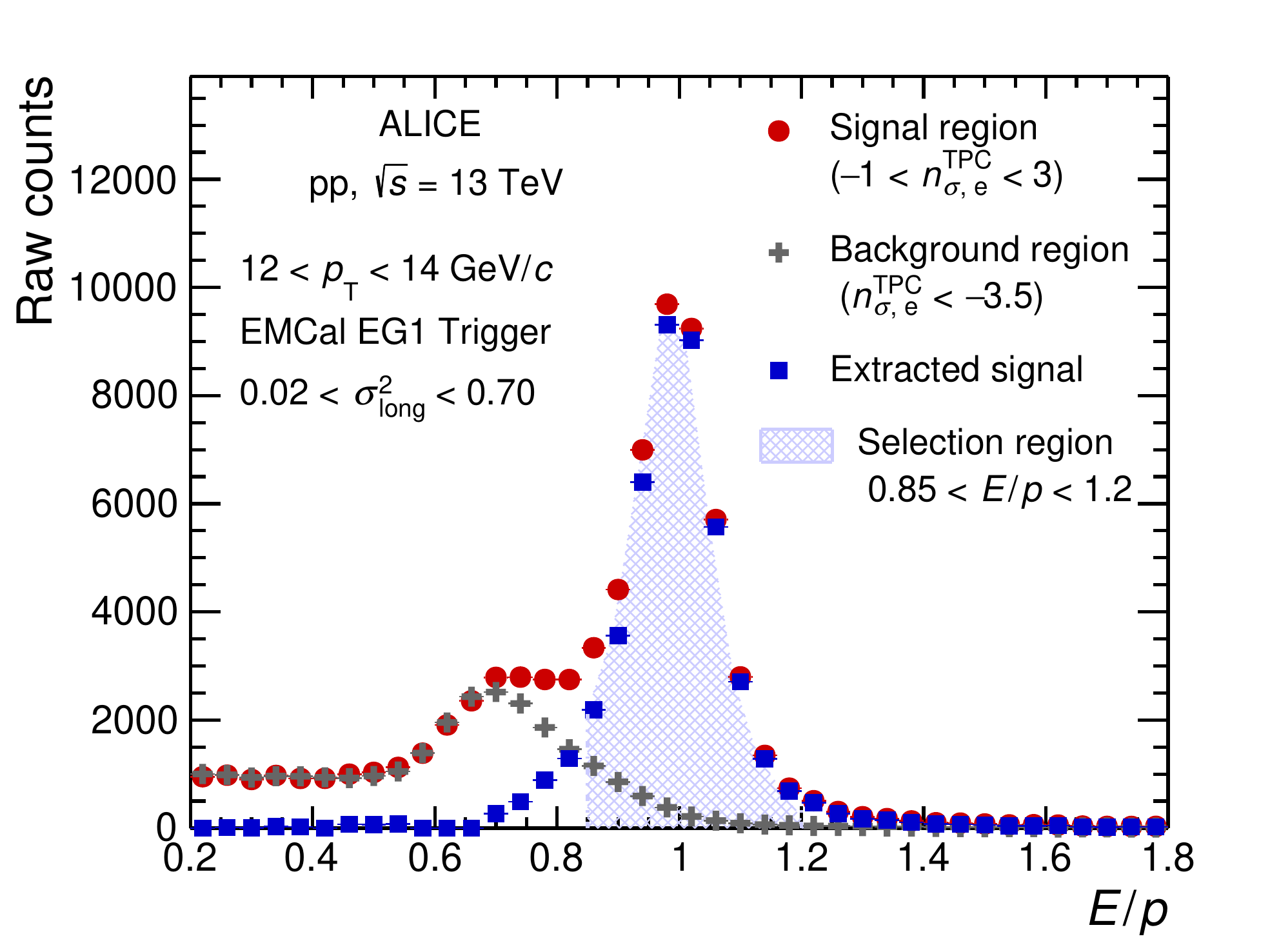}
    \caption{The $E/p$ distribution measured in pp collisions at $\sqrt{s}$ = 13 TeV for  EG1 triggered events.}
    \label{fig:pp_13_EbyP}
\end{figure}

Analyses using TPC and EMCal detectors were performed by spatially matching reconstructed charged tracks in the ITS
and TPC with EMCal clusters. This is implemented by extrapolating the reconstructed charged tracks with ITS and TPC to the EMCal, taking into account the energy loss of the particle when it traverses the detector materials, and matching within 
 the $\Delta \eta$ and $\Delta \varphi$ as given by Eq.~(\ref{eqn:deltaetaphi}). At low \pt, the position resolution of tracks and the EMCal clusters gets worse which leads to a \pt-dependent matching criteria. The \pt-dependent selection window is approximately one EMCal cell size at high \pt and few cell sizes below 1 \GeVc~\cite{ALICE:2022qhn}. This matching criterion removes the contribution from photon signals and from wrong associations of EMCal clusters to charged-particle tracks. 
\begin{align}
\label{eqn:deltaetaphi}
\begin{split}
  |\Delta \eta| &\leq 0.010 + (p_{\rm T,track} (\GeVc) +4.07)^{-2.5} ,
\\
  |\Delta \varphi| &\leq 0.015 + (p_{\rm T,track} (\GeVc) +3.65)^{-2} .
\end{split}
\end{align}
Candidate tracks matched with EMCal clusters with   $-1 < n^{\rm{TPC}}_{\sigma,\rm{e}}$ $<$ 3 were selected. Electrons were identified and separated from hadrons using the $E/p$ information, where, $E$ is the energy deposited by the particle in the EMCal detector and $p$ is the momentum of the track. It was required that the measured $E/p$ is around unity, $0.85 < E/p < 1.2$, as expected for electrons, while hadrons have lower $E/p$ values.
To further reduce the amount of hadron contamination, a condition on the shape of the electromagnetic shower, $\sigma_{\rm{long}}^{2}$,~\cite{Alessandro:2006yt} was applied. 
The quantity $\sigma_{\rm{long}}^{2}$ stands for the eigenvalues of the dispersion matrix of the shower shape ellipse defined by the energy distribution within the EMCal cluster~\cite{Awes:1992yp, Acharya:2017hyu}.  A \pt-dependent selection criterion was applied, $0.02 <  \sigma_{\rm{long}}^{2} < 0.9$ at low \pt and a more stringent selection up to $0.02 <\sigma_{\rm{long}}^{2} < 0.5$ at higher \pt, in both pp and p--Pb collisions.
The lower threshold on $\sigma_{\rm{long}}^{2}$ removes contamination caused by neutrons hitting the readout electronics. The remaining hadron contamination in the electron sample was estimated by fitting the measured $E/p$ distributions of electron candidates in momentum slices. For this purpose, the shape of the $E/p$ spectrum for hadrons was obtained by selecting hadrons in the TPC with $n^{\rm{TPC}}_{\sigma,\rm{e}} < -3.5 $. The obtained hadron $E/p$ distribution was then scaled to match the $E/p$ distribution of electron candidates in a region within $E/p< 0.7$, as shown in Fig.~\ref{fig:pp_13_EbyP}. The electron yield was calculated by integrating the $E/p$ distributions of electron candidates in the range $0.85 < { E}/{p} < 1.2$ after the subtraction of the hadron contamination. In the pp (\pPb) analysis, the hadron contamination was negligible at low $p_{\rm{T}}$, increasing up to $23\%$ ($25\%$) at $p_{\rm{T}} $ $=$ 35 (26) GeV$/c$.

\subsection{Subtraction of electrons from non heavy-flavour sources}\label{subsection:Nonheavyflavour}
The selected inclusive electron sample contains electrons from open heavy-flavour hadron decays and from  different sources of background:

\begin{itemize}
    \item dielectrons originating from Dalitz decays of light-neutral mesons such as $\pi^{0}$, $\eta$ as well as conversions of photons in the detector material, named as photonic electrons in the text,
    \item dielectrons from decays of J$/\psi$ (J$/\psi$ $\rightarrow$ $\rm e^{+}e^{-}$) and low-mass vector mesons ($\rho$ $\rightarrow$ $\rm e^{+}e^{-}$, $\omega$ $\rightarrow$ $\rm e^{+}e^{-}$, $\phi$ $\rightarrow$ $\rm e^{+}e^{-}$),
\item electrons from kaon weak decays $\rm K^{0,\pm}$ $\rightarrow$ $\rm e^{\pm}$ $\rm \pi^{\mp,0}$ $\rm \overset{(-)}{\nu_{e}}$ ($\rm K_{e3}$),
\item electrons from W and Z decays.
\end{itemize}

The dominant sources of background electrons are photon conversions in the detector material and Dalitz decays of light-neutral mesons. These contributions were estimated using an invariant mass technique~\cite{Adam:2015qda} of electron--positron pairs. Unlike-signed electron--positron pairs (ULS) were defined by pairing the selected electrons with opposite-charge electron partners.
To increase the efficiency of finding the partner, associated electrons were selected applying similar but looser track quality and particle identification criteria than those used for selecting signal electrons. The selection criteria are summarised in Table~\ref{Table:AssocatedTrackSelection}. The electron--positron pairs from photonic background have a small invariant mass ($ m_{\rm e^{+}e^{-}}$). Heavy-flavour decay electrons can form
ULS pairs mainly through random combinations with other electrons.
The combinatorial contribution was estimated from the invariant mass distribution of like-signed electron (LS) pairs. The photonic background contribution was then evaluated by subtracting the LS distribution from the ULS one in the invariant mass region $m_{\rm e^{+}e^{-}} < 0.14$ GeV$/c$. The efficiency of finding the partner electron, called tagging efficiency ($\epsilon_{\rm tag}$) from hereon, was estimated using MC simulations. In the pp and \pPb analyses, the MC sample was obtained using PYTHIA 6~\cite{Sjostrand:2006za} and HIJING~\cite{Wang:1991hta} generators, respectively. The generated particles were propagated through the ALICE apparatus using GEANT 3~\cite{Brun:1073159}. In order to increase the statistical precision of
$\epsilon_{\rm tag}$ using the invariant mass method, $\pi^0$ and $\eta$ mesons were embedded in the simulated events.
The simulated $\pi^0$ and $\eta$ \pt distributions
were reweighted to match the measured spectra. For pp collisions, the $\pi^0$ spectrum was estimated as the average of the spectra of $\pi^{+}$ and $\pi^{-}$~\cite{Acharya:2020uxl}, whereas the $\eta$ spectrum was obtained using $m_{\rm T}$ scaling, as in~\cite{Gatoff:1992cv, Khandai:2011cf, Altenkamper:2017qot}. For the \pPb analysis, the measured transverse momentum spectra of $\pi^{0}$ and $\eta$ were used~\cite{alicecollaboration2021nuclear}. 
In the pp analysis, the tagging efficiency at low $p_{\rm{T}}$ ($p_{\rm{T}} < 1.0$ GeV$/c$) is 55--65\% for the low-$B$ field data set, whereas for the nominal-$B$ field data set it is around 45--55\% at low $p_{\rm{T}}$ ($p_{\rm{T}} < 1.0$ GeV$/c$) increasing to about 85\% at high $p_{\rm{T}}$ ($p_{\rm{T}} > 15$ GeV$/c$). In the \pPb analysis, the tagging efficiency  varies between 40\% at low \pt ($p_{\rm{T}} < 3.0$ GeV$/c$) and 80\% at high \pt ($p_{\rm{T}} > 7.0$ GeV$/c$).

\begin{table}[h]
\caption{Summary of the track selection criteria imposed on the associated electron candidates for different data sets and electron identification strategies.}

\centering
\small
	\renewcommand{\arraystretch}{2}
\begin{tabular}{c |c c c | c c}
\hline\hline
 & \multicolumn{3}{c|}{pp $\sqrt{s} = 13$ TeV} & \multicolumn{2}{c}{\pPb $\sqrt{s_{\rm NN}} = 8.16$ TeV}  \\ 
  \hline
 \pt interval (\GeVc) & 0.2--4.0 & 0.5--4.0  & 3.0--35.0 & 0.5--4.0 & 3.0--26.0 \\ 
\makecell{Track and PID\\cuts}  & \makecell{Low-$B$\\TPC--TOF} & \makecell{Nominal-$B$\\TPC--TOF}  & \makecell{Nominal-$B$\\TPC--EMCal} & \makecell{Nominal-$B$\\TPC--TOF} & \makecell{Nominal-$B$\\TPC--EMCal} \\
\hline \hline
$\pt^{\rm min}$ & 0.0 GeV/c   &{ 0.1 GeV/c}&   { 0.1 GeV/c}& 0.1 GeV/c &  0.1 GeV/c\\\hline
$|\textrm{\it y}|$ & $<$ 0.8  &  { $<$ 0.9 } &  { $<$0.9 } & $<$ 0.8 & $<$ 0.8 \\\hline
\makecell{No. of TPC\\d$E/$d$x$ clusters for PID} & $\geq$ 60  &$\geq$ 60 & $\geq$ 60 & $\geq$ 60 & $\geq$ 60\\\hline
Number of ITS hits & $\geq$ 2 & $\geq$ 2& $\geq$ 2 & $\geq$ 2& $\geq$ 2\\\hline
$\chi^{2} / N^{\rm{cls}}_{\rm{TPC}}$ & $<$ 4 &$<$ 4 & $<$ 4 &$<$ 4&$<$ 4\\ \hline
$|\rm DCA_{xy}|$ & $<$ 1 cm &$<$ 1 cm  & $<$ 1 cm & $<$ 1 cm& $<$ 1 cm\\ \hline
$|\rm DCA_{z}|$ & $<$ 2 cm & $<$ 2 cm & $<$ 2 cm & $<$ 2 cm & $<$ 2 cm\\ \hline
\hline
\end{tabular}
\label{Table:AssocatedTrackSelection}
\end{table}

{Due to the requirement of hits in the SPD layers, the contribution of electrons from $\rm K_{e3}$ decays was found to be negligible with respect to the heavy-flavour signal for $p_{\rm{T}} > 0.5$ GeV$/c$. At lower $p_{\rm{T}}$, the relative contribution of electrons from $\rm K_{e3}$ decays becomes non-negligible and hence it was subtracted from the  \pt-differential cross section of electrons from heavy-flavour hadron decays.} 
The $\rm K_{e3}$ contribution was estimated using a parameterisation of the ratio of $\rm K_{e3}$ to photonic electrons obtained from previous analyses using the so-called cocktail approach~\cite{Acharya:2018upq,Abelev:2014gla,ALICE:2012mzy}. The same parametrisation was used in the analysis of pp collisions at $\sqrt{s}=13~\rm{TeV}$ and in p–Pb collisions at $\sqrt{s_{\rm NN}} = 8.16$ TeV. 

Other background contributions of $\rm e^{+}e^{-}$ pairs from J/$\psi$ and low-mass vector mesons were negligible~\cite{Acharya:2018upq, ALICE:2020try} compared to the signal and were therefore not subtracted. Electrons from $\rm{W}^{\pm}$ and $\rm{Z}^0$ boson decays form a significant background at high $p_{\rm{T}}$ ($p_{\rm{T}}>20$ GeV$/c$), which was estimated with the next-to-leading order event generator POWHEG~\cite{Oleari:2010nx}, interfaced with PYTHIA as a decayer, and subtracted from the \pt-differential cross section of electrons from heavy-flavour hadron decays. This contribution increases from 1\% at $p_{\rm{T}} = 15$ GeV$/c$ to
about {3\%} at $p_{\rm{T}} = 20$ GeV$/c$ and up to 25\% at $p_{\rm{T}} = 35$ GeV$/c$ with respect to the heavy-flavour decay electron yield in both pp and \pPb collisions.

In pp collisions, the ratio of signal over background electrons is about 0.08 at \pt $=$ 0.2 GeV$/c$,  3 at $\mbox{\pt $=$ 0.5  GeV$/c$}$ increasing to $\sim$ 10.5 at \pt $=$ 25 \GeVc and reaches 12.3 at \pt $=$ 35 GeV$/c$, whereas in \pPb collisions it is about 2.65 at \pt $=$ 0.5 GeV$/c$ and increases up to 8.39  at \pt $=$ 26 GeV$/c$. 

\subsection{EMCal trigger rejection factor} \label{ssec:TriggerRejectionFactor}
\begin{figure}[h!]
    \centering
    \includegraphics[width=0.48\linewidth]{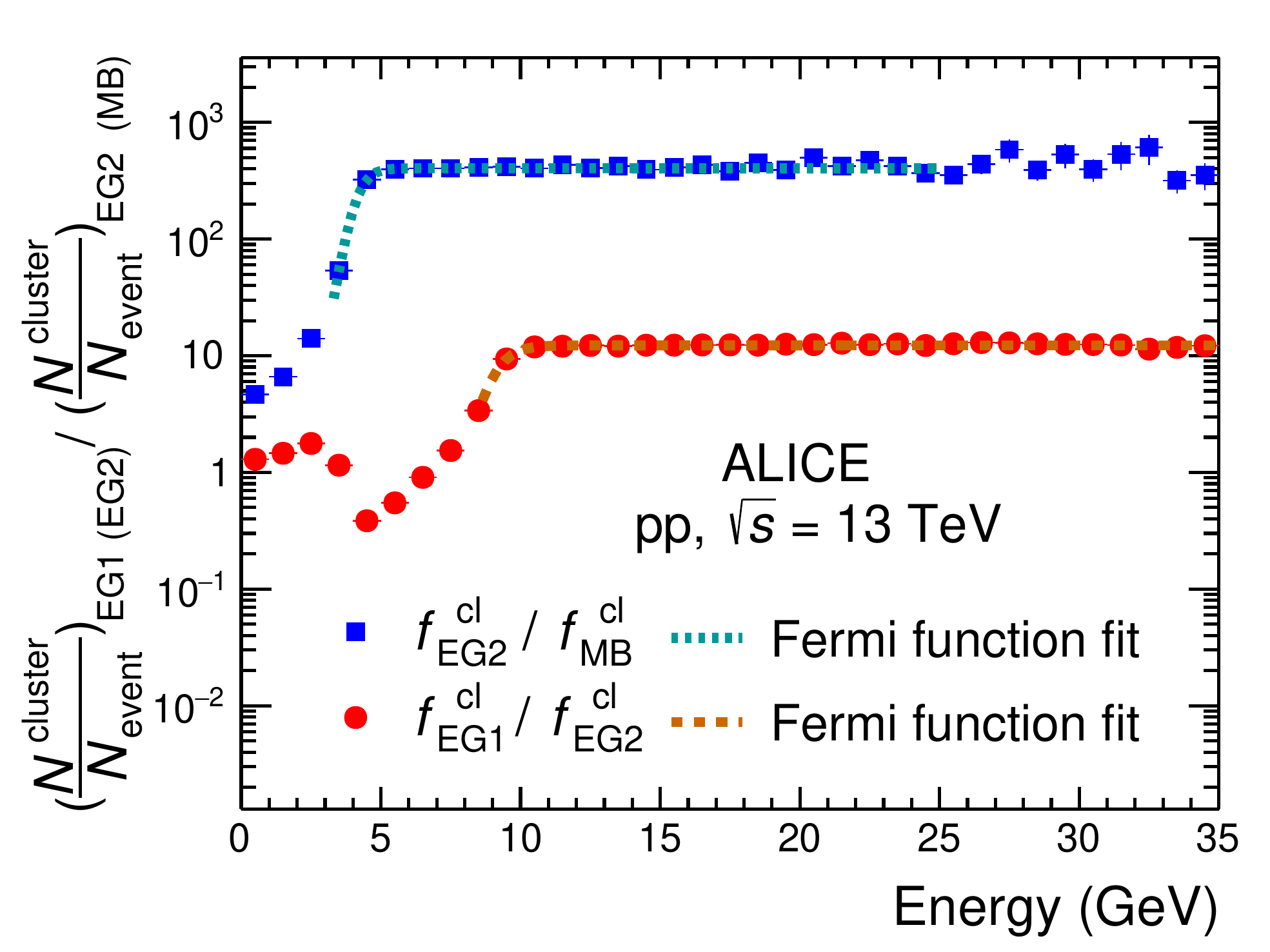}
    \includegraphics[width=0.48\linewidth]{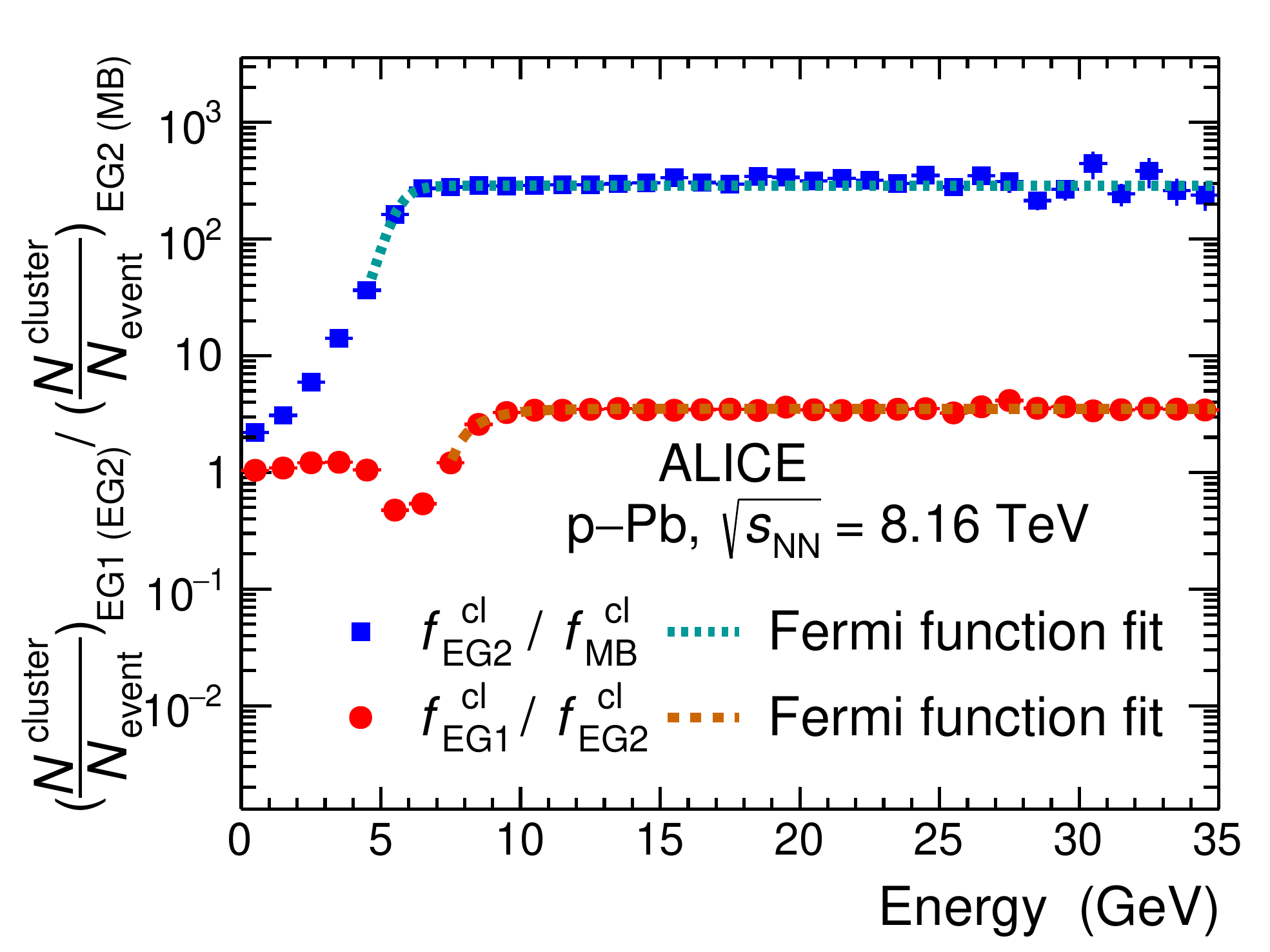}
    \caption{Trigger RF for EG2 and EG1 triggers in pp collisions at $\sqrt{s}=13$ TeV (left panel) and in \pPb collisions at $\sqrtsNN = 8.16$ TeV (right panel).}
   \label{fig:RF}
\end{figure}

The EMCal triggered events are reported in terms of $N_{\rm evt}\times \rm RF$, where $N_{\rm evt}$ is the number of triggered collisions and the Rejection Factor (RF) is the average number of rejected MB events per EMCal triggered event. 
The RFs were estimated with a data-driven method. For the EG2 trigger the RF was calculated as the ratio of the cluster energy distribution in EG2 triggered data to the one in MB triggered data ($f^{\rm{cl}}_{\rm{EG2}}/f^{\rm{cl}}_{\rm{MB}}$), which gives the EG2 turn-on curve. In order to reduce the effect of  poor statistics in the MB sample at high \pt, the EG1 trigger turn-on curve was obtained using the ratio of the EG1 triggered data cluster energy distribution to the one in EG2 triggered data ($f^{\rm{cl}}_{\rm{EG1}}/f^{\rm{cl}}_{\rm{EG2}}$). The RF of EG1 trigger is then the product of $f^{\rm{cl}}_{\rm{EG2}}/f^{\rm{cl}}_{\rm{MB}}$ and $f^{\rm{cl}}_{\rm{EG1}}/f^{\rm{cl}}_{\rm{EG2}}$.
The turn-on curve was determined for the multiplicity-integrated interval and for different multiplicity intervals in pp and \pPb collisions.
The turn-on curves are shown for both trigger energy thresholds (EG1 and EG2) in multiplicity integrated pp and \pPb collisions in Fig.~\ref{fig:RF}.

A Fermi function~\cite{Fermi:1934hr, Wilson:1968pwx} was used to fit the trigger turn-on curves and determine the RF above the trigger threshold. 
The fit range is from the beginning of the turn-on region i.e. near trigger threshold to the highest energy where the distribution remains flat.
The EG2/MB and the EG1/EG2 values correspond to the constant values determined by the plateau of the fitted Fermi function above the trigger thresholds. 
The final values of the RF used in the analyses are summarised in Table ~\ref{table:RF}.
The systematic uncertainty on the trigger RF was estimated by varying the fit region of the trigger turn-on curve and the fit function, i.e. using a linear function in the plateau region. A systematic uncertainty of 3\% (2\%) for the EG2 trigger and 4\% (3\%) for EG1 trigger was obtained for pp (\pPb) collisions.

\begin{table*}[tb!]
\caption{Multiplicity-integrated values of the EMCal trigger RF with their uncertainties for the EG2 and EG1 triggered data sets in pp and \pPb collisions.}
    \centering
    	\renewcommand{\arraystretch}{1.2}
     \resizebox{\textwidth}{!}{%
    \begin{tabular}{c| c c | c c}
    \hline
    \multicolumn{1}{c|}{}&
    \multicolumn{2}{c|}{pp $\sqrt{s} = 13\ \TeV$}& \multicolumn{2}{c}{\pPb $\sqrt{s_{\rm NN}} = 8.16\ \TeV$}\\
    Trigger   & EG2 & EG1 & EG2 & EG1 \\
    \hline
     RF value & { 406  } & { 4985  } &  283.4 & 1020.7 \\
       & {  $\pm$ 1 (stat.) $\pm$  12 (syst.) } & {  $\pm$ 11 (stat.) $\pm$ 200 (syst.) } &  $\pm$ 1.5 (stat.) $\pm$ 5.9 (syst.) & $\pm$ 3.6 (stat.)  $\pm$ 30.4 (syst.) \\
     \hline
    \end{tabular}}
    \label{table:RF}
\end{table*}

\subsection{Efficiency correction and normalisation}\label{section:corrections}
The raw number of electrons and positrons from heavy-flavour hadron decays, $N_{\rm raw}$, was obtained by subtracting the hadron contamination, photonic electrons, and the other background electron contributions. The $\mbox{\pt-differential}$ cross section of electrons from heavy-flavour hadron decays at midrapidity was then calculated using the formula
 \begin{equation}
 \frac{{\rm d}^2\sigma}{{\rm d} p_{\rm T}{\rm d} y} = \frac{1}{2}  \frac{1}{\Delta y \Delta p_{\rm T}} \frac{{N_{\rm{raw}}}} {\epsilon^{\rm{geo}} \times \epsilon^{\rm reco} \times  \epsilon^{\rm {eID}} } 
 \frac{1}{\mathcal{L}_{\text{int}}}
\end{equation}

, where $\mathcal{L}_{\rm int}$ is the integrated luminosity, and $\Delta p_{\rm T}$ and $\Delta y$ the width of the $p_{\rm T}$ and rapidity intervals, respectively. The integrated luminosity was calculated using the number of analysed events and the measured MB trigger cross sections ($\sigma_{\rm MB}$), as $N_{\rm evt}/\sigma_{\rm MB}$ for minimum bias triggered events and {$\mbox{$N_{\rm evt}\times \rm{RF}/\sigma_{\rm MB}$}$} for EMCal triggered events, and are listed in Table~\ref{table:EventStat}. The trigger bias was studied in MC simulations and was found to be negligible for the selections applied to tracks and clusters. The measured $\sigma_{\rm MB}$ values are 58.44 $\pm$ 1.11 mb, 58.10  $\pm$  1.57 mb, and 57.52 $\pm $ 1.21 mb for the pp collisions at $\sqrt{s}$ $=13~\rm TeV$ collected in the years 2016, 2017, and 2018, respectively~\cite{ALICE-PUBLIC-2021-005}, and 2100 $\pm$ 60 mb~\cite{ALICE-PUBLIC-2018-002} in \pPb collisions at $\sqrtsNN$ $=$ 8.16 TeV. The raw number of electrons from heavy-flavour hadron decays was corrected for the geometrical acceptance ($\epsilon^{\rm geo}$), the track reconstruction ($\epsilon^{\rm reco}$), and electron identification ($\epsilon^{\rm eID}$) efficiencies. The factor of two accounts for the charged averaged contribution of electrons and positrons. The trigger and event selection criteria were found to be fully efficient for electrons from heavy-flavour hadron decays.

The above mentioned acceptance and track reconstruction efficiencies are computed by means of MC simulations using PYTHIA 6~\cite{Sjostrand:2006za} and HIJING~\cite{Wang:1991hta} event generators for pp and \pPb collisions, respectively. For pp simulations PYTHIA 6 generated events with at least one ${\rm c{\overline{c}}}$ or ${\rm b{\overline{b}}}$ pair were selected for propagation through the apparatus with GEANT 3~\cite{Brun:1073159} and subsequent reconstruction. In the case of \pPb collisions, to have an efficient generation of heavy-flavour signals and reproduce the detector occupancy, one PYTHIA 6 event with a ${\rm c{\overline{c}}}$ or ${\rm b{\overline{b}}}$ pair was embedded in each HIJING simulated event.

The electron identification (eID) efficiencies for the TOF, TPC, and EMCal detectors were obtained separately, and then multiplied according to the detectors used in the analysis to compute the full electron identification efficiency $\varepsilon^{\rm eID}$. The TPC--TOF track matching and  electron identification efficiency of the TOF detector was calculated with the above mentioned MC sample and was found to be 60--70$\%$ (40--65$\%$) for  0.5 $<$ $p_{\rm T}$ $<$ 1.5 GeV$/c$ increasing up to 75$\%$ (70$\%$) at 4 GeV$/c$ in the low (nominal)-$B$ field analysis. A better track matching between the TPC and the TOF detectors is achieved at a given \pt with the low-$B$ field compared to the nominal-$B$ field, due to the smaller curvature of the tracks. Therefore, a higher reconstruction efficiency is observed in the low-$B$ field data compared to the nominal-$B$ field sample. The TPC  electron identification efficiency was determined using a data-driven approach based on the $n^{\rm{TPC}}_{\sigma,\rm{e}}$ distribution~\cite{ALICE:2012mzy}. It is about 88\% at $p_{\rm T}= 0.2$ GeV$/c$ and increases to $89\%$ for $\mbox{$p_{\rm T} > 0.5$ GeV$/c$}$, for the low-$B$ field data sample.

\begin{figure}[tb!]
 \begin{center}
      \includegraphics[width=0.48\linewidth]{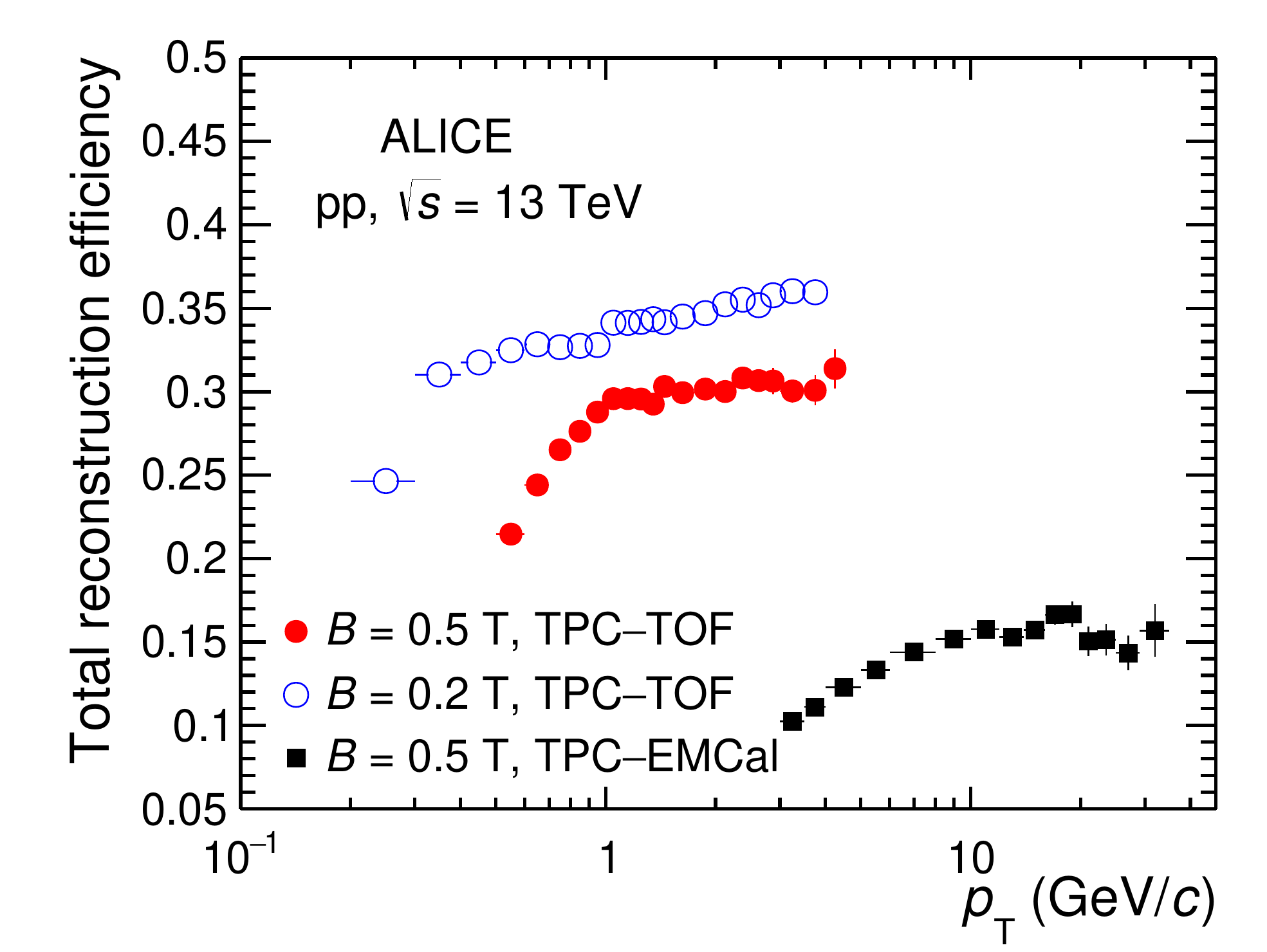}
      \includegraphics[width=0.48\linewidth]{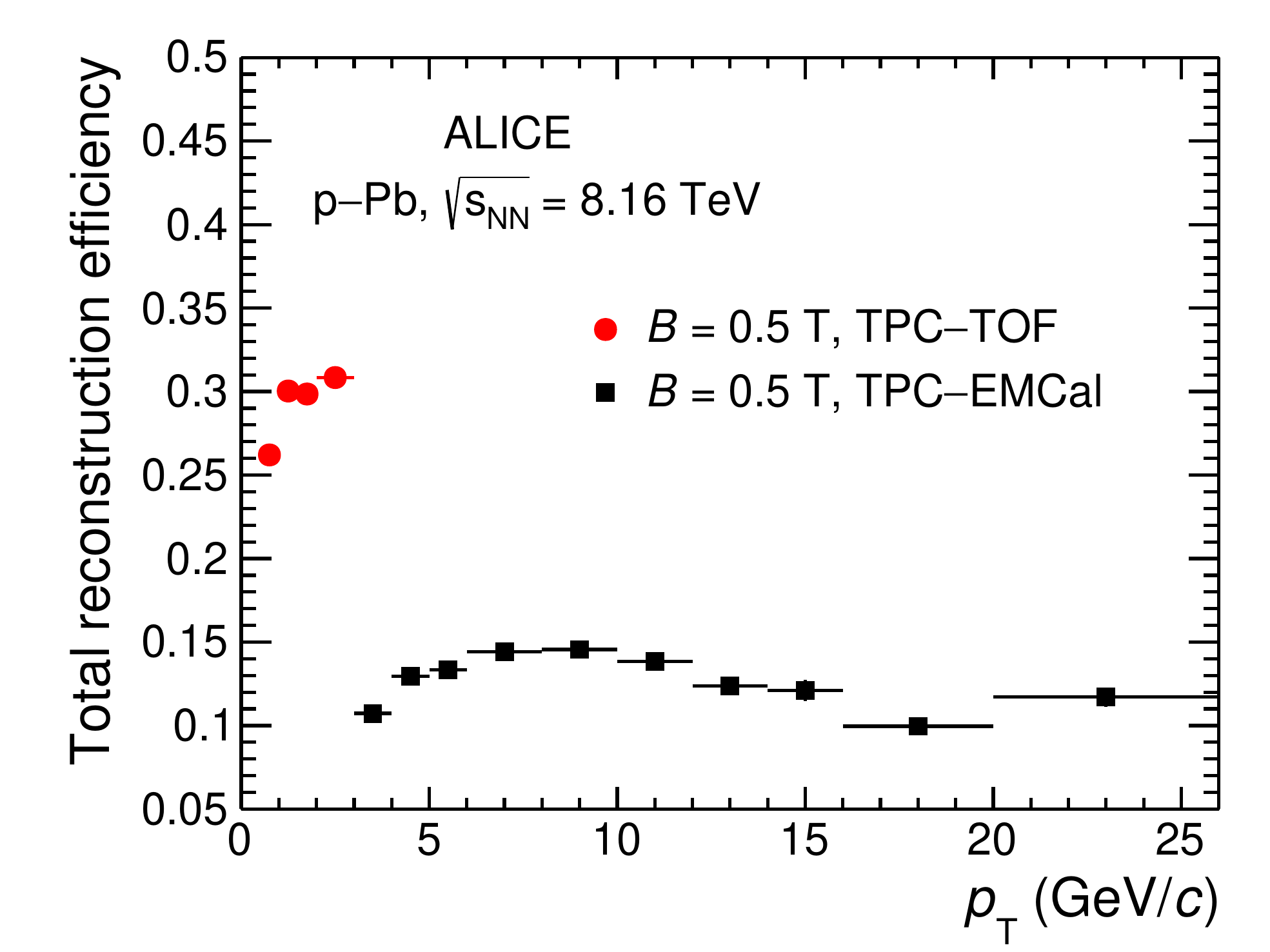}
      \end{center}

\caption{Total reconstruction efficiency of electrons from heavy-flavour hadron decays using the TPC and TOF or the TPC and EMCal detectors in pp collisions at $\sqrt{s}= 13$ TeV with nominal and low magnetic field (left panel) and in \pPb collisions at $\sqrtsNN=8.16$ TeV (right panel).}        
\label{Fig:ppHFEEff}
\end{figure}

In the nominal-$B$ field data set, it was found to be around $86\%$ at $p_{\rm T}= 0.5$ GeV$/c$ increasing to $88\%$ for $p_{\rm T} > 4$ GeV$/c$. 
The  electron identification efficiency with EMCal was estimated using MC simulations and was found to be about 60\% at 3 GeV$/c$, increasing up to 80\% for \pt larger than 10 GeV/$c$ for pp collisions. 
The total reconstruction efficiency ($\epsilon^{\rm{geo}} \times \epsilon^{\rm reco} \times  \epsilon^{\rm {eID}}$) for different data sets and with different detectors is presented in Fig.~\ref{Fig:ppHFEEff}. 

The production of electrons from heavy-flavour hadron decays was further studied as a function of the charged-particle pseudorapidity density in pp or \pPb collisions using the self-normalised yield of heavy-flavour hadron decay electrons. The  differential yield measured in a given multiplicity class was divided by its average over all INEL $>$ 0 events $({\rm d}^{2} {N}/{\rm d}p_{\rm T} {\rm d}{\eta} / \langle {\rm d}^{2}{N}/{\rm d}p_{\rm T}{\rm d}{\eta}\rangle_{\rm INEL>0})$ in pp or \pPb collisions.  All efficiency were obtained as a function of multiplicity. At low \pt, the tagging efficiency
and the total reconstruction efficiency of electrons from heavy-flavour hadron decays were observed to be multiplicity dependent, while no dependencies were seen at high \pt, so the efficiencies cancelled out in the self-normalised ratios at high \pt. 

\section{Systematic uncertainties}\label{sec:systematics}
The systematic uncertainties on the measured cross sections in pp and \pPb collisions were obtained separately for the different $p_{\rm{T}}$ intervals and for the different analyses performed using the TPC--TOF and TPC--EMCal detector combinations. 
For the self-normalised yield measurements, the systematic uncertainties were estimated directly on the self-normalised yield for each multiplicity class and $p_{\rm{T}}$ interval. The different sources of systematic uncertainties are discussed in this section and the assigned values are summarised in Tables~\ref{tab:SystematicSummaryTPCTOC}, ~\ref{tab:SystematicSummaryTPCEMCal}, and ~\ref{tab:summarysystSN}.

\begin{table}[th!]
 \centering
 \caption{Sources of systematic uncertainties and their assigned values in pp collisions at $\sqrt{s} = 13~\rm TeV$ for {$\mbox{$B = 0.2$ T}$} (0.2 $<p_{\rm T}$ $<$ 0.5 \GeVc) and $B = 0.5$ T (0.5 $<p_{\rm T}$ $<$ 4 \GeVc) data sets with the TPC and TOF  detectors, as well as with the TPC and EMCal detectors  (4 $<p_{\rm T}$ $<$ 35 \GeVc). The values presented as a range correspond to the lowest- and highest-\pt intervals. } 
 \renewcommand{\arraystretch}{1.5}
\begin{tabular*}{\textwidth}{@{\extracolsep{\fill}}  c| ccc}
   \hline
    \multirow{2}{*}{\makecell{Sources of systematic\\ uncertainties}} & { 0.2 $<p_{\rm T}$  $<$ 0.5 \GeVc} & { 0.5 $<p_{\rm T}$ $<$ 4 \GeVc} & { 4 $<p_{\rm T}$ $<$ 35 \GeVc}\\
 & \makecell{low-$B$\\TPC--TOF} & \makecell{nominal-$B$\\TPC--TOF} & \makecell{ nominal-$B$\\TPC--EMCal}\\

\hline
\hline
{Track selection } & {negl.} & 1$\%$ & 3$\%$ \\
\hline
{TPC--TOF matching} & { 4$\%$--2$\%$} & 2\% & N/A\\\hline  
{TPC--ITS matching } & { 2\%} & 3\%  &  3\%\\ \hline
{TPC--EMCal matching } & { N/A} & N/A  &   negl.\\ \hline
{SPD hit requirement } & {25$\%$--15$\%$} &  10\%--3\% & { 5\%--negl. }\\\hline
{Electron identification} & {5$\%$} &  {5$\%$--negl.}& 6\%--12\% \\\hline
{Hadron contamination } & {negl.} & negl.--2\%  &  negl.--7\%  \\\hline 
\makecell{Photonic electron subtraction} & {20$\%$--11$\%$} & 7\%--1\%  &  negl.\\\hline  

{$\rm K_{e3}$ subtraction } & {15\%--1$\%$} & N/A & N/A \\\hline
 {$\rm{W}^{\pm}/\rm{Z}^0 \rightarrow$ e } & {N/A} &N/A &  negl.--8\%\\ \hline
 {$\pi^{0}$, $\eta$ weights } & { 3\%--1$\%$} & negl. &  negl.\\
\hline
{RF } & {N/A} & N/A & 3\%--4\% \\ \hline
{Luminosity } & {2.3\%} & {2.3\%} & {2.3\%--5\%} \\ \hline
\hline
{Total systematic} & {36\%--20$\%$} & {14\%--7\%} & {11\%--18\%}\\
\hline
  \end{tabular*}
   \label{tab:SystematicSummaryTPCTOC}
\end{table}

\begin{table}[h!]
 \centering
 \caption{Sources of systematic uncertainties and their assigned values in \pPb collisions at {$\mbox{ $\sqrtsNN =8.16$ TeV}$} with TPC and TOF detectors (0.5 $<p_{\rm T}$ $<$ 4 \GeVc), as well as with the TPC and EMCal detectors  
 {$\mbox{(4 $<$ \pt $<$ 26 \GeVc)}$}. The values presented as a range correspond to the lowest- and highest-\pt intervals.}
  \renewcommand{\arraystretch}{1.4}
\begin{tabular*}{\textwidth}{@{\extracolsep{\fill}}  c| cc}
    \hline
   \multirow{2}{*}{\makecell{Sources of systematic\\ uncertainties}} & { 0.5 $<p_{\rm T}$ $<$ 4 \GeVc} & { 4 $<p_{\rm T}$ $<$ 26 \GeVc}\\
    & \makecell{nominal-$B$\\TPC--TOF} & \makecell{nominal-$B$\\TPC--EMCal} \\
\hline
\hline
{Track selection } & 1$\%$ & 5\%--1\% \\
\hline
{TPC--TOF matching} & 2\% & N/A \\
\hline  
{TPC--EMCal matching } & { N/A}  &  negl.\\ \hline
{TPC--ITS matching }&2\%&2\% \\ 
\hline
 {SPD hit requirement } & 10\%--3\% & 5\%--negl.\\
\hline
{Electron identification} & 3\%--1\% & 1\%--5\%\\
\hline 
{Hadron contamination } & negl. & negl.--5\% \\
\hline 
{Photonic electron subtraction} & 7\%--1\% &  negl. \\
 \hline  
 {$\rm K_{e3}$ subtraction }&N/A &N/A \\
\hline
{$\rm{W}^{\pm}/\rm{Z}^0 \rightarrow$ e } & N/A &  negl.--1\%\\ \hline
{$\pi^{0}$, $\eta$ weights } & negl. & negl. \\ \hline
{RF }& N/A & 2\%--4\% 
\\ \hline
{Luminosity } & {3\%} & {3\%} 
\\ 
\hline
\hline
{Total systematic}& 13$\%$--5$\%$ & 8$\%$--9$\%$ \\
\hline
  \end{tabular*}
   \label{tab:SystematicSummaryTPCEMCal}
\end{table}

The systematic uncertainty on the track reconstruction and selection efficiency was obtained by multiple variations of the track selection criteria, namely, the minimum number of space points in the TPC, the number of TPC crossed rows, the number of TPC d$E/$d$x$ clusters and the number of hits in the ITS.

The uncertainty due to an imperfect description in the simulation of the TPC--TOF (and the TPC--ITS) track matching was estimated by calculating the difference between efficiencies of the TPC--TOF (and the TPC--ITS) track matching in data and MC. To obtain the matching efficiency, the abundances of primary and secondary particles in data were estimated via template fits to the track impact-parameter distributions, and the relative abundances in the simulation were weighted to match those in data~\cite{ALICE:2017olh, ALICE-PUBLIC-2017-005}. For the low-$B$ field sample in pp collisions, the uncertainty on the track matching between the ITS and TPC is 2$\%$ at $\pt = 0.2$ \GeVc increasing up to $4\%$ at 4 \GeVc, whereas, the uncertainty on the track matching between the TPC and the TOF detector is $4\%$ at $\pt = 0.2$ \GeVc and $2\%$ at 4 \GeVc. In case of the nominal-$B$ field data set,  the uncertainty is about 2$\%$ for the TPC--TOF track matching and about 3$\%$ for the TPC--ITS track matching in the whole \pt range. In the \pPb analysis, the uncertainty for the TPC--TOF track matching, as well as the TPC--ITS track matching, was found to be  around 2$\%$ in the whole $\pt$ range.

The systematic uncertainty on the SPD hit requirement was obtained by varying the condition on the minimum number of hits and the specific layer of the SPD on which a hit was required for both the TPC--TOF and the TPC--EMCal analyses. For the low-$B$ field sample in pp collisions,  the systematic uncertainty due to the SPD hit requirement was about 25\% at \pt $= 0.2$ \GeVc decreasing to 15\% at $\mbox{0.5 \GeVc}$, whereas for nominal-$B$ field data sets in pp and \pPb collisions, the uncertainty is about 10\% at $\mbox{\pt $= 0.5$ \GeVc}$ decreasing to 5\% at 4 GeV/c and becomes negligible in the highest-\pt interval.

The uncertainty on the procedure of track matching to EMCal clusters was obtained by varying the $\Delta \eta - \Delta \varphi$ selection using \pt-independent thresholds ranging from 0.015 to 0.05 rad in $\eta$ and $\varphi$. The resulting uncertainty was found to be negligible.

The uncertainty on the electron identification originates from imprecisions in the description of the detector response in the MC, as well as from potential biases in the procedure employed to select electron candidates and to estimate the hadron contamination. It was studied by varying the electron identification selection  criteria on $n^{\rm{TPC}}_{\sigma,\rm{e}}$, $E/p$, and $\sigma^2_{\rm{long}}$. The assigned systematic uncertainties are listed as ``Electron identification" in the Tables~\ref{tab:SystematicSummaryTPCTOC},~\ref{tab:SystematicSummaryTPCEMCal}, and~\ref{tab:summarysystSN}.
The assigned systematic uncertainties vary from 5$\%$ to 12$\%$ depending on the \pt and the analysis method.

 Additionally, the robustness of the fit procedure used to extract the hadron contamination in both  electron identification strategies was checked. In the TPC--TOF analysis, different analytical functions were utilised to parameterise the TPC ${\rm{d}}E/{{\rm{d}}x}$, which had negligible effects on the estimated hadron contamination up to a \pt of 3 \GeVc. In the TPC--EMCal analysis, the scaling region of the hadron $E/p$ distribution was varied. The
resulting uncertainties were found to be negligible at low \pt and of the order of 5\% at high \pt.

The uncertainty on the subtraction of photonic electrons is related to the efficiency of finding the partner electron and was studied by varying the selection of partner tracks, i.e. the number of TPC clusters used for (d$E/$d$x$) calculation and the minimum $p_{\rm{T}}$ requirement, as well as the selection on the invariant mass of $\rm e^{+}e^{-}$ pairs. 

The subtraction of $\rm K_{e3}$ decay electrons in pp collisions for $p_{\rm{T}} < 0.5$ GeV$/c$ can be affected by the uncertainty on the parameterisation of the ratio of $\rm K_{e3}$ to photonic electrons, and was found to result in an uncertainty of 15\% at $p_{\rm{T}}=0.2$~\GeVc and to be negligible at $p_{\rm{T}} \geq  0.5$ GeV$/c$. 

The uncertainty on the contribution of electrons from $\rm{W}^{\pm}$ and $\rm{Z}^0$ boson decays was estimated by
varying the yield of electrons from $\rm{W}^{\pm}$ and $\rm{Z}^0$ boson decays by 25\%. The strategy is imported from the most recent measurements from ALICE~\cite{Acharya:2019hao,ALICE:2019nuy}. The resulting uncertainty was found to be negligible for \pPb collisions and only relevant at very high $p_{\rm{T}}$ for pp collisions, where it amounts to about 8\%.

In the MC simulations, the $\pi^{0}$ and $\eta$ meson \pt distributions were weighted such that their measured \pt spectra are reproduced. The uncertainty from the measurements was propagated to the efficiency of finding the partner electron by parameterising the data along the upper and lower ends of their statistical and systematic uncertainties added in quadrature. The uncertainty was found to be about 3\% at $\mbox{$p_{\rm{T}} = 0.2~\GeVc$}$ in pp collisions and to be negligible at $p_{\rm{T}}$ $ \geq $ 0.5 GeV$/c$ in both pp and \pPb collisions. 

The systematic uncertainty on the trigger RF, as explained in Sec.~\ref{ssec:TriggerRejectionFactor}, was propagated on the $p_{\rm{T}}$-differential cross section of electrons from heavy-flavour hadron decays. The uncertainty was of the order of 4\% at the highest \pt. 

The systematic uncertainty on the luminosity was propagated on the $p_{\rm{T}}$-differential cross section of electrons from heavy-flavour hadron decays. The uncertainty was 2.3\% to 5\% in pp collisions depending on the \pt, as the uncertainty from the rejection factors for triggered samples were taken into consideration, and 3\% in \pPb collisions, where the uncertainty from the rejection factors contributed negligibly to the uncertainty on luminosity. 
 
As the geometrical acceptance and reconstruction efficiencies are essentially independent of \dnchdeta in the measured multiplicity range, these corrections and their corresponding systematic uncertainties largely cancel in the ratio to the multiplicity-integrated yield, thus resulting in a lower systematic uncertainty for self-normalised yields compared to the one for the $p_{\rm{T}}$ spectra.

The total systematic uncertainties on the $p_{\rm{T}}$ spectra and the self-normalised yields were calculated by summing the different contributions in quadrature, as they are considered to be uncorrelated.

\begin{table}[h!]
\centering
\caption{Systematic uncertainty on self-normalised yield in pp collisions at $\sqrt{s} = 13~\rm TeV$ and \pPb collisions at $\sqrt{s_{\rm NN}} = 8.16~\rm TeV$.  }
	\renewcommand{\arraystretch}{1.7}
 
  \begin{tabular*}{\textwidth}{@{\extracolsep{\fill}} l|c|ccc|ccc}
    \hline
     \multicolumn{1}{c}{ }& \multicolumn{1}{c|}{ }& 
      \multicolumn{3}{c|}{pp, $\sqrt{\rm s}$ = 13 TeV}&
       \multicolumn{3}{c}{\pPb, $\sqrt{s_{\rm NN}}$ = 8.16 TeV}\\
      
    \hline
 
\multicolumn{1}{c|}{} & \multicolumn{1}{c|}{Multiplicity} &
   \multicolumn{3}{c|}{\pt interval (\GeVc)} & \multicolumn{3}{c}{\pt interval (\GeVc)} \\
   [-0.5ex]
 \multicolumn{1}{c|}{} &   \multicolumn{1}{c|}{intervals} &
   \multicolumn{1}{c}{0.5--6} & \multicolumn{1}{c}{6--12} & \multicolumn{1}{c|}{15--30} & \multicolumn{1}{c}{0.5--6} & \multicolumn{1}{c}{6--8} & \multicolumn{1}{c}{14--26} \\
 \hline\hline
 \multirow{3}{*}{Track selection}& {I} & { negl. }  & {2$\%$} & {2$\%$} & {negl.}& {3$\%$}& {3$\%$} \\
& {III}  & {negl.} & {2$\%$} & {2$\%$} & {negl.}& {3$\%$}& {3$\%$} \\
{}& {V}  & {negl.} & {2$\%$} & {2$\%$} & {negl.}& {4$\%$}& {4$\%$} \\
 \hline
 \multirow{3}{*}{SPD hit requirement} & {I} &
{ 10$\%$}  &{6$\%$} & {14$\%$} & {10$\%$}& {2$\%$}& {10$\%$} \\
& {III}  & {3$\%$} & {6$\%$} & {6$\%$} & {2$\%$}& {2$\%$}& {2$\%$} \\
{ }& {V}  & {4$\%$} & {6$\%$} & {6$\%$} & {2$\%$}& {2$\%$}& {2$\%$} \\
\hline
\multirow{3}{*}{Electron identification}& {I} &
{ 1$\%$}  & {3$\%$} &  {3$\%$} &  {1$\%$}&  {2$\%$}&  {2$\%$} \\
 {}& {III}  &  {1$\%$} &  {3$\%$} &  {3$\%$} &  {1$\%$}&  {2$\%$}&  {2$\%$} \\
 {}&  {V}  &  {1$\%$} &  {3$\%$} &  {3$\%$} &  {1$\%$}&  {4$\%$}&  {4$\%$} \\
 \hline
 \multirow{3}{*}{\makecell{Photonic electron\\subtraction}}& {I} &
{ 1$\%$}  & {1$\%$} &  {2$\%$} &  {1$\%$}&  {1$\%$}&  {1$\%$} \\
&  {III}  &  {1$\%$} &  {2$\%$} &  {2$\%$} &  {1$\%$}&  {1$\%$}&    {1$\%$} \\
& {V}  & {1$\%$} & {2$\%$} & {2$\%$} & {1$\%$}& {1$\%$}& {1$\%$} \\
 \hline\hline
 \multirow{3}{*}{Total systematics}&{I} &
{ 10$\%$}  &{7$\%$} & {15$\%$} & {10$\%$}& {4$\%$}& {11$\%$} \\
& {III}  & {3$\%$} & {7$\%$} & {7$\%$} & {2$\%$}& {4$\%$}&{4$\%$} \\
{ }& {V}  & {4$\%$} & {7$\%$} & {7$\%$} & {2$\%$}&{6$\%$}&{6$\%$} \\
   \hline
  \end{tabular*}
   \label{tab:summarysystSN}
\end{table}

\section{Results}\label{section:results}

\subsection{\pt-differential cross section of heavy-flavour hadron decay electrons in pp and \pPb collisions}
\begin{figure}[!ht]
\centering

\includegraphics[width=0.48\linewidth]{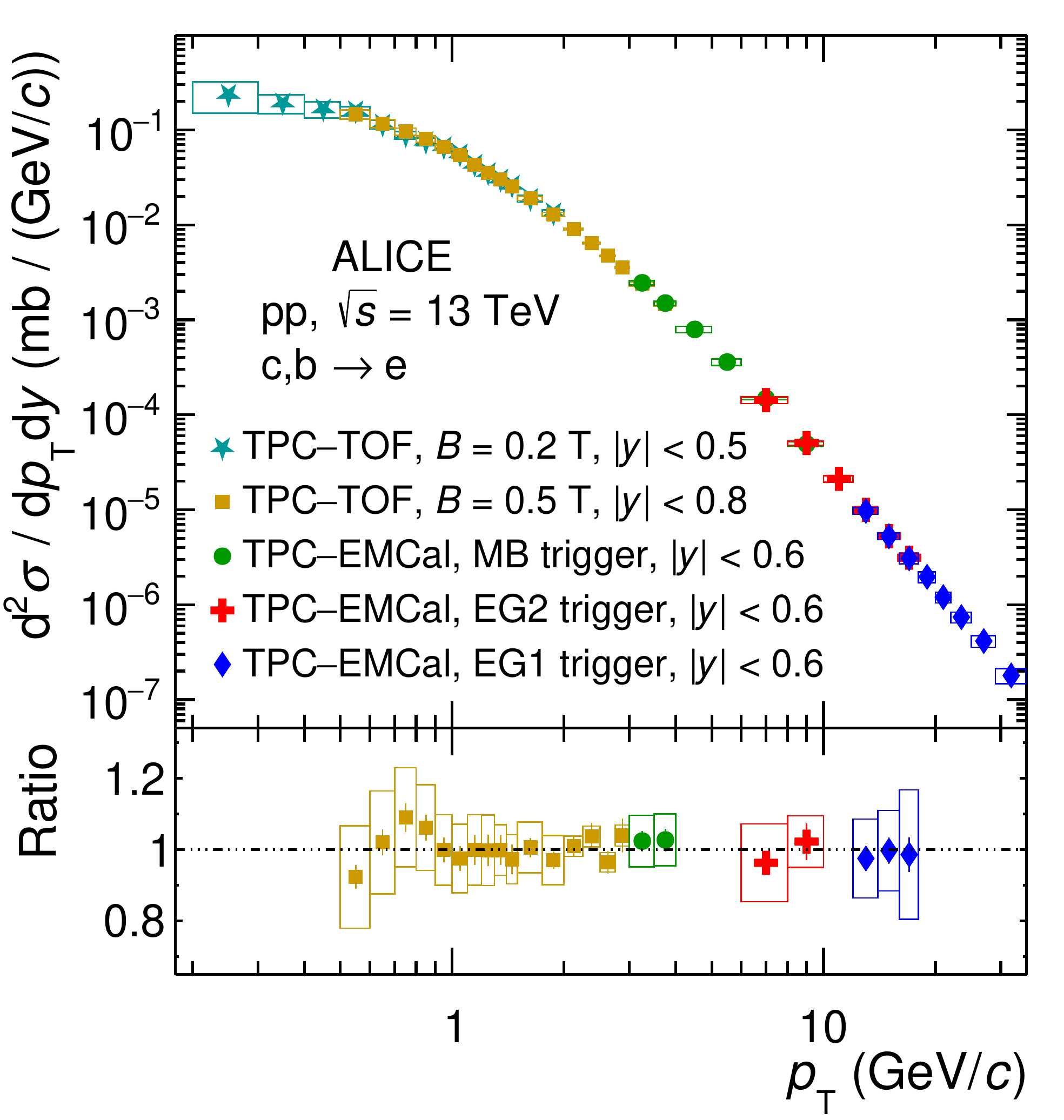}
\includegraphics[width=0.48\linewidth]{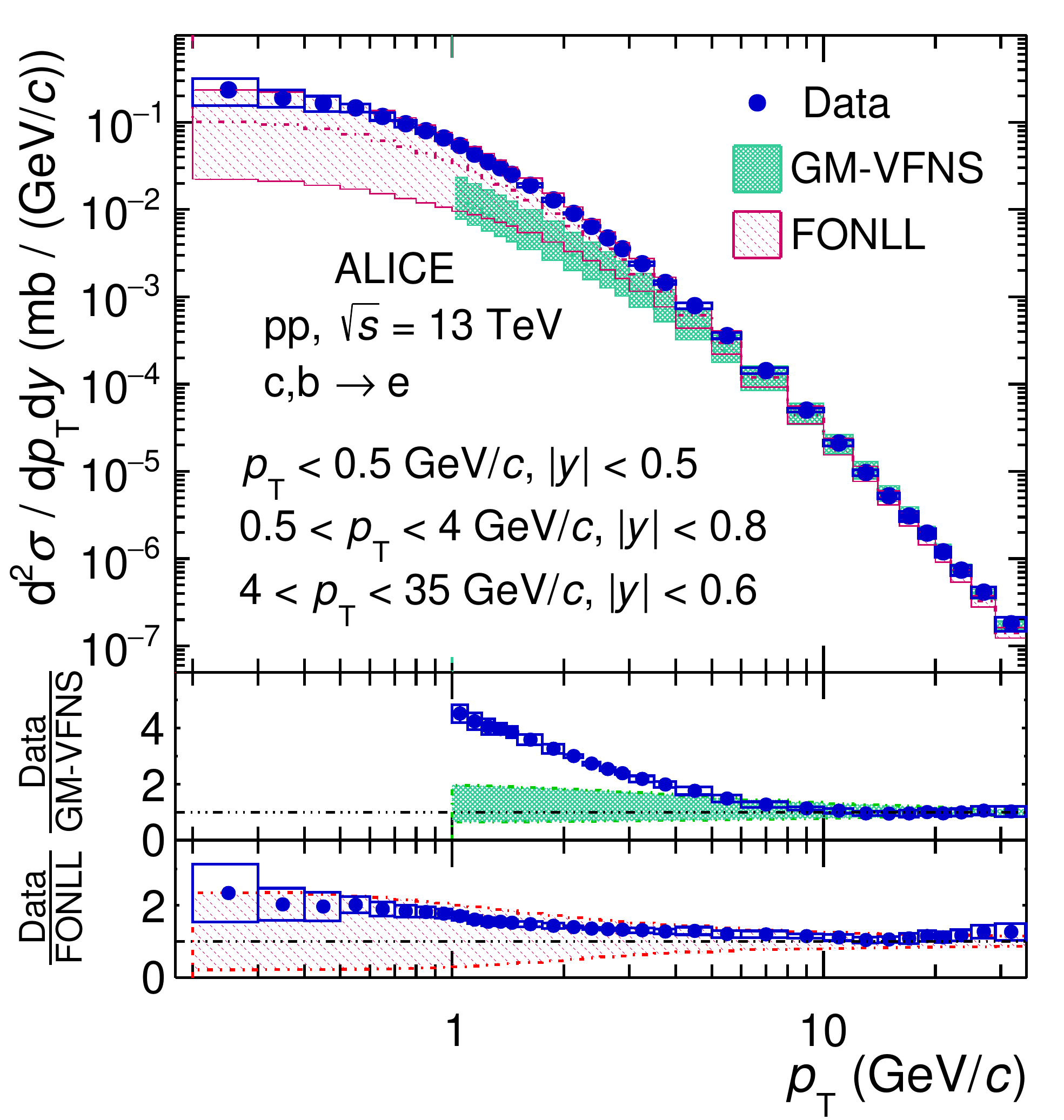}
\caption{Left, top: $p_{\rm T}$-differential cross section of electrons from heavy-flavour hadron decays in pp collisions at $\sqrt{s} =$ 13 TeV measured at midrapidity with different detectors and data sets. Left, bottom: Ratios of the different measurements in the overlapping \pt intervals. Right: $p_{\rm{T}}$-differential cross section compared with Fixed Order with Next-to-Leading-Log resummation (FONLL)~\cite{Cacciari:2012ny} and General-mass-variable-flavour-number-Scheme (GM-VFNS)~\cite{Bolzoni_2014} predictions and its ratios with respect to FONLL and GM-VFNS central values in the two lower panels. Vertical bars and boxes denote statistical and systematical uncertainties, respectively.}        
\label{Fig:ppHFESpectra}
\end{figure}

The \pt-differential production cross section of electrons from semileptonic decays of heavy-flavour hadrons at midrapidity
in pp collisions at $\sqrt{s} = 13$ TeV measured in the transverse momentum interval $\mbox{$0.2 < p_{\rm{T}} < 35$ GeV$/c$}$ is shown in Fig.~\ref{Fig:ppHFESpectra}. The statistical uncertainties are represented as vertical lines while the total systematic uncertainties are displayed as boxes. 
In the top left panel of Fig.~\ref{Fig:ppHFESpectra}, the cross sections measured with the TPC--TOF detectors and the two different data sets collected with different magnetic fields are plotted together with the spectra obtained using the TPC--EMCal detectors with MB triggered events, as well as with EMCal triggered events, EG1, and EG2. 
The ratios of the different analyses in the overlapping \pt~intervals are shown in the bottom left panel of Fig.~\ref{Fig:ppHFESpectra}.  
For $\mbox{0.5 $<$ \pt $<$ 4~\GeVc}$, the ratio of the result from the TPC--TOF analyses with $B = $ 0.5 T to the one obtained with $B = $ 0.2 T is displayed. In 3 $<$ \pt $<$ 4 \GeVc, the ratio of the cross section obtained from the TPC--TOF analysis to that obtained from the TPC--EMCal analysis is shown for MB triggered events. At higher \pt, namely 6 $<$ \pt $<$ 10 \GeVc (12 $<$ \pt $<$ 18 \GeVc), the ratio of the TPC--EMCal results for MB and EG2 (EG2 and EG1) triggered events is reported. All ratios are consistent with unity within statistical and systematic uncertainties, which demonstrates that the different analyses are in agreement with each other. The final cross section in the \pt intervals 0.2--0.5 GeV/$c$, 0.5--4 GeV$/c$, 4--6 GeV$/c$, 6--12 GeV$/c$, and 12--35 GeV$/c$ was obtained from the TPC--TOF low-$B$ field analysis, the TPC--TOF nominal-$B$ field analysis, and from the results obtained with the TPC--EMCal detectors using MB, EG2 and EG1 triggered events, respectively. In this way, for each \pt range, the measurement with the smallest total uncertainty (quadratic sum of statistical and systematic uncertainty) is used.

The $p_{\rm{T}}$-differential cross section measurement was compared with FONLL~\cite{Cacciari:2012ny} and GM-VFNS~\cite{Bolzoni:2012kx} pQCD calculations
\footnote{${\rm Central~values:}~~{\rm FONLL:}~\mu_{\rm F}~=~\mu_{\rm R}~=~\sqrt{m_{\rm Q}^2+p_{\rm T}^2},~m_{\rm b}~=~4.75~{\rm GeV},~m_{\rm c}~=~1.5~{\rm GeV};~~{\rm GM-VFNS:}~\mu_{\rm F}~=~0.49~\mu_{\rm R},\\\mu_{\rm R}~=~\sqrt{4m_{\rm Q}^2+p_{\rm T}^2},~m_{\rm b}~=~4.5~{\rm GeV},~m_{\rm c}~=~1.5~{\rm GeV};~~{\rm where}~\mu_{\rm R}~=~{\rm renormalization~scale,}~\mu_{\rm F}~=~{\rm factorisation~scale}$},
as shown in the right panel of Fig.~\ref{Fig:ppHFESpectra}.
 The uncertainties of the FONLL calculations reflect different choices for the charm- and beauty-quark masses, and for the factorisation and renormalisation scales as well as the uncertainty on the set of parton distribution functions (PDF) (CTEQ6.6~\cite{Nadolsky:2008zw}). The FONLL calculations describe the measurements within the uncertainties, although the theoretical uncertainties are large, up to a factor of two. The data are found to be close to the upper edge of the FONLL prediction, which can be clearly seen in the right bottom panel of Fig.~\ref{Fig:ppHFESpectra}, where the ratio of the data points to the FONLL calculations is shown. 
Similar observations were made for the measurements of electrons from heavy-flavour hadron decays in pp collisions at lower energies at the LHC~\cite{Abelev:2014gla, ALICE:2012mzy, Acharya:2018upq, Acharya:2019mom} and at RHIC~\cite{STAR:2011bqq, PHENIX:2006tli}. The measurement of the cross section of D mesons is also consistent with upper bound of FONLL pQCD calculations in pp collisions at LHC~\cite{Abelev:2012vra,ALICE:2017olh,ALICE:2019nxm,Acharya:2021cqv,ATLAS:2015igt,LHCb:2015swx, LHCb:2016ikn} and RHIC~\cite{STAR:2012nbd}, as well as in $\rm{p \bar{p}}$ collisions at Tevatron energies~\cite{CDF:2003vmf}. 
The FONLL calculations use fragmentation functions tuned on e$^{+}$e$^{-}$ data and assume that all charm quarks fragment only into  D$^+$ and D$^0$ mesons (and their antiparticles). 
Recent measurements of charm-baryon production at midrapidity in pp and \pPb collisions from ALICE show a baryon-to-meson ratio significantly higher than that in e$^{+}$e$^{-}$ collisions, suggesting that the fragmentation of charm quark is not universal across different collision systems~\cite{ALICE:2021dhb,ALICE:2021npz}.
As a consequence, calculations taking properly into account the latest open-cham baryon measurements at midrapidity to constrain the charm fragmentation are expected to predict a smaller yield of heavy-flavour hadron decays by about 9\% compared to the FONLL spectrum.
The largest source of uncertainties in the GM-VFNS prediction is due to scale variation, and hence PDF related uncertainties and variations of the bottom and charm mass are not considered. The GM-VFNS framework includes leptoproduction from the following three steps: beauty quark to beauty hadrons (b $\rightarrow$ B), transition from  beauty quark to  charm hadrons (b $\rightarrow$ B $\rightarrow$ D), and charm quark to charm hadrons (c $\rightarrow$ D). The GM-VFNS calculations describe the data within the uncertainties for \pt greater than 5 \GeVc, but largely underestimate the cross section for lower \pt, up to a factor of five at 1 \GeVc, as seen in the right middle panel of Fig.~\ref{Fig:ppHFESpectra}. Similar observations were reported for the non-prompt D meson measurements at $\sqrt{s}~=~5.02~{\rm TeV}$~\cite{Acharya:2021cqv}. For prompt D mesons at $\sqrt{s}~=~5.02~{\rm TeV}$, however, the GM-VFNS predictions describe the cross section within the uncertainties~\cite{Acharya:2021cqv}. Electrons from heavy-flavour hadron decays are dominated by semileptonic decays of beauty hadrons for \pt~$>5$ ~GeV$/c$~\cite{Abelev:2012sca, ALICE:2014aev}. Therefore, the cross section measured up to 35 GeV$/c$ can provide important information to beauty hadron production. 

\begin{figure}[!ht]
\centering
\includegraphics[width=0.5\linewidth]{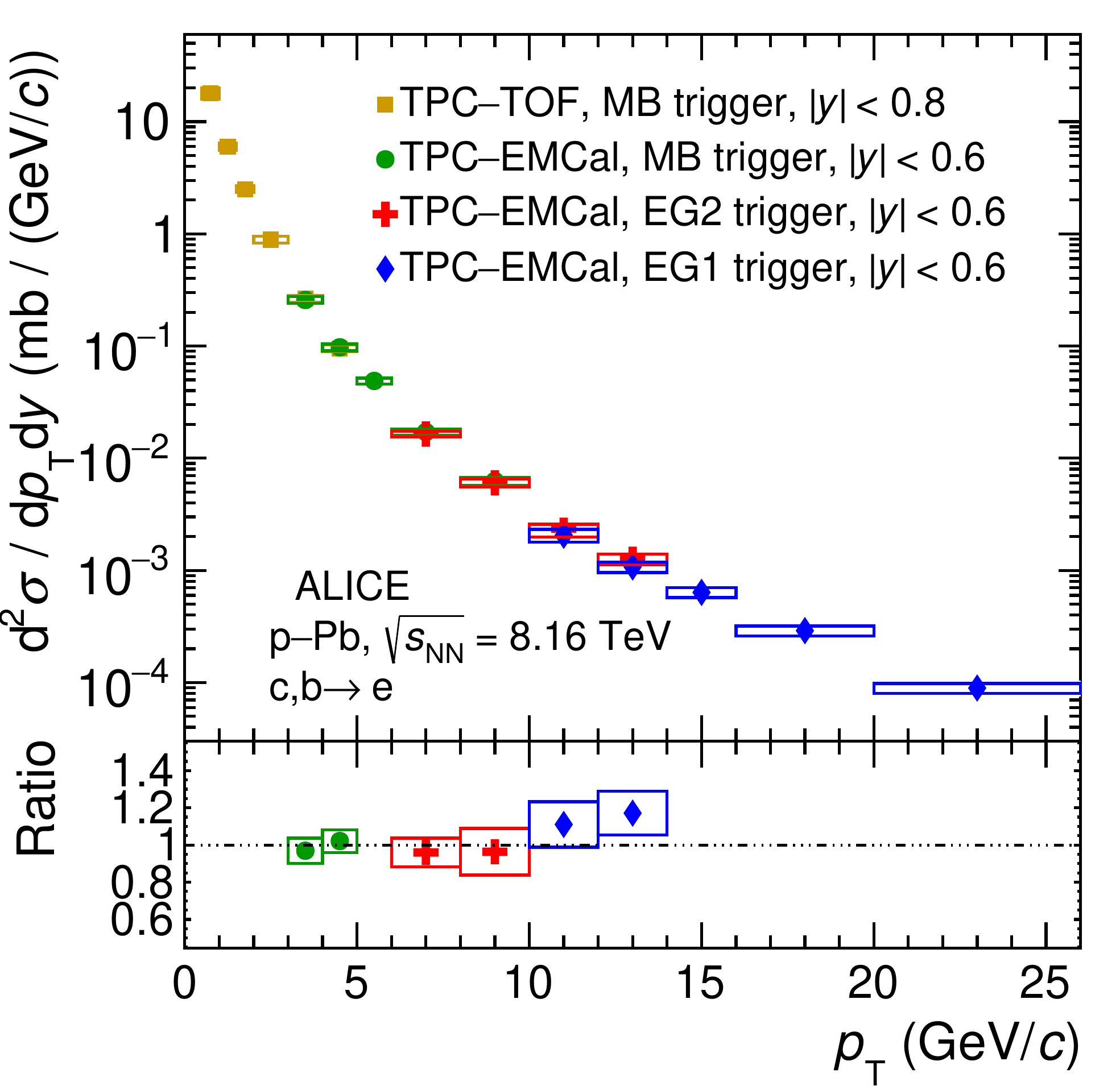}
\caption{Top: $p_{\rm T}$-differential cross section of electrons from heavy-flavour hadron decays in \pPb  collisions at \sqrtsNN $= 8.16~\rm TeV$ measured at midrapidity with different detectors. Bottom: Ratios of the different measurements in the overlapping \pt intervals.}        
\label{Fig:pPbHFESpectra}
\end{figure}

 The $p_{\rm{T}}$-differential production cross section of electrons from semileptonic heavy-flavour hadron decays at midrapidity in \pPb collisions at $\sqrt{s_{\rm{NN}}} = 8.16$ TeV measured in the transverse momentum interval $0.5 < p_{\rm{T}} < 26$~GeV$/c$ is shown in Fig.~\ref{Fig:pPbHFESpectra}. In the upper panel of Fig.~\ref{Fig:pPbHFESpectra}, the cross sections measured with the TPC--TOF detectors are plotted together with the measurements obtained using the TPC--EMCal detectors with MB and EMCal EG2 and EG1  triggered events. On the bottom left panel of Fig.~\ref{Fig:pPbHFESpectra}, the ratios of the cross sections obtained from the different measurements are calculated in the overlapping \pt intervals. For 3 $<$ \pt $<$ 5 \GeVc, the ratio of the result obtained from the TPC--TOF analysis with respect to that from the TPC--EMCal is shown for  MB triggered events. For 6 $<$ \pt $<$ 10 \GeVc (12 $<$ \pt $<$ 14 \GeVc) the ratio of the TPC--EMCal results obtained with MB and EG2   triggered events (EG2 and EG1) is reported. All ratios are consistent with unity within statistical and systematic uncertainties. The same strategy as in pp collisions was used to get the final cross section in  \pPb collisions. The final cross section in the \pt intervals 0.5--4 GeV$/c$, 4--6 GeV$/c$, 6--9 GeV$/c$, and 9--26 GeV$/c$ was obtained from the TPC--TOF nominal-$B$ field analysis and from the results using the TPC--EMCal detectors with MB, EG2, and EG1 triggered events, respectively.

\subsection{Nuclear modification factor of electrons from heavy-flavour hadron decays in \pPb collisions}

\begin{figure}[!ht]
\centering
\includegraphics[width=0.5\linewidth]{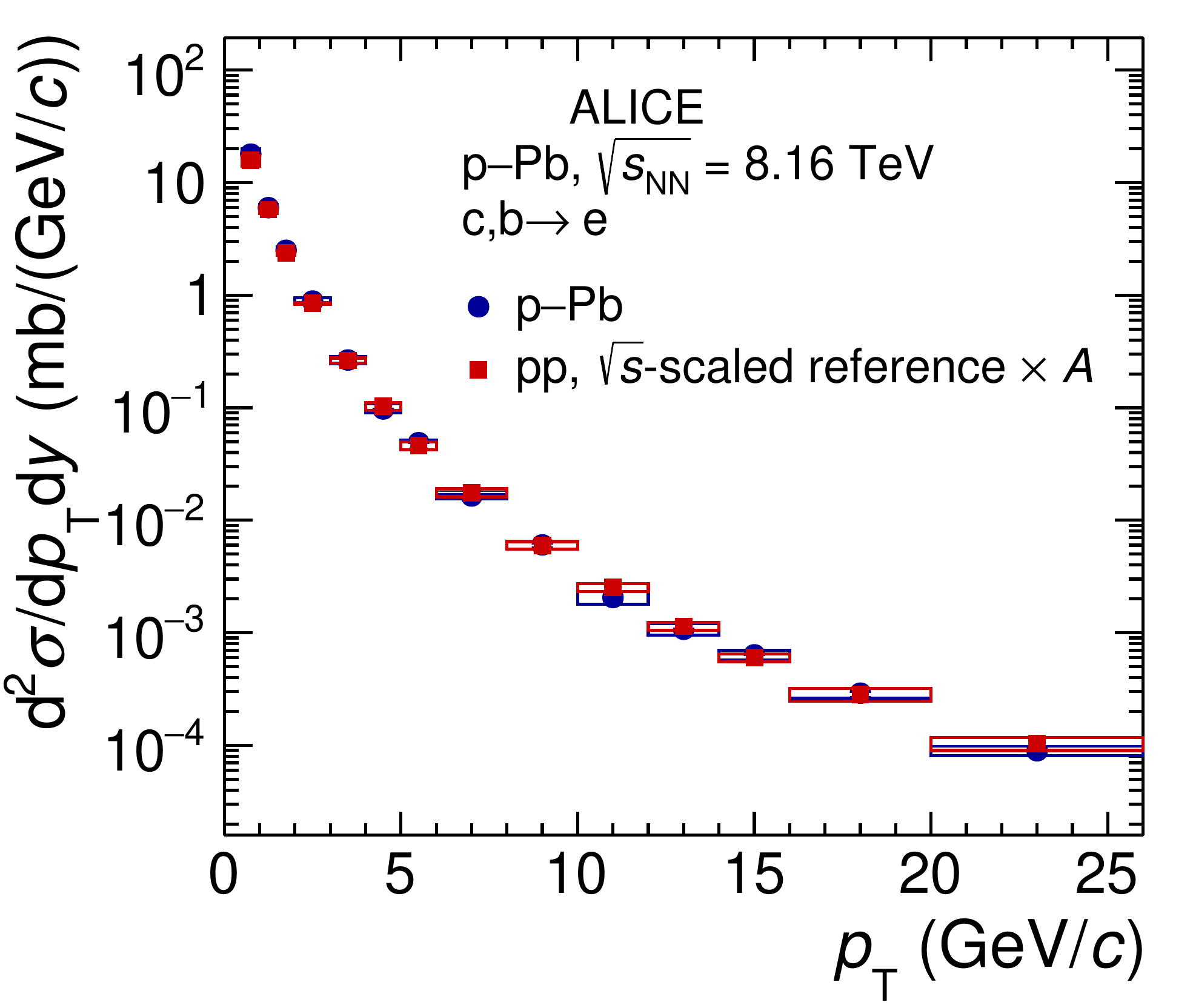}
\caption{$p_{\rm{T}}$-differential cross section of electrons from heavy-flavour hadron decays measured in \pPb collisions at \sqrtsNN $= 8.16~\rm TeV$ compared with the pp reference at the same centre-of-mass energy obtained from the measurement in pp collisions at \sqrts $= 13~\rm TeV$ scaled to \sqrts $= 8.16~\rm TeV$.}        
\label{Fig:HFESpectraComparison}
\end{figure}

The nuclear modification factor of electrons from heavy-flavour hadron decays, $R_{\rm{pPb}}$, is defined as
\begin{equation}
    R_{\rm{pPb}}(p_{\rm{T}},\it{y}) = \frac{\rm 1}{ A} \frac{{\rm {d}}^2 \sigma_{\rm{pPb}}/{\rm{d}} p_{\rm{T}}\rm{d}\it{y}}{{\rm{d}}^2\sigma_{\rm{pp}}/{\rm{d}} p_{\rm{T}}\rm{d}\it{y}},
\end{equation}
where  ${\rm {d}}^2 \sigma_{\rm{pPb}}/{\rm{d}} p_{\rm{T}}\rm{d}\it{y}$ is the cross section of electrons from heavy-flavour hadron decays measured in {$\mbox{\pPb}$} collisions at \sqrtsNN $= 8.16~\rm TeV$ and ${\rm {d}}^2 \sigma_{\rm{pp}}/{\rm{d}} p_{\rm{T}}\rm{d}\it{y}$ is the cross section of electrons from heavy-flavour hadron decays in pp collisions at the same centre-of-mass energy, scaled with the number of nucleons ($A$) in the lead ion.
The reference cross section in pp collisions was obtained using the measurement at \sqrts $= 13~\rm TeV$, presented here. The cross section at \sqrts $= 13~\rm TeV$ was scaled to \sqrts $= 8.16~\rm TeV$ using pQCD calculations. The $p_{\rm{T}}$-dependent scaling factor was obtained by calculating the ratio of the production cross sections of electrons from heavy-flavour hadron decays from FONLL calculations~\cite{Cacciari:2012ny} at $\sqrt{s}= 8.16$ TeV to $\sqrt{s}=13$ TeV. The systematic uncertainty on the pp reference includes the systematic uncertainties on the measured cross section at $\sqrt{s}=13$ TeV, which was described above, and the ones on the $p_{\rm{T}}$-dependent scaling factor. The uncertainty on the scaling factor ranges between 11\% and 1\% going from \pt $= 0.2$ \GeVc to \pt $= 26$ \GeVc. This includes the uncertainties on the PDFs, quark masses, and factorisation and renormalisation scales, as described in Ref.~\citenum{Averbeck:2011ga}. The two contributions were added in quadrature leading to a total systematic uncertainty of 5-15\%, depending on $p_{\rm{T}}$. In addition, a global normalisation systematic uncertainty of 2.3\% from the pp analysis at $\sqrt{s}=13$ TeV was also considered.
The $p_{\rm{T}}$-differential cross section of electrons from heavy-flavour hadron decays in pp collisions at \sqrts $= 13~\rm TeV$ scaled to \sqrts $= 8.16~\rm TeV$ using the aforementioned procedure is shown together with the $p_{\rm{T}}$-differential cross section of electrons from heavy-flavour hadron decays in \pPb collisions at \sqrtsNN $= 8.16~\rm TeV$ in Fig.~\ref{Fig:HFESpectraComparison}.

\begin{figure}[!ht]
      \begin{center}
      \includegraphics[width=0.48\linewidth]{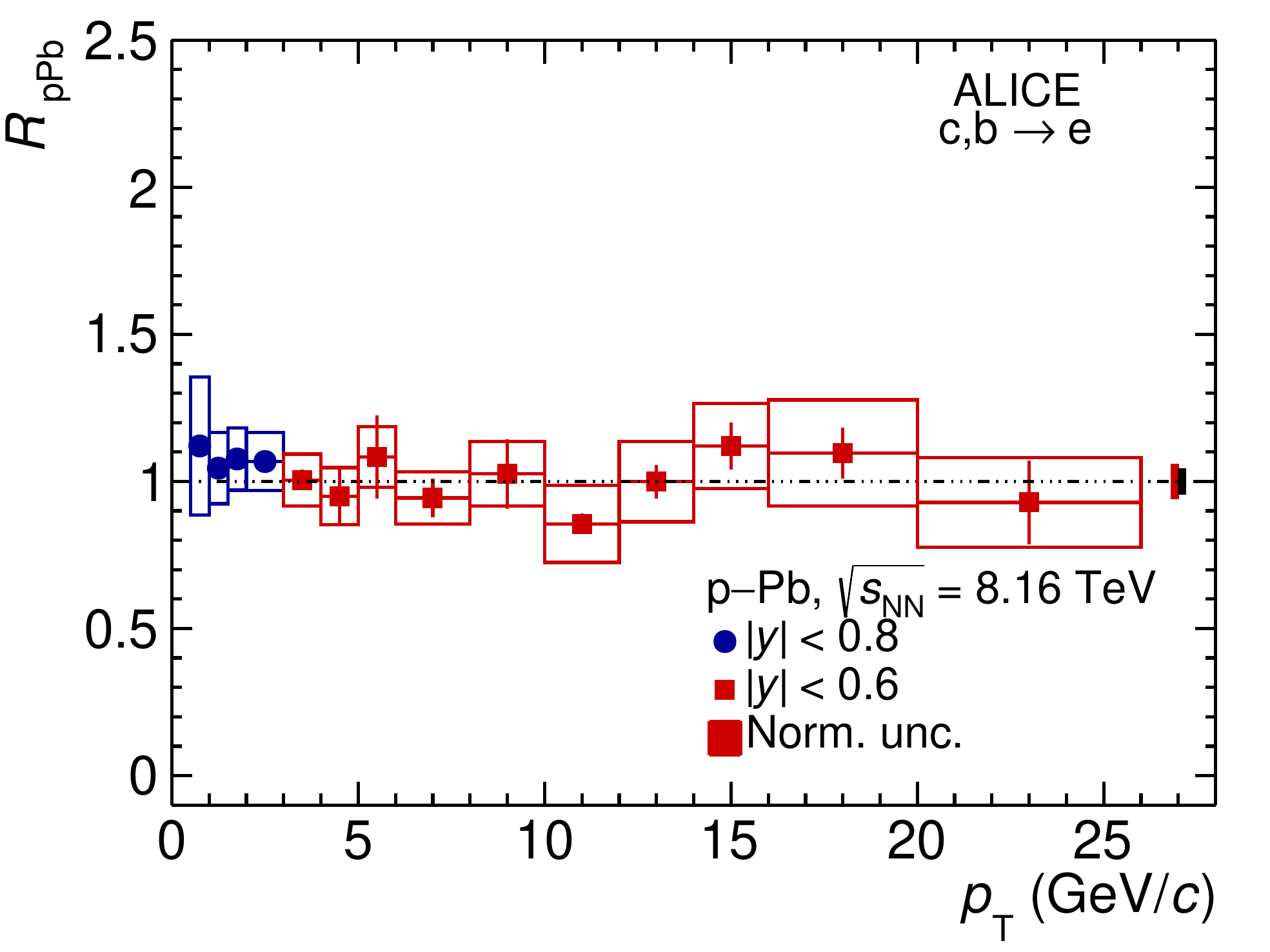}
      \includegraphics[width=0.48\linewidth]{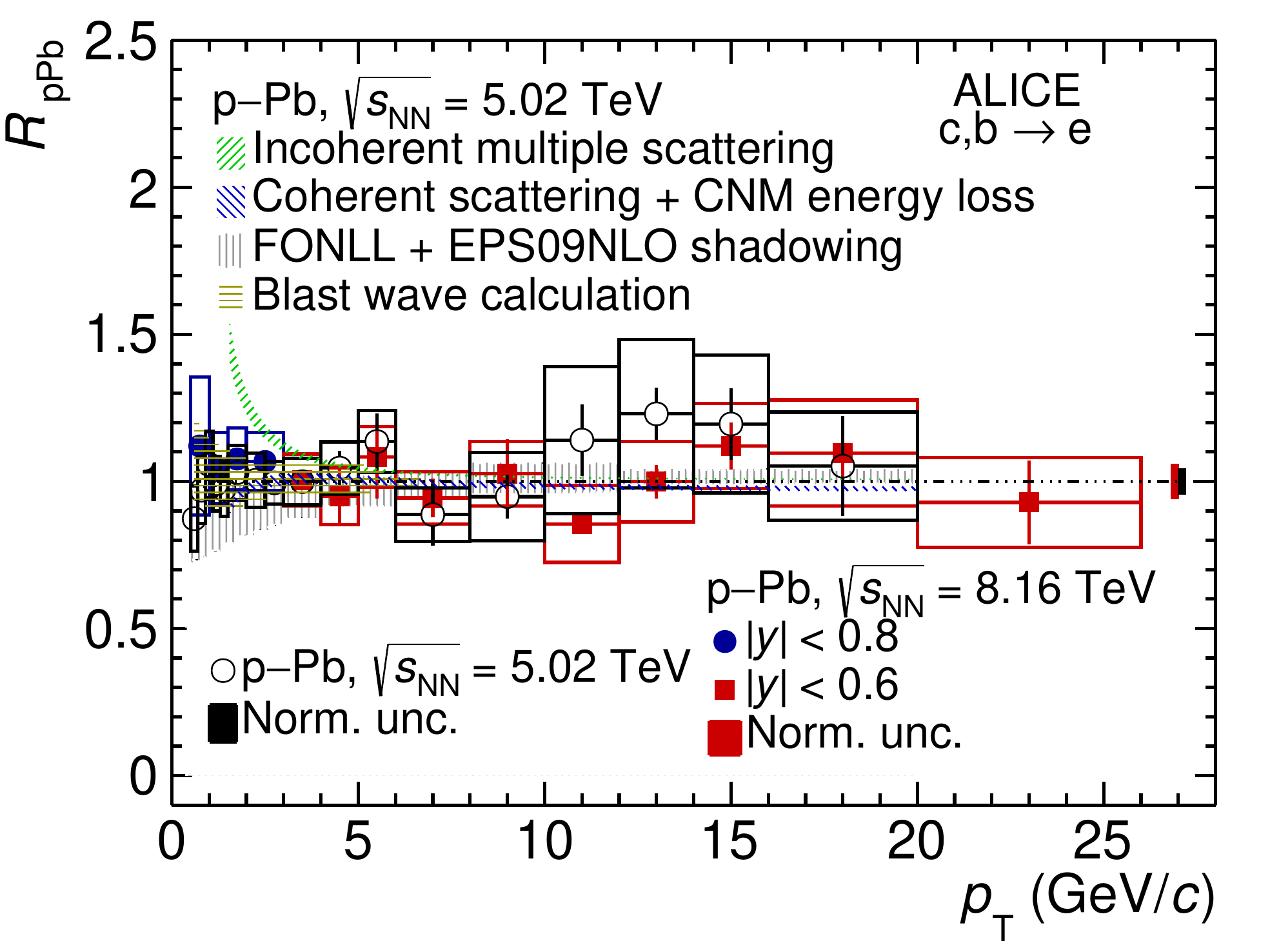}
      \end{center}
\caption{ The nuclear modification factor $R_{\rm{pPb}}$ of electrons from heavy-flavour hadron decays in \pPb collisions at
\sqrtsNN $= 8.16~\rm TeV$ (left) compared with that at \sqrtsNN $= 5.02~\rm TeV$ and theoretical models at \sqrtsNN $= 5.02~\rm TeV$ (right) ~\cite{Acharya:2019hao}.}    
\label{Fig:RpPb}
\end{figure}

The nuclear modification factor of electrons from heavy-flavour hadron decays as a function of transverse momentum at $\sqrt{s_{\rm{NN}}} = 8.16$ TeV is presented in Fig.~\ref{Fig:RpPb}. The statistical and systematic uncertainties of the spectra in p–Pb and pp collisions were propagated as uncorrelated. The normalisation uncertainties are shown as a solid box at $R_{\rm{pPb}} = 1$. The $R_{\rm{pPb}}$ is consistent with unity within statistical and systematic uncertainties over the whole $p_{\rm{T}}$ range of the
measurement. Modifications of the cross section of electrons from
heavy-flavour hadron decays in \pPb collisions due to different cold nuclear matter effects, are small compared to the current uncertainties of the measurement in the probed $p_{\rm{T}}$ range.
The sample of electrons from heavy-flavour hadron decays is dominated by beauty-hadron decays for  $\mbox{$p_{\rm{T}} > 5$ GeV$/c$ }$ ~\cite{Abelev:2012sca, Abelev:2014hla}. The $R_{\rm{pPb}}$ was fitted with a constant function above 5 \GeVc and the value was $\rm 0.95 \pm 0.02(stat.) \pm 0.13(sys.)$, thus consistent with unity within 13\%. The $R_{\rm{pPb}}$ of unity indicates that the beauty production is not modified in \pPb collisions within the kinematic range of this measurement, which is also consistent with the measurement of $R_{\rm{pPb}}$ of beauty-decay electrons up to $p_{\rm{T}} = 8$ GeV$/c$ at $\sqrt{s_{\rm{NN}}} = 5.02$ TeV~\cite{Adam:2016wyz}. In the right panel of Fig.~\ref{Fig:RpPb}, the $R_{\rm{pPb}}$ at $\sqrt{s_{\rm{NN}}} = 8.16$ TeV is compared with that at $\sqrt{s_{\rm{NN}}} = 5.02$ TeV and different theoretical models provided for $\sqrt{s_{\rm{NN}}} = 5.02$ TeV ~\cite{Acharya:2019hao}. The $R_{\rm{pPb}}$ is observed to be independent of the centre-of-mass energy. The data disfavour the enhancement trend at low \pt predicted by the model calculations which are based on incoherent multiple scatterings ~\cite{KANG201523}. Model predictions which are based on coherent multiple scattering and energy loss in the CNM, pQCD calculations using FONLL 
framework and EPS09NLO for the nuclear modification of the PDF, as well as calculations which assume the formation of a hydrodynamical expanding medium in \pPb collisions at $\sqrt{s_{\rm{NN}}} = 5.02$~TeV within the Blast wave framework predict an $R_{\rm{pPb}}$ close to unity and are in agreement with the measurements.

\subsection[Self-normalised yield of electrons from heavy-flavour hadron decays vs. normalised multiplicity]{Self-normalised yield of electrons from heavy-flavour hadron decays vs. normalised multiplicity in pp and \pPb collisions}
\begin{figure}[!ht]
\centering
\includegraphics[width=0.6\linewidth]{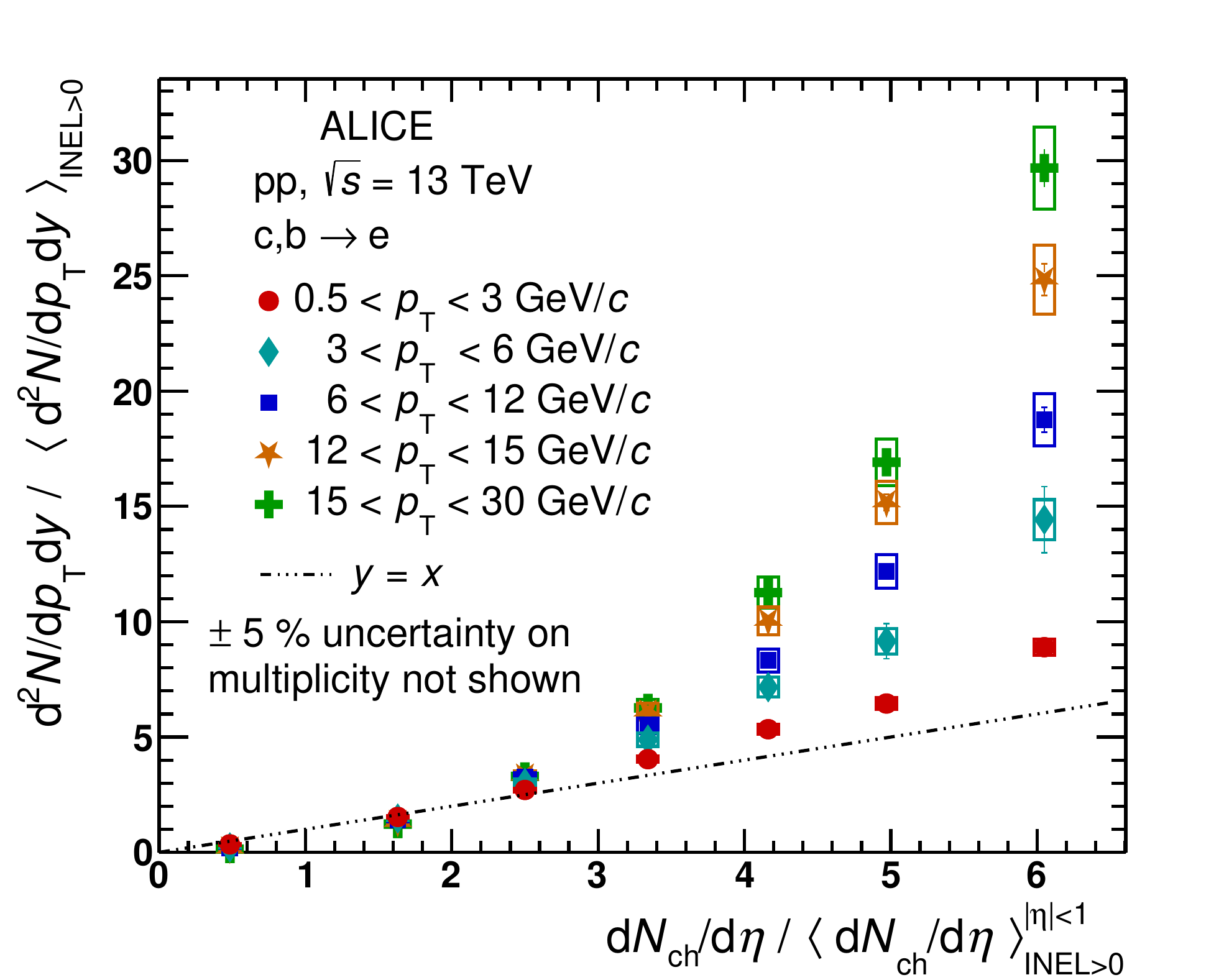}
\caption{Self-normalised yield of electrons from heavy-flavour hadron decays as a function of normalised charged-particle pseudorapidity density at midrapidity computed in \pp collisions at \sqrts $= 13~\rm TeV$ in different \pt intervals.}
\label{Fig:SelfnormalisedYield}
\end{figure}

The self-normalised yield of electrons from heavy-flavour hadron decays as a function of the self-normalised charged-particle pseudorapidity density  at midrapidity, i.e., ${\rm d}^{2} {N}/{\rm d}\pt {\rm d}{y} / \langle {\rm d}^{2}{N}/{\rm d}\pt{\rm d}{y}\rangle_{{\rm INEL>0}}$ vs. $\dnchdeta/\left<\dnchdeta\right>$, in pp collisions at $\sqrt{s} =$ 13 TeV is presented in Fig.~\ref{Fig:SelfnormalisedYield}.  The results are self-normalised to the INEL $>$ 0 event class. The measurements were performed in five \pt intervals from 0.5 to 30 GeV$/c$. 
The dashed line shown in the figure is a linear function with a slope of unity. The available data samples allow us to examine events with a multiplicity more than six times larger than the average multiplicity in pp collisions. The self-normalised yield of electrons from heavy-flavour hadron decays grows faster than linear with the self-normalised multiplicity. The measurement in intervals of \pt shows that this increase is more pronounced for high-\pt electrons. The yield of heavy-flavour decay electrons increases by approximately a factor of nine with respect to its multiplicity-integrated value for the lowest measured \pt~interval ($0.5 < \pt < 1.5$ \GeVc) and a factor of 29 for the highest measured \pt interval $\mbox{$(20 < \pt < 35\ \GeVc)$}$ for multiplicities of six times the average multiplicity. 

\begin{figure}[!h]
\includegraphics[width=0.48\linewidth]{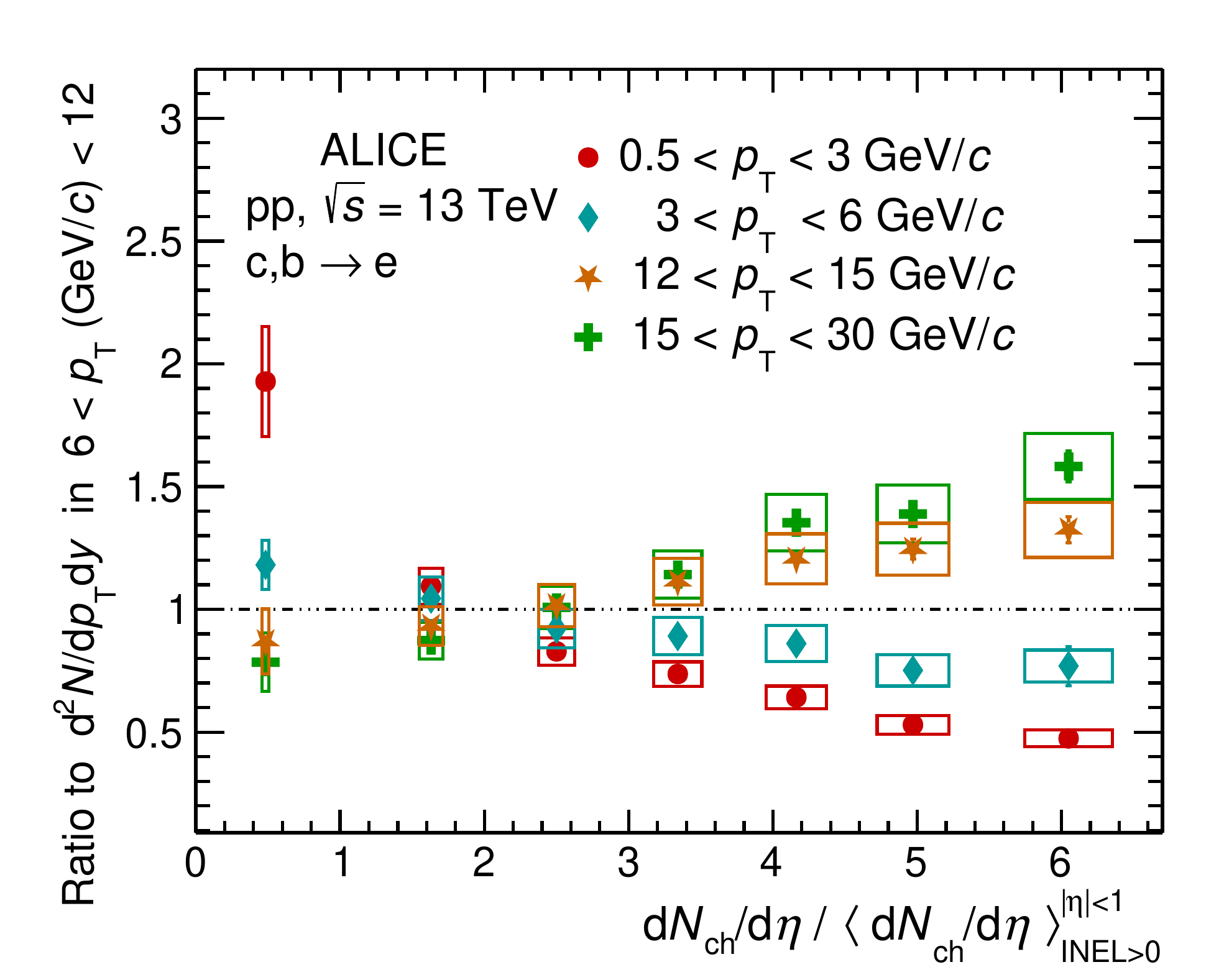}
\includegraphics[width=0.48\linewidth]{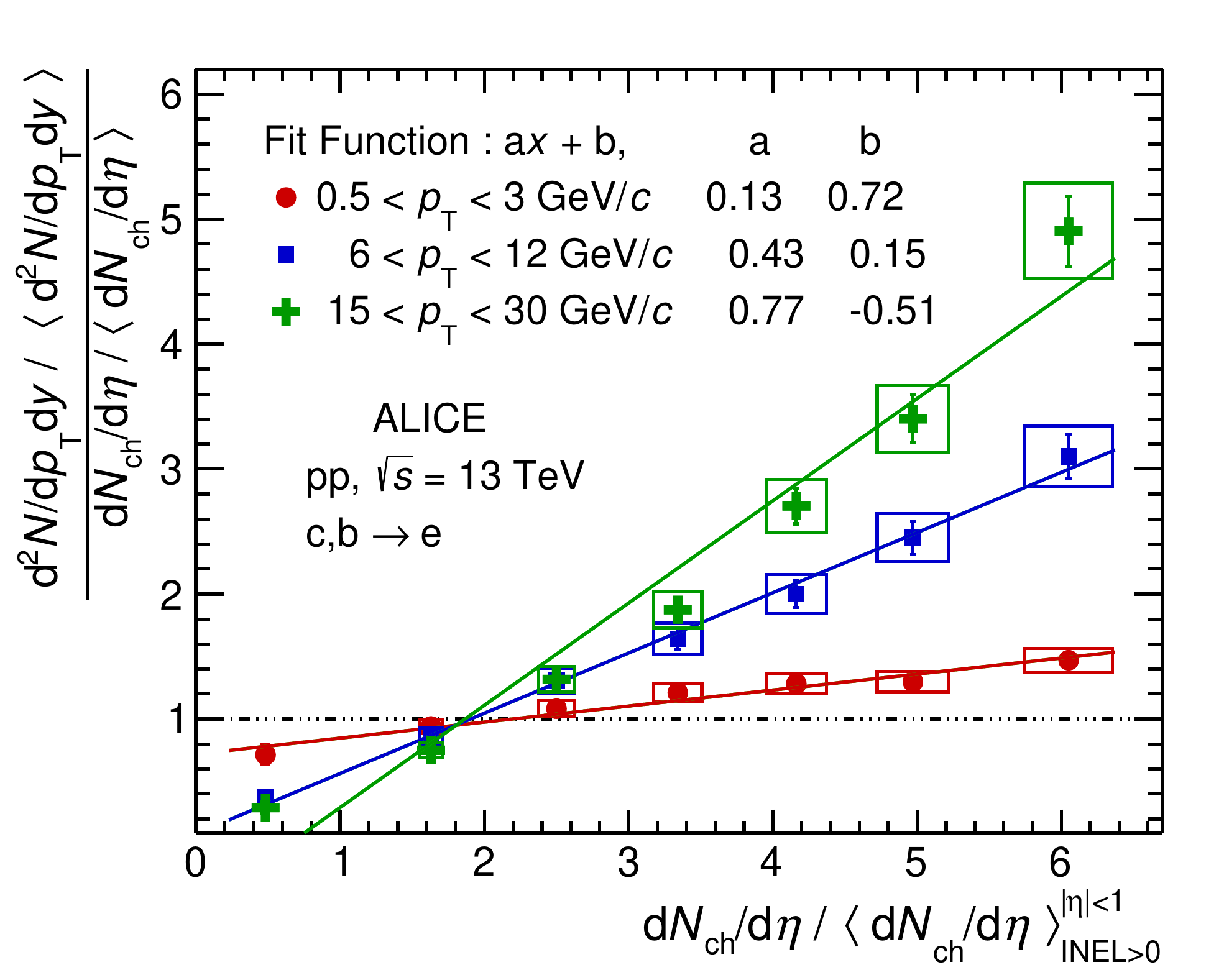}
\caption{ Ratio of the  self-normalised yields
in different \pt intervals with respect to that in the $6 < \pt < 12~{\rm GeV}/c$ interval (left) and double ratio of the self-normalised yields of  electrons to the self-normalised multiplicity (right) in pp collisions at $\sqrts=13$ TeV for three \pt ranges.}
\label{Fig:SNY_Ratio_pp}
\end{figure}

In the left panel of Fig.~\ref{Fig:SNY_Ratio_pp}, the ratios of the self-normalised yields of electrons from heavy-flavour hadron decays in various \pt intervals with respect to the one measured in the $6 < \pt < 12$ GeV$/c$ interval are shown. The yield of lower-$\pt$ electrons is higher in low  multiplicity events, while it decreases in higher multiplicity events. 
An opposite trend is observed for electrons at higher \pt, where the yield is lower in low  multiplicity events and increases at higher multiplicities.
The increase of the slope with \pt is influenced by the momentum dependence of jet fragmentation affecting the measured multiplicity at midrapidity, and the momentum dependence of the fraction of electrons from charm and beauty hadron decays. The relative fraction of electrons from beauty hadron decays increases with $\pt$ and becomes the main source of heavy-flavour hadron decay electrons at high \pt ($\pt > 5$ GeV$/c$)~\cite{Adam:2015ota,ALICE:2020msa,Weber:2018ddv}.

In the right panel of Fig.~\ref{Fig:SNY_Ratio_pp}, the double ratio of the self-normalised  electron yield to the self-normalised multiplicity in pp collisions is presented. The double ratio is observed to increase with multiplicity. The increase is weaker for low-\pt electrons than for high-\pt electrons. A linear function was used to fit the multiplicity dependence of the double ratio, 
which was found to describe the data reasonably well for all \pt intervals. This indicates that in the measured \pt range the yield grows approximately with the square of the multiplicity with a slope increasing with \pt.

\begin{figure}[!h]
\includegraphics[width=0.5\linewidth]{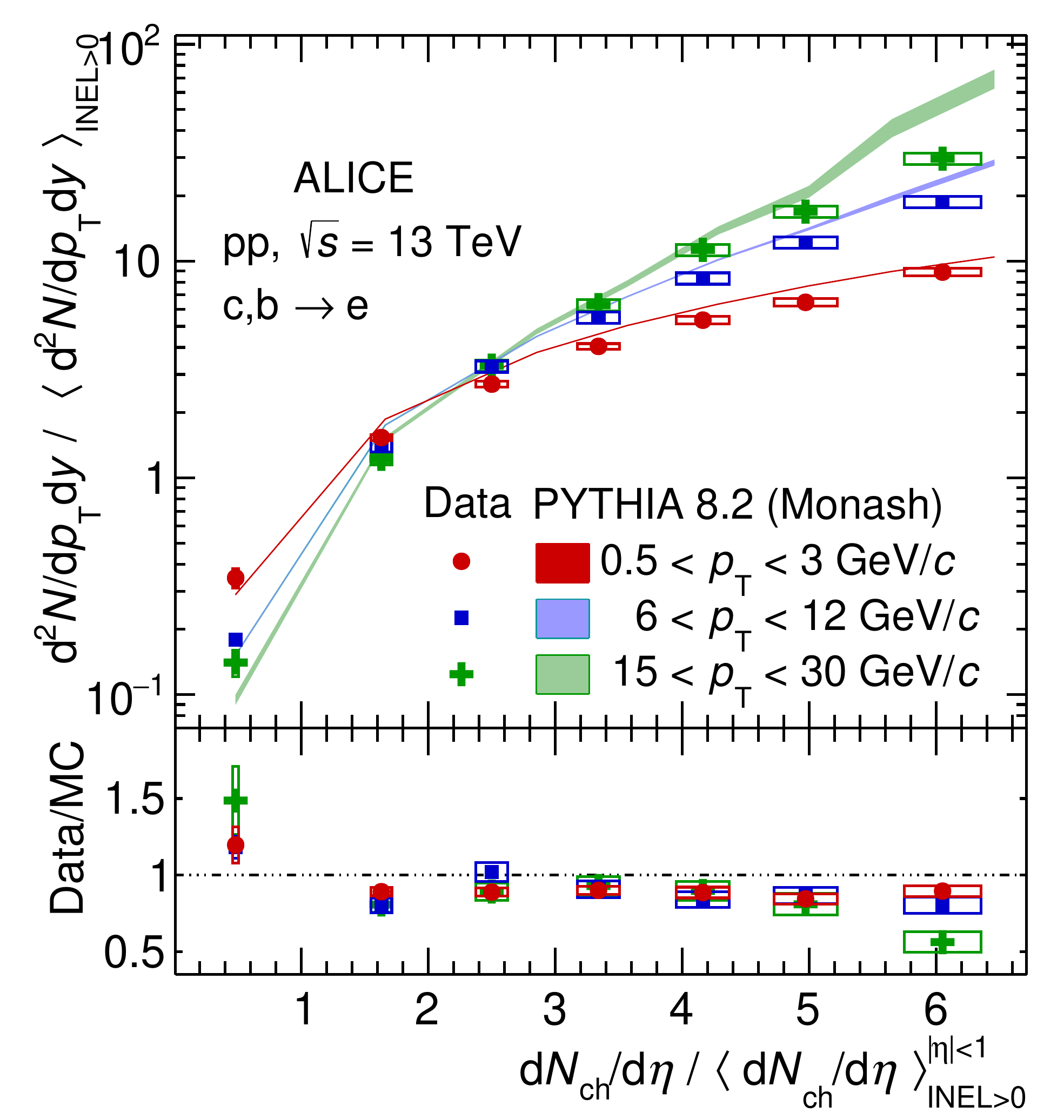}
\includegraphics[width=0.5\linewidth]{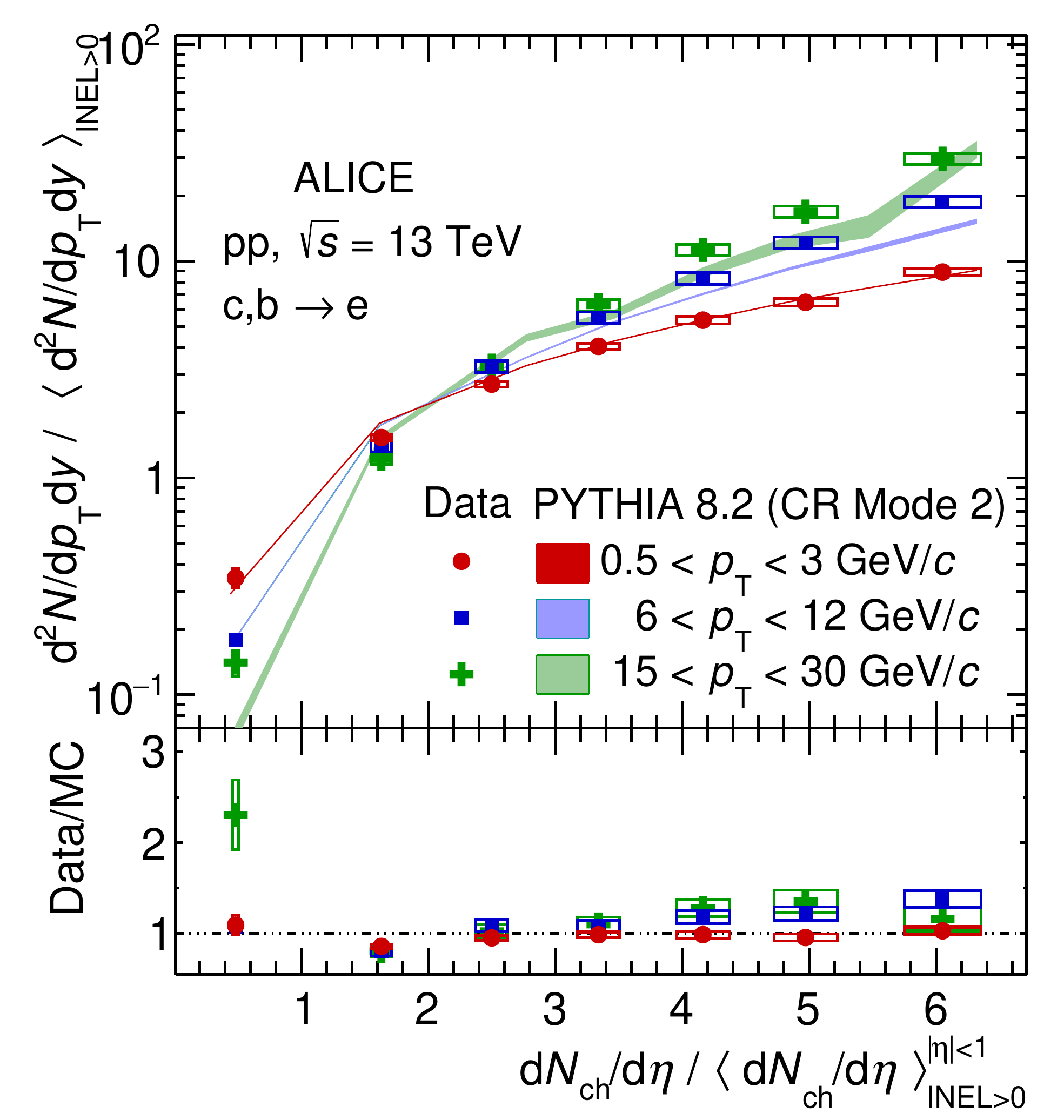}
\caption{Comparison of the self-normalised yield of electrons from heavy-flavour hadron decays as a function of multiplicity measured in \pp collisions at \sqrts $= 13~\rm TeV$ for different \pt intervals with PYTHIA 8.2 Monash tune (left) and PYTHIA 8.2 with CR mode 2 (right). The width of the band is the statistical uncertainty from PYTHIA simulations. The bottom panel shows the ratio of data with respect to the MC predictions. The vertical bars correspond to the propagated statistical error from the data and the MC predictions, and the boxes correspond to systematical uncertainties from the data. }
     \label{Fig:SelfnormalisedYield_pp_PYTHIA}
\end{figure}

The self-normalised yield of electrons from heavy-flavour hadron decays is compared in Fig.~\ref{Fig:SelfnormalisedYield_pp_PYTHIA} with PYTHIA 8.2 simulations using different tunes. 
In the PYTHIA~8.2 framework, multiparton interactions (MPI) and the colour reconnection (CR) mechanism are implemented, which reproduce the charged-particle multiplicity distribution measured at the LHC~\cite{Adam:2016mkz, ATL-PHYS-PUB-2017-008}. These mechanisms are important in order to describe the stronger than linear increase of charm and beauty production with multiplicity as demonstrated in ~\cite{Adam:2015ota}. 
The charged-particle multiplicity also includes particles directly produced in the same hard partonic scattering process in which the heavy quark is created, making them strongly related. These dependencies come from the initial- and final-state radiations, decays of heavy-flavour hadrons, and charged particles produced in the jet fragmentation and are known as auto-correlation effects.
A study of the self-normalised yield of heavy-flavour particles using the PYTHIA 8.2 generator shows that the stronger than linear increase of the yield of heavy-flavour particles is mainly driven by auto-correlation effects. In the absence of auto-correlation effects the increase of the yield of particles produced in hard scattering processes is weaker than linear for multiplicities exceeding about three times the mean multiplicity~\cite{Weber:2018ddv}.
In PYTHIA 8.2, the \pt dependence of the increase of the self-normalised yield with multiplicity is also due to auto-correlation effects introduced by the parton fragmentation because high momentum partons are accompanied by a larger number of fragments which contribute to the multiplicity. In the case of electrons from heavy-flavour hadron decays, the high-\pt part of the spectra is dominated by beauty decay electrons, whose yield was demonstrated to have a more pronounced increase with multiplicity due to the larger jet activity~\cite{Weber:2018ddv}.
In the left panel of Fig.~\ref{Fig:SelfnormalisedYield_pp_PYTHIA}, the measured self-normalised yield of electrons from heavy-flavour hadron decays is compared to calculations with the PYTHIA 8.2 Monash tune that describe the overall trend in data, but the slope is overestimated at high \pt.
In the right panel of Fig.~\ref{Fig:SelfnormalisedYield_pp_PYTHIA}, an improved tune which includes string formation beyond the leading-colour approximation i.e. PYTHIA 8.2 with  CR mode 2~\cite{Sjostrand:2014zea, Christiansen:2015yqa}, is shown to reproduce the \pt dependence,  however the slope is underestimated at high \pt.

\begin{figure}[!h]
\centering
\includegraphics[width=0.5\linewidth]{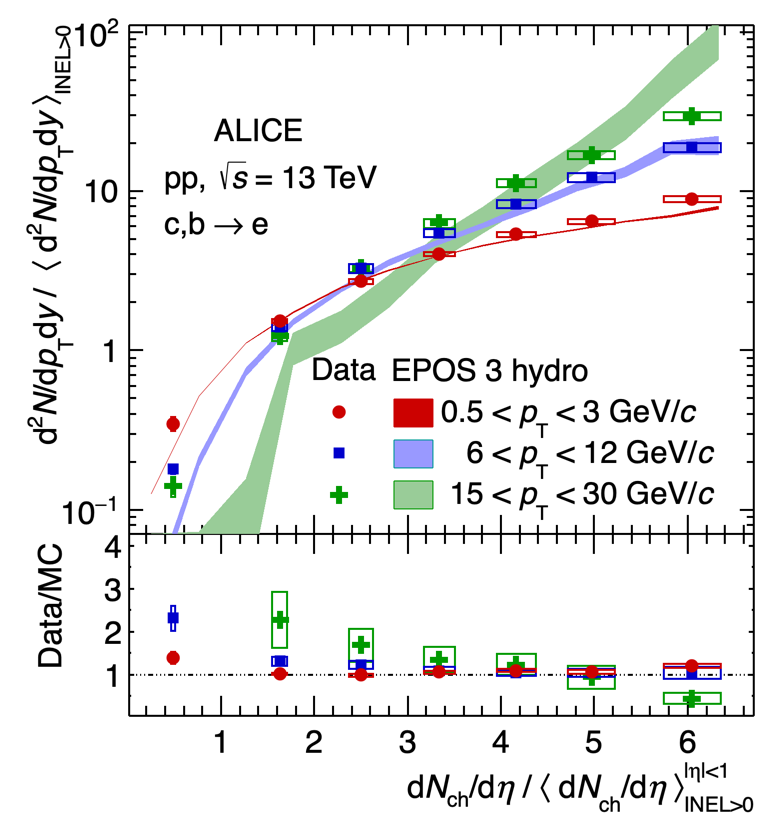}
\caption{Comparison of self-normalised yield of electrons from heavy-flavour hadron decays as a function of multiplicity measured in \pp collisions at \sqrts $= 13~\rm TeV$ for different \pt intervals with EPOS 3 hydro calculations. The width of the band is the statistical uncertainty from EPOS simulations. The bottom panel shows the ratio of data with respect to the MC predictions. The vertical bars correspond to the propagated statistical error from the data and the MC predictions, and the boxes correspond to systematical uncertainties from the data. The ratio for the lowest multiplicity point  for 15 $<$ \pt $<$ 30 \GeVc~(not shown in the figure) is 13$\pm$18.}
     \label{Fig:SelfnormalisedYield_pp_EPOS}
\end{figure}

Calculations with the EPOS 3 event generator~\cite{Werner:2013tya} are able to reproduce the data well, except for the highest measured \pt interval, as can be seen in Fig.~\ref{Fig:SelfnormalisedYield_pp_EPOS}. In the EPOS 3 model, the elementary scattering objects are pomerons, which are exchanged between the partons participating in the collision. The pomerons consist of a hard pQCD scattering vertex, accompanied by initial (space-like) and final (time-like) state parton emission.
The production of a hard probe is more likely from events with hard pomeron exchanges. 
This implies that for a given charged-particle multiplicity the presence of heavy-flavour hadrons favours events with fewer but harder pomerons, which leads to a stronger than linear increase of heavy-flavour production with charged-particle multiplicity. The increase also gets stronger with the increasing \pt, which, as discussed above for the case of PYTHIA 8.2 simulations, is related to the hardness of the partonic scattering and the accompanying  jet activity in the event. The subsequent hydrodynamic evolution of the system then amplifies the increase because the charged-particle multiplicity is reduced by the hydrodynamic expansion, in contrast to the heavy-flavour production.  The charged-particle multiplicity is reduced because part of the available energy goes into flow rather than particle production~\cite{Werner:2016nsq}.

\begin{figure}[!h]
\centering
\includegraphics[width=0.45\linewidth]{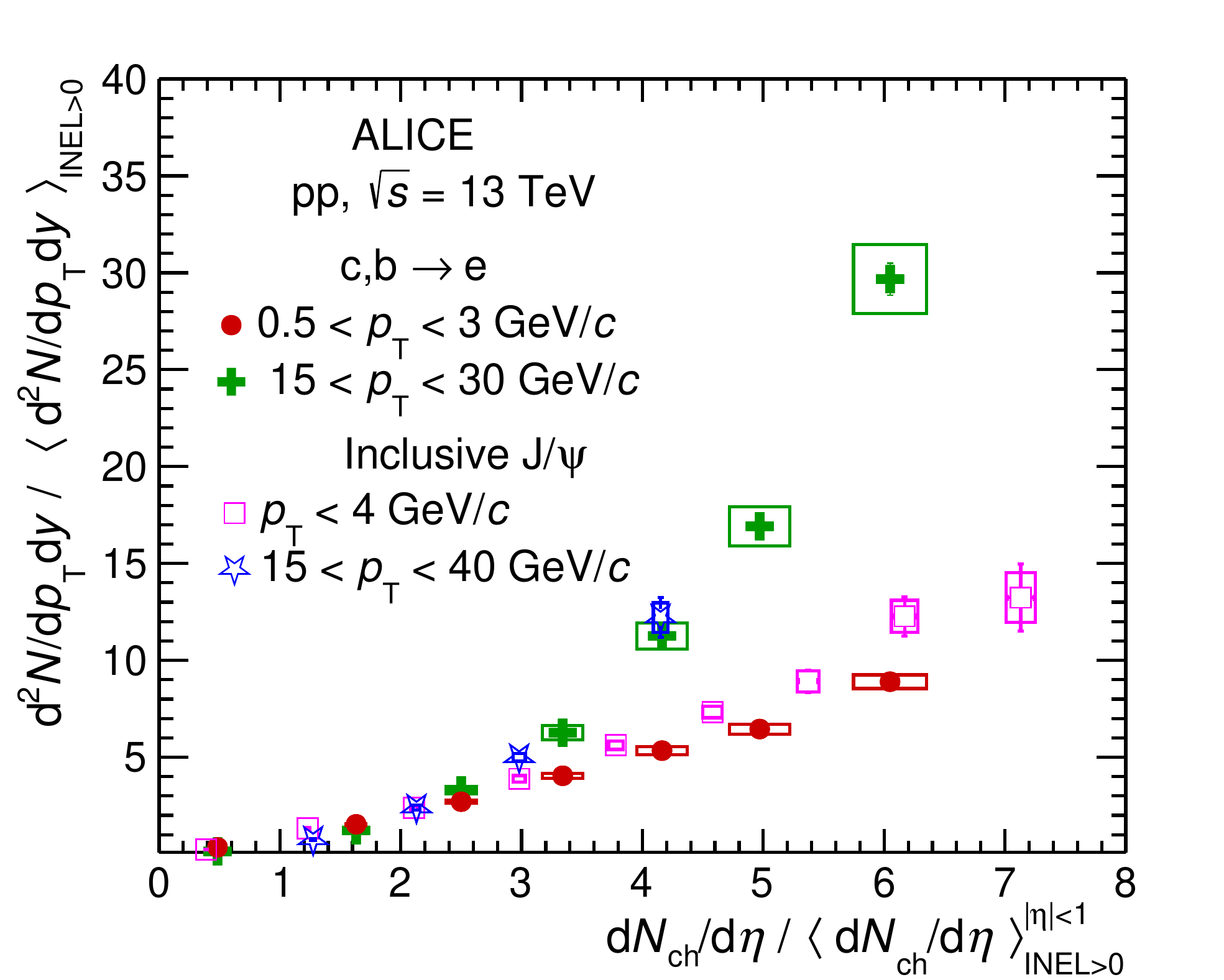}
\includegraphics[width=0.45\linewidth]{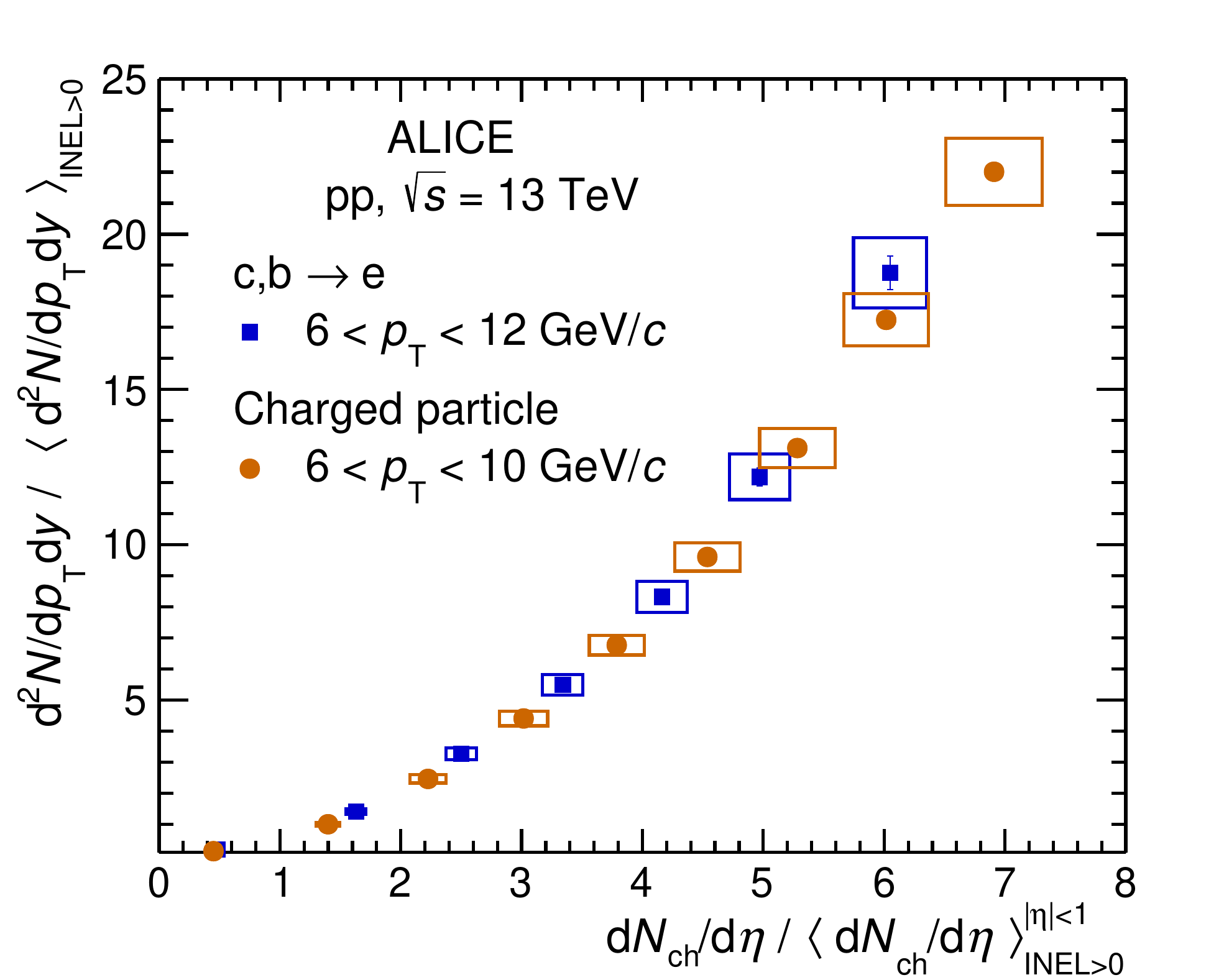}
\includegraphics[width=0.45\linewidth]{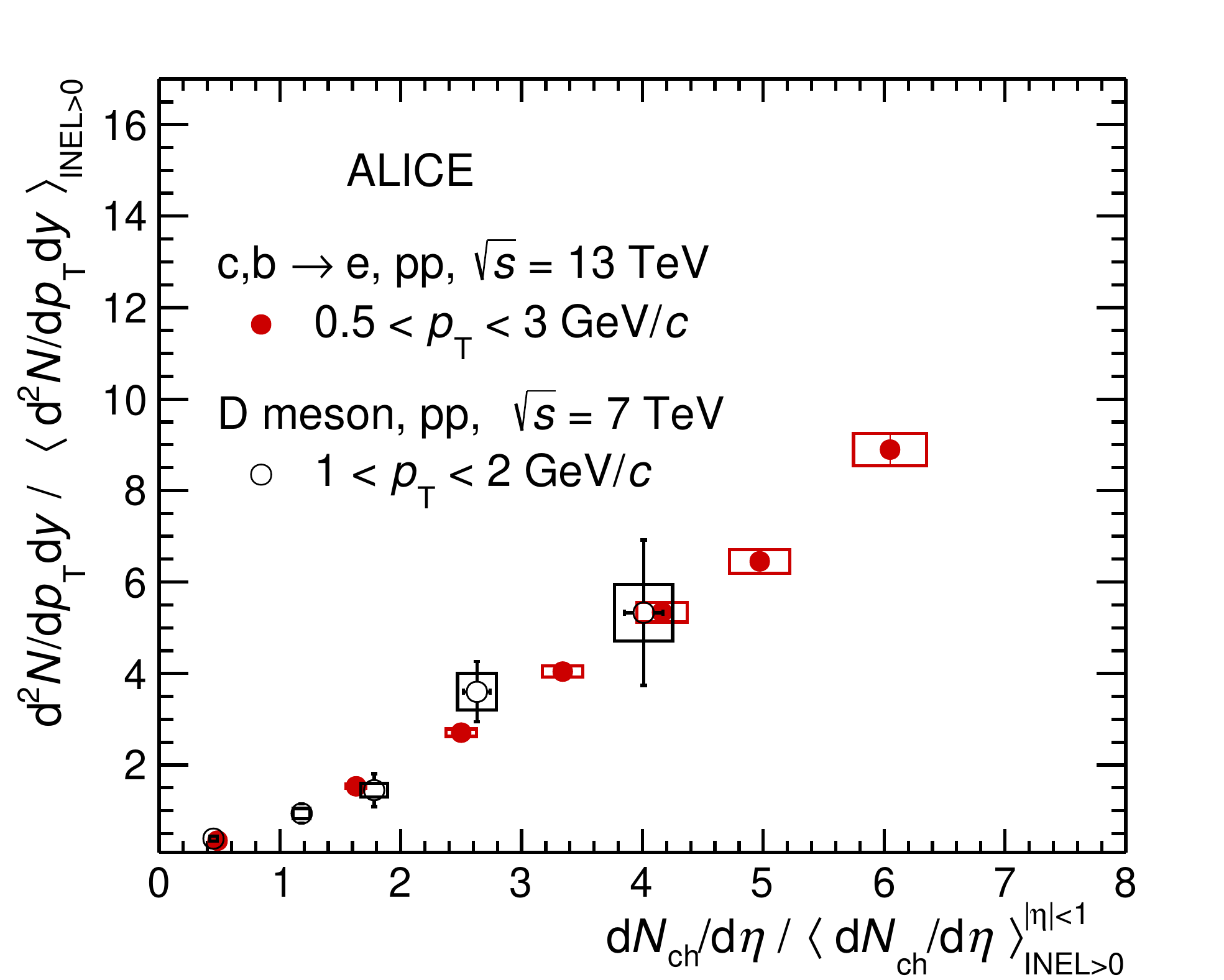}
\includegraphics[width=0.45\linewidth]{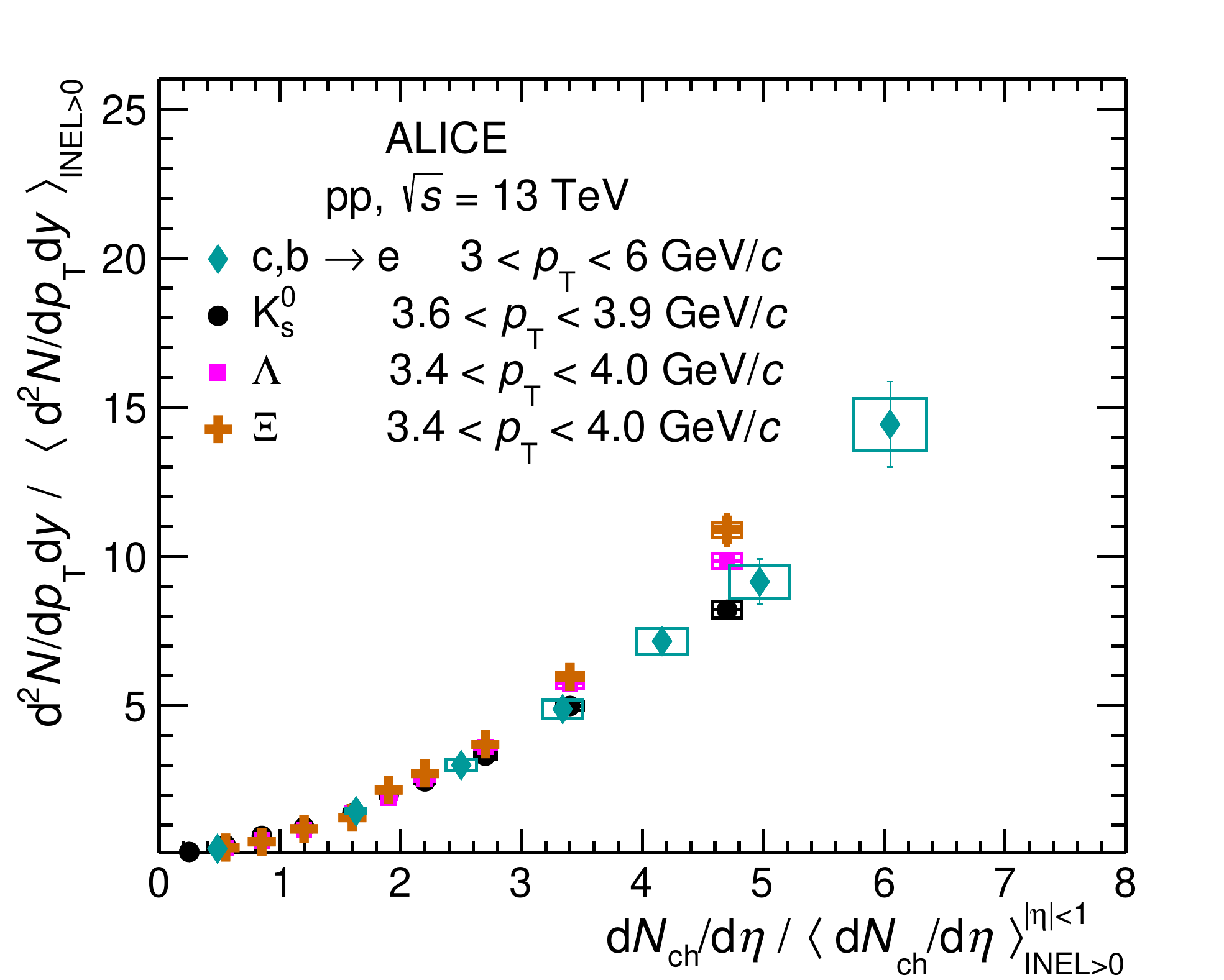}
\caption{Comparison of the self-normalised yield of electrons from heavy-flavour hadron decays measured in \pp collisions at \sqrts $= 13~\rm TeV$  with the self-normalised yields of J/$\psi$ in \pp collisions at \sqrts $= 13~\rm TeV$ (top left), charged particles in \pp collisions at \sqrts $= 13~\rm TeV$ (top right), D mesons in \pp collisions at \sqrts $= 7~\rm TeV$ (bottom left) and strange particles in \pp collisions at \sqrts $= 13~\rm TeV$ (bottom right), in comparable \pt bins.}
\label{Fig:SNY_CompOtherPart_pp}
\end{figure}

The trend of the self-normalised yield of electrons in pp collisions as a function of self-normalised multiplicity is compared in Fig~\ref{Fig:SNY_CompOtherPart_pp} with the self-normalised yield of other particles measured by the ALICE Collaboration, namely J/$\psi$~\cite{ALICE:2020msa}, charged particles~\cite{Acharya:2019mzb}, strange hadrons~\cite{Acharya:2019kyh} in pp collisions at $\sqrt{s}$ $= 13~\rm TeV$, and D mesons~\cite{Adam:2015ota} in pp collisions at $\sqrt{s}$ $= 7~\rm TeV$. The self-normalised yields for strange hadrons were calculated using the multiplicity-dependent cross section measurements reported in~\cite{Acharya:2019kyh}. These self-normalised yields allow a direct comparison of multiplicity-dependent production of different particle species, with the advantage that the charged-particle pseudorapidity density is measured using the same detector and procedure. 
The \pt ranges of electrons are selected to be similar to the measured \pt range of the compared particles, with a caveat that the \pt interval of electron parents (heavy-flavour hadrons) is considerably broader and shifted towards higher \pt values compared to the
one of the electrons. The slope of the increase of the self-normalised yield of electrons from heavy-flavour hadron decays as a function of self-normalised multiplicity at midrapidity is similar to that measured for J/$\psi$, charged particles, strange mesons, and D mesons in similar \pt ranges. 
At high and intermediate \pt, the production of hadrons is dominated by hard partonic scattering processes, independent of the particle species, accompanied by jet activity in the event. For heavy-flavour particles this is also true at low \pt due to the large charm and beauty quark masses. As it was discussed above, in PYTHIA 8.2, the particle production associated with jet activity leads to strong auto-correlation effects, which give rise to the observed stronger than linear increase of particle yields, making the self-normalised yield of the different particles reported here compatible with each other.

\begin{figure}[!ht]
\centering
\includegraphics[width=0.6\linewidth]{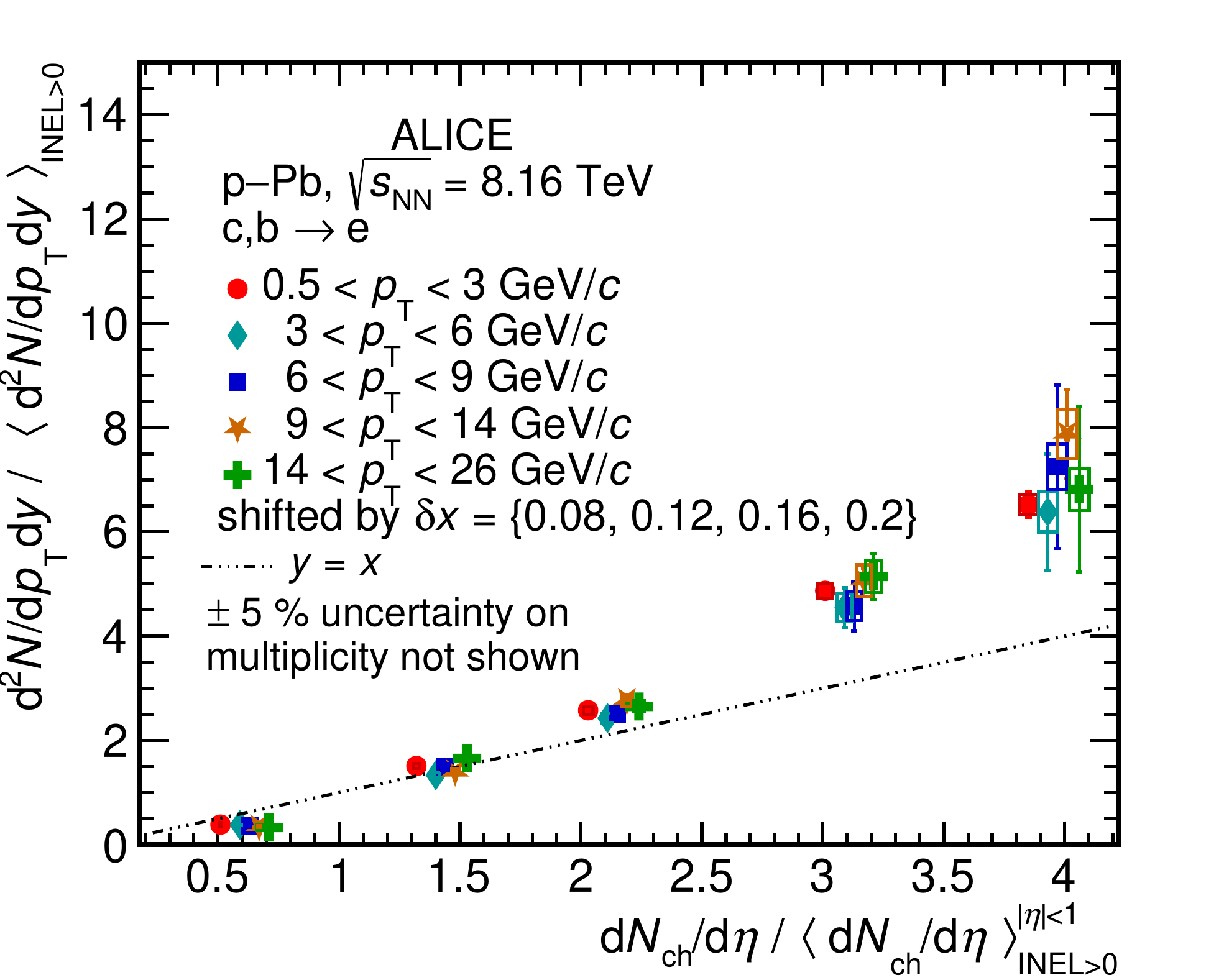}

\caption{Self-normalised yield of electrons from heavy-flavour hadron decays as a function of self-normalised charged-particle pseudorapidity density at midrapidity measured in \pPb collisions at \sqrtsNN $= 8.16~\rm TeV$ in different \pt intervals. The position of the points on the $x$-axis are shifted horizontally by $\rm \delta x$  to improve the visibility.}
\label{Fig:SelfnormalisedYield_pPb}
\end{figure}

The self-normalised yield of electrons from heavy-flavour hadron decays 
as a function of the self-normalised charged-particle pseudorapidity density 
  for \pPb collisions at $\mbox{\sqrtsNN $= 8.16~{\rm TeV}$}$ is presented in Fig~\ref{Fig:SelfnormalisedYield_pPb}. The results are self-normalised to the INEL$>0$ event class, similarly to pp collisions. The dashed line is a linear function with a slope of unity as shown in the figure. The measurements were performed in five $p_{\rm T}$ intervals from 0.5 \GeVc to 26 \GeVc. 
Events with multiplicity more than four times larger than the average multiplicity in \pPb collisions are studied. 
The self-normalised yield of electrons from heavy-flavour hadron decays grows faster than linear with the self-normalised multiplicity. The measurements in $\pt$ intervals show no $\pt$ dependence within the uncertainties of the measurement. The yield increase is approximately a factor of seven for multiplicities four times larger than the average multiplicity.

\begin{figure}[!h]
\includegraphics[width=0.48\linewidth]{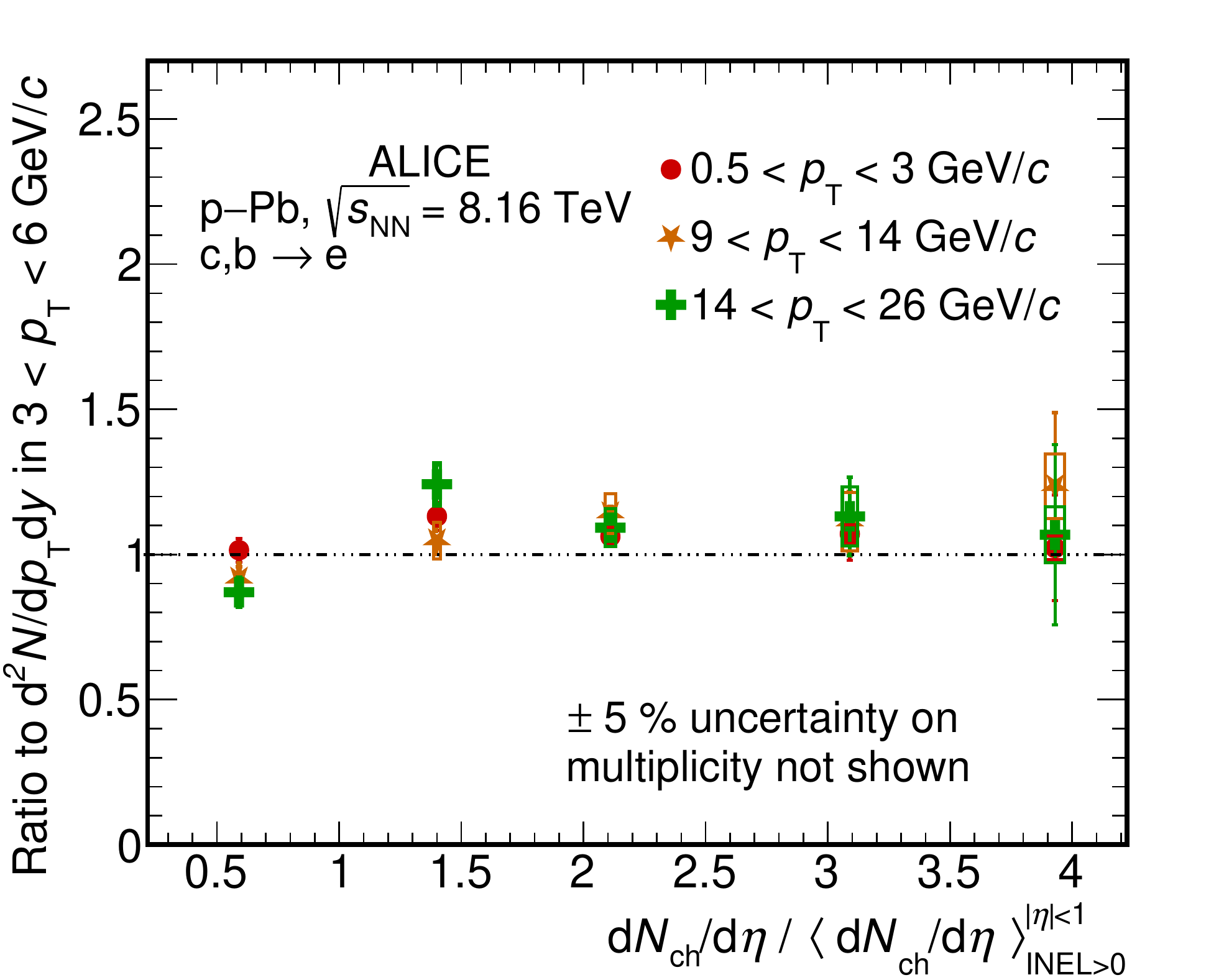}
\includegraphics[width=0.48\linewidth]{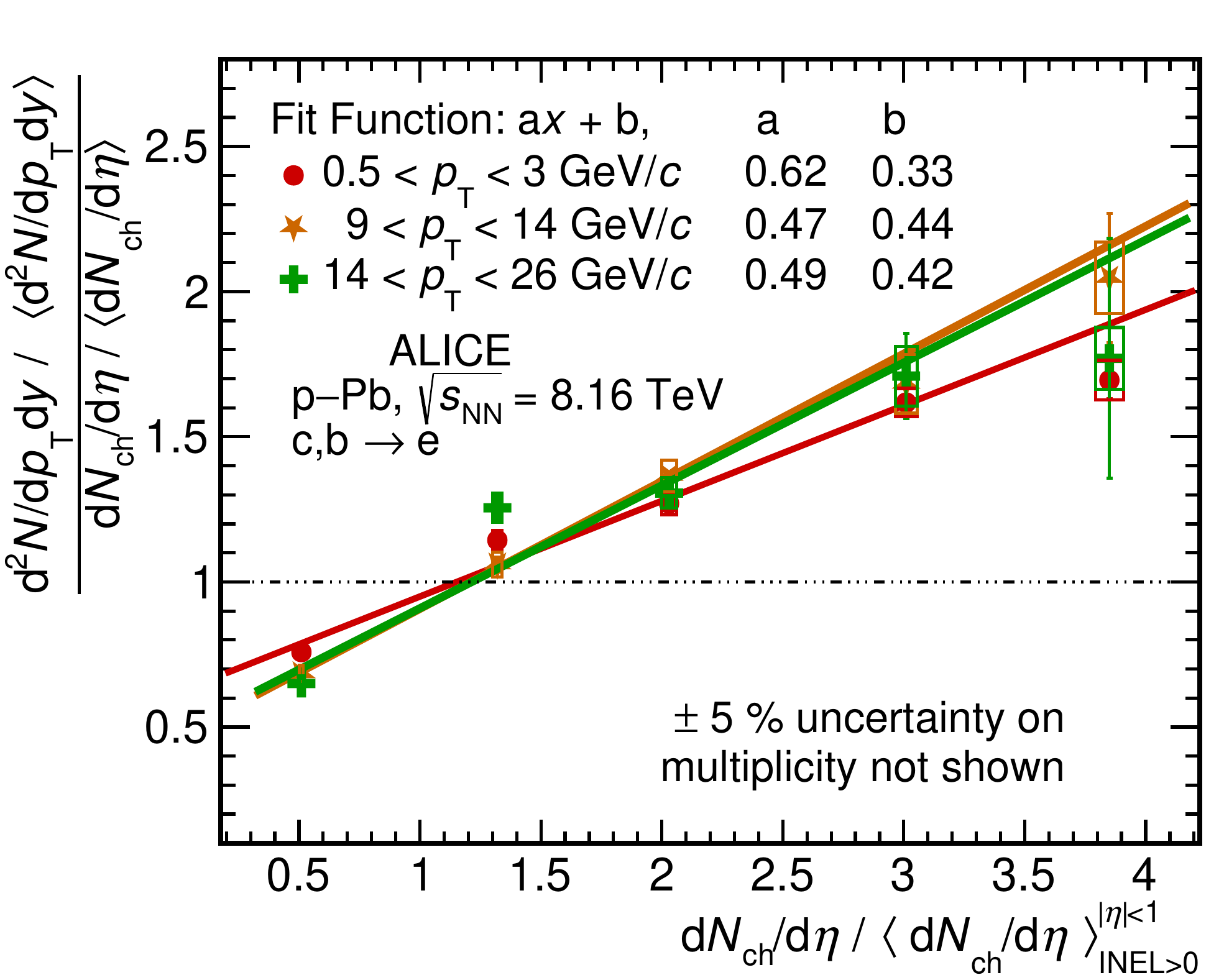}

\caption{Ratio of the  self-normalised yield
in different \pt intervals with respect to that in the $3 < \pt < 6~{\rm GeV}/c$ interval (left). Double ratio of the self-normalised yield of heavy-flavour hadron decay electrons to the self-normalised multiplicity in \pPb collisions at \sqrtsNN $= 8.16~\rm TeV$ in three \pt ranges (right).}
\label{Fig:SNY_Ratio_pPb}
\end{figure}

In the left panel of Fig.~\ref{Fig:SNY_Ratio_pPb}, the ratios of the self-normalised yield of electrons from heavy-flavour hadron decays 
in various \pt intervals with respect to the one measured in the 3 $ < \pt < $ 6 \GeVc interval are shown. Contrary to the pp collision case,  within the uncertainties no \pt dependence is observed. The right panel of Fig.~\ref{Fig:SNY_Ratio_pPb} shows the double ratio of the self-normalised heavy-flavour hadron decay electron yield to the self-normalised multiplicity. The double ratio increases with multiplicity, with no dependence on \pt. The double ratio was fitted with a linear function, which reasonably describes the data for all \pt intervals. This indicates that in the measured \pt range the yield increases approximately with the square of the multiplicity with a similar coefficient for all \pt intervals.

Though the self-normalised yields of electrons from heavy-flavour hadron decays  in pp and \pPb collisions show similar features in their increase with multiplicity, a quantitative comparison of the measurements between the two systems is not straightforward. In pp collisions, a high multiplicity event arises mostly from hard events, with multiparton interactions and jets fragmenting in multiple hadrons.  In \pPb collisions, the multiplicity dependence of heavy-flavour production is also driven by the presence of multiple binary nucleon--nucleon interactions, which make the contribution from possible auto-correlation effects smaller in such collisions. In \pPb collisions, an event with a high multiplicity value similar to those in pp collisions can come from the superposition of a few soft nucleon--nucleon collisions. Therefore, for similar multiplicity, the hardness of the event is not the same in the two systems.

\begin{figure}[!h]
\centering
\includegraphics[width=0.55\linewidth]{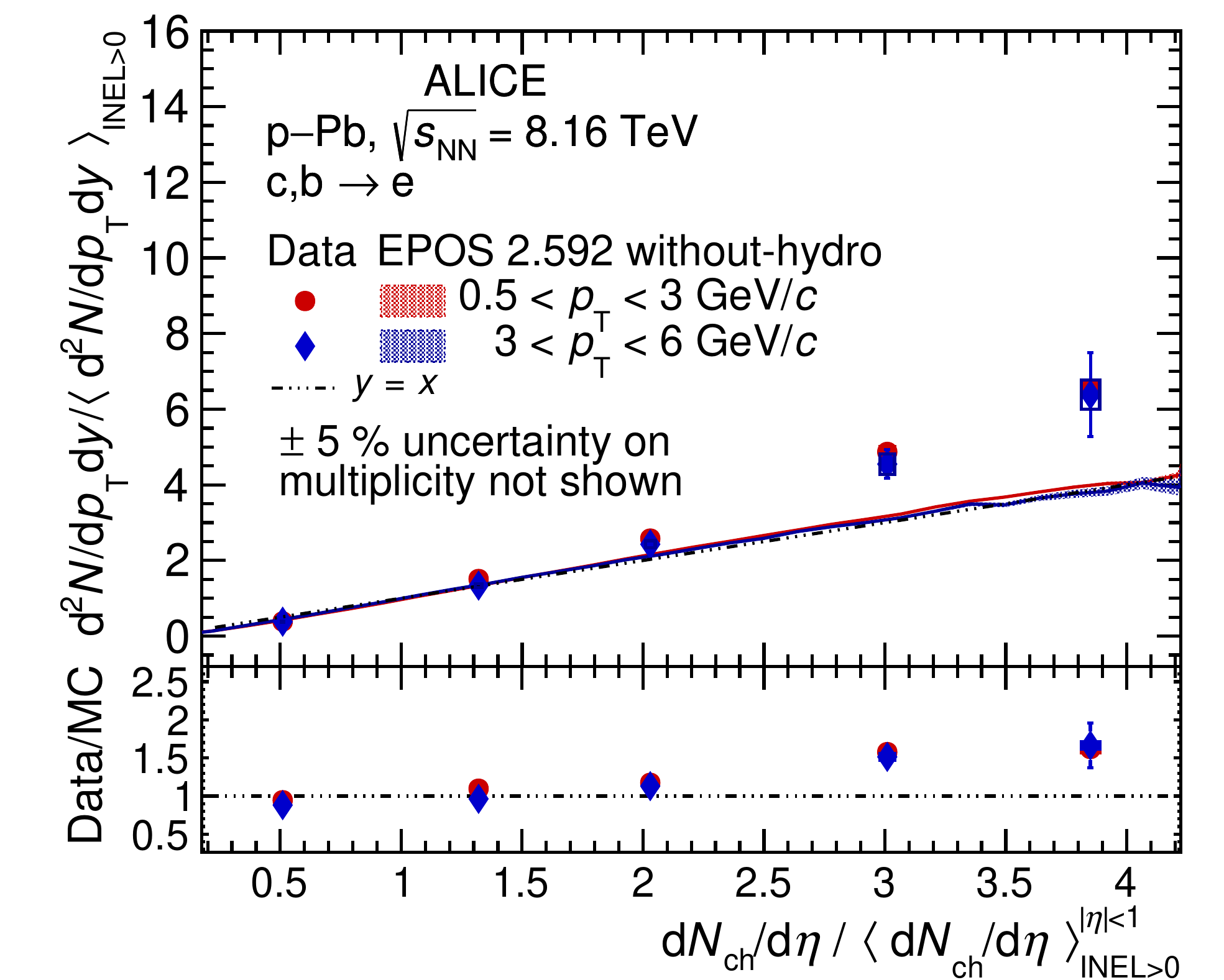}
\caption{ Self-normalised yields of electrons from heavy-flavour hadron decays as a function of self-normalised charged-particle pseudorapidity density at midrapidity measured in \pPb collisions at \sqrtsNN $= 8.16~\rm TeV$ compared with EPOS 2.592 without-hydrodynamics in two \pt intervals 0.5 $< \pt <$ 3 \GeVc and 3 $< \pt <$ 6 \GeVc. The width of the band is the statistical uncertainty from EPOS simulations. The bottom panel shows the ratio of data with respect to the MC predictions. The vertical bars correspond to the propagated statistical uncertainties from the data and the MC predictions, and the boxes correspond to systematical uncertainties from the data.}
\label{Fig:SNYComparisonWithEPOS_pPb8TeV}
\end{figure}

The self-normalised yield of electrons from heavy-flavour hadron decays is compared in Fig.~\ref{Fig:SNYComparisonWithEPOS_pPb8TeV} with EPOS 2.592 simulations ~\cite{Werner:2013tya, Werner:2010aa}. The measurements in two \pt intervals are compared with the EPOS model without the hydrodynamic component, as provided by the authors. The EPOS model shows no \pt dependence similar to the observations in the data, but underpredicts the data at high multiplicity, showing an almost linear increase. 

\begin{figure}[!ht]
\centering
\includegraphics[width=0.55\linewidth]{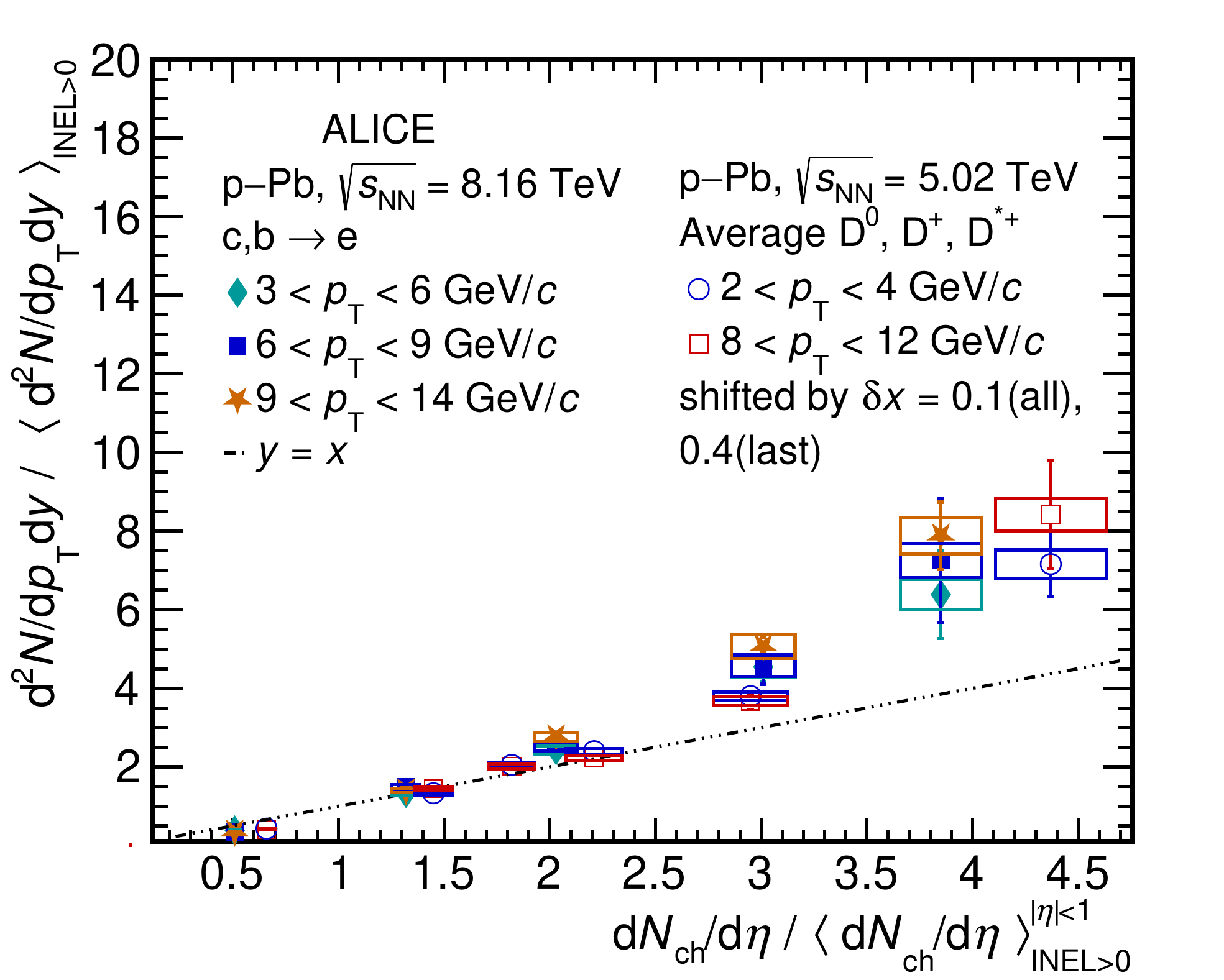}
\caption{Self-normalised yields of electrons from heavy-flavour hadron decays measured in \pPb collisions at \sqrtsNN $= 8.16~\rm TeV$ for different \pt intervals compared with self-normalised yields of D mesons in \pPb collisions at \sqrtsNN $= 5.02~\rm TeV$. The position of the points for \pPb collisions at \sqrtsNN $= 5.02~\rm TeV$ on the $x$-axis are shifted horizontally by $\rm \delta x$ to improve the visibility.}
      \label{Fig:SNP_CompOtherPart_pPb}
\end{figure}

The self-normalised electron yields in \pPb collisions in different \pt ranges are also compared with the normalised yields of D mesons~\cite{Adam:2016mkz} in \pPb collisions at \sqrtsNN $= 5.02~\rm TeV$ in Fig.~\ref{Fig:SNP_CompOtherPart_pPb}. Similar to the observation in pp collisions, the self-normalised yield of electrons from heavy-flavour hadron decays as a function of the self-normalised multiplicity shows a trend compatible with the one of D mesons. Also the multiplicity dependence of D meson yields in \pPb collisions does not show a \pt dependence, which gives a hint that the production mechanisms of charm and beauty as a function of the multiplicity in \pPb collisions are similar.

\section{Summary}\label{section:summary}
Heavy-flavour production at midrapidity was studied using electrons from heavy-flavour hadron decays in pp collisions at \sqrts $= 13~\rm \TeV$ and in \pPb collisions at \sqrtsNN $= 8.16~\rm TeV$ with the ALICE detector at the LHC. The \pt-differential production cross section of electrons from heavy-flavour hadron decays in pp collisions was compared with FONLL and GM-VFNS (b $\rightarrow$ B $\rightarrow$ D $\rightarrow$ e, b $\rightarrow$ B $\rightarrow$ e, c  $\rightarrow$ D $\rightarrow$ e) pQCD calculations. The data are observed to lie on the upper edge of the FONLL uncertainties. The GM-VFNS calculation underestimates the cross section at low \pt but describes the data within the uncertainties for \pt $>$ 5 \GeVc. The nuclear modification factor in \pPb collisions, $R_{\rm{pPb}}$, was computed and is consistent with unity within the statistical and systematic uncertainties. The $R_{\rm{pPb}}$ measurement shows no effects that could signal the formation of a hot medium and no significant cold nuclear matter effects within the uncertainties of the data in the measured \pt range. The $R_{\rm{pPb}}$ at \sqrtsNN $= 8.16~\rm TeV$ is consistent with that measured at \sqrtsNN $= 5.02~\rm TeV$. 

The multiplicity-dependent production of electrons from heavy-flavour hadron decays was measured using the self-normalised yield as a function of self-normalised charged-particle pseudorapidity density at midrapidity in pp and \pPb collisions as a function of transverse momentum. A faster than linear increase was observed in both pp and p--Pb collisions. While in \pPb collisions, no \pt dependence is observed within uncertainties, in pp collisions a strong \pt dependence is seen with high-\pt electrons showing a faster increase as a function of the self-normalised multiplicity. The measurement of self-normalised yield of electrons from heavy-flavour hadron decays in pp collisions was compared with PYTHIA 8.2 and EPOS 3 simulations, which describe the data. The measurement in \pPb collisions was compared with the EPOS 2.592 model without hydrodynamics, which underestimates the data. The comparison of self-normalised yields of heavy-flavour and light-flavour particles show a similar stronger than linearly increasing trend in both colliding systems. In pp collisions the stronger than linear increase of heavy-flavour particles is mainly driven by auto-correlation effects which are independent of particle species, whereas in the case of \pPb collisions, it is difficult to draw any conclusion since the multiplicity dependence of heavy-flavour production is also largely affected by the presence of multiple binary nucleon–nucleon interactions.


\newenvironment{acknowledgement}{\relax}{\relax}
\begin{acknowledgement}
\section*{Acknowledgements}

The ALICE Collaboration would like to thank all its engineers and technicians for their invaluable contributions to the construction of the experiment and the CERN accelerator teams for the outstanding performance of the LHC complex.
The ALICE Collaboration gratefully acknowledges the resources and support provided by all Grid centres and the Worldwide LHC Computing Grid (WLCG) collaboration.
The ALICE Collaboration acknowledges the following funding agencies for their support in building and running the ALICE detector:
A. I. Alikhanyan National Science Laboratory (Yerevan Physics Institute) Foundation (ANSL), State Committee of Science and World Federation of Scientists (WFS), Armenia;
Austrian Academy of Sciences, Austrian Science Fund (FWF): [M 2467-N36] and Nationalstiftung f\"{u}r Forschung, Technologie und Entwicklung, Austria;
Ministry of Communications and High Technologies, National Nuclear Research Center, Azerbaijan;
Conselho Nacional de Desenvolvimento Cient\'{\i}fico e Tecnol\'{o}gico (CNPq), Financiadora de Estudos e Projetos (Finep), Funda\c{c}\~{a}o de Amparo \`{a} Pesquisa do Estado de S\~{a}o Paulo (FAPESP) and Universidade Federal do Rio Grande do Sul (UFRGS), Brazil;
Bulgarian Ministry of Education and Science, within the National Roadmap for Research Infrastructures 2020-2027 (object CERN), Bulgaria;
Ministry of Education of China (MOEC) , Ministry of Science \& Technology of China (MSTC) and National Natural Science Foundation of China (NSFC), China;
Ministry of Science and Education and Croatian Science Foundation, Croatia;
Centro de Aplicaciones Tecnol\'{o}gicas y Desarrollo Nuclear (CEADEN), Cubaenerg\'{\i}a, Cuba;
Ministry of Education, Youth and Sports of the Czech Republic, Czech Republic;
The Danish Council for Independent Research | Natural Sciences, the VILLUM FONDEN and Danish National Research Foundation (DNRF), Denmark;
Helsinki Institute of Physics (HIP), Finland;
Commissariat \`{a} l'Energie Atomique (CEA) and Institut National de Physique Nucl\'{e}aire et de Physique des Particules (IN2P3) and Centre National de la Recherche Scientifique (CNRS), France;
Bundesministerium f\"{u}r Bildung und Forschung (BMBF) and GSI Helmholtzzentrum f\"{u}r Schwerionenforschung GmbH, Germany;
General Secretariat for Research and Technology, Ministry of Education, Research and Religions, Greece;
National Research, Development and Innovation Office, Hungary;
Department of Atomic Energy Government of India (DAE), Department of Science and Technology, Government of India (DST), University Grants Commission, Government of India (UGC) and Council of Scientific and Industrial Research (CSIR), India;
National Research and Innovation Agency - BRIN, Indonesia;
Istituto Nazionale di Fisica Nucleare (INFN), Italy;
Japanese Ministry of Education, Culture, Sports, Science and Technology (MEXT) and Japan Society for the Promotion of Science (JSPS) KAKENHI, Japan;
Consejo Nacional de Ciencia (CONACYT) y Tecnolog\'{i}a, through Fondo de Cooperaci\'{o}n Internacional en Ciencia y Tecnolog\'{i}a (FONCICYT) and Direcci\'{o}n General de Asuntos del Personal Academico (DGAPA), Mexico;
Nederlandse Organisatie voor Wetenschappelijk Onderzoek (NWO), Netherlands;
The Research Council of Norway, Norway;
Commission on Science and Technology for Sustainable Development in the South (COMSATS), Pakistan;
Pontificia Universidad Cat\'{o}lica del Per\'{u}, Peru;
Ministry of Education and Science, National Science Centre and WUT ID-UB, Poland;
Korea Institute of Science and Technology Information and National Research Foundation of Korea (NRF), Republic of Korea;
Ministry of Education and Scientific Research, Institute of Atomic Physics, Ministry of Research and Innovation and Institute of Atomic Physics and University Politehnica of Bucharest, Romania;
Ministry of Education, Science, Research and Sport of the Slovak Republic, Slovakia;
National Research Foundation of South Africa, South Africa;
Swedish Research Council (VR) and Knut \& Alice Wallenberg Foundation (KAW), Sweden;
European Organization for Nuclear Research, Switzerland;
Suranaree University of Technology (SUT), National Science and Technology Development Agency (NSTDA), Thailand Science Research and Innovation (TSRI) and National Science, Research and Innovation Fund (NSRF), Thailand;
Turkish Energy, Nuclear and Mineral Research Agency (TENMAK), Turkey;
National Academy of  Sciences of Ukraine, Ukraine;
Science and Technology Facilities Council (STFC), United Kingdom;
National Science Foundation of the United States of America (NSF) and United States Department of Energy, Office of Nuclear Physics (DOE NP), United States of America.
In addition, individual groups or members have received support from:
European Research Council, Strong 2020 - Horizon 2020 (grant nos. 950692, 824093), European Union;
Academy of Finland (Center of Excellence in Quark Matter) (grant nos. 346327, 346328), Finland;
DST-DAAD Project-based Personnel Exchange Programme, India;
Programa de Apoyos para la Superaci\'{o}n del Personal Acad\'{e}mico, UNAM, Mexico.

\end{acknowledgement}

\bibliographystyle{utphys}   
\bibliography{bibliography}

\newpage
\appendix

%
%

\section{The ALICE Collaboration}
\label{app:collab}
\begin{flushleft} 
\small

S.~Acharya\,\orcidlink{0000-0002-9213-5329}\,$^{\rm 125}$, 
D.~Adamov\'{a}\,\orcidlink{0000-0002-0504-7428}\,$^{\rm 86}$, 
A.~Adler$^{\rm 69}$, 
G.~Aglieri Rinella\,\orcidlink{0000-0002-9611-3696}\,$^{\rm 32}$, 
M.~Agnello\,\orcidlink{0000-0002-0760-5075}\,$^{\rm 29}$, 
N.~Agrawal\,\orcidlink{0000-0003-0348-9836}\,$^{\rm 50}$, 
Z.~Ahammed\,\orcidlink{0000-0001-5241-7412}\,$^{\rm 132}$, 
S.~Ahmad\,\orcidlink{0000-0003-0497-5705}\,$^{\rm 15}$, 
S.U.~Ahn\,\orcidlink{0000-0001-8847-489X}\,$^{\rm 70}$, 
I.~Ahuja\,\orcidlink{0000-0002-4417-1392}\,$^{\rm 37}$
% ...

\section*{Affiliation Notes}

$^{\rm I}$ Deceased\\
$^{\rm II}$ Also at: Max-Planck-Institut f\"{u}r Physik, Munich, Germany\\
$^{\rm III}$ Also at: Italian National Agency for New Technologies, Energy and Sustainable Economic Development (ENEA), Bologna, Italy\\
$^{\rm IV}$ Also at: Department of Applied Physics, Aligarh Muslim University, Aligarh, India\\
$^{\rm V}$ Also at: Institute of Theoretical Physics, University of Wroclaw, Poland\\
$^{\rm VI}$ Also at: An institution covered by a cooperation agreement with CERN\\

\section*{Collaboration Institutes}

$^{1}$ A.I. Alikhanyan National Science Laboratory (Yerevan Physics Institute) Foundation, Yerevan, Armenia\\
$^{2}$ AGH University of Science and Technology, Cracow, Poland\\
$^{3}$ Bogolyubov Institute for Theoretical Physics, National Academy of Sciences of Ukraine, Kiev, Ukraine\\
$^{4}$ Bose Institute, Department of Physics  and Centre for Astroparticle Physics and Space Science (CAPSS), Kolkata, India\\
$^{5}$ California Polytechnic State University, San Luis Obispo, California, United States\\
$^{6}$ Central China Normal University, Wuhan, China\\
$^{7}$ Centro de Aplicaciones Tecnol\'{o}gicas y Desarrollo Nuclear (CEADEN), Havana, Cuba\\
$^{8}$ Centro de Investigaci\'{o}n y de Estudios Avanzados (CINVESTAV), Mexico City and M\'{e}rida, Mexico\\
$^{9}$ Chicago State University, Chicago, Illinois, United States\\
$^{10}$ China Institute of Atomic Energy, Beijing, China\\
$^{11}$ Chungbuk National University, Cheongju, Republic of Korea\\
$^{12}$ Comenius University Bratislava, Faculty of Mathematics, Physics and Informatics, Bratislava, Slovak Republic\\
$^{13}$ COMSATS University Islamabad, Islamabad, Pakistan\\
$^{14}$ Creighton University, Omaha, Nebraska, United States\\
$^{15}$ Department of Physics, Aligarh Muslim University, Aligarh, India\\
$^{16}$ Department of Physics, Pusan National University, Pusan, Republic of Korea\\
$^{17}$ Department of Physics, Sejong University, Seoul, Republic of Korea\\
$^{18}$ Department of Physics, University of California, Berkeley, California, United States\\
$^{19}$ Department of Physics, University of Oslo, Oslo, Norway\\
$^{20}$ Department of Physics and Technology, University of Bergen, Bergen, Norway\\
$^{21}$ Dipartimento di Fisica, Universit\`{a} di Pavia, Pavia, Italy\\
$^{22}$ Dipartimento di Fisica dell'Universit\`{a} and Sezione INFN, Cagliari, Italy\\
$^{23}$ Dipartimento di Fisica dell'Universit\`{a} and Sezione INFN, Trieste, Italy\\
$^{24}$ Dipartimento di Fisica dell'Universit\`{a} and Sezione INFN, Turin, Italy\\
$^{25}$ Dipartimento di Fisica e Astronomia dell'Universit\`{a} and Sezione INFN, Bologna, Italy\\
$^{26}$ Dipartimento di Fisica e Astronomia dell'Universit\`{a} and Sezione INFN, Catania, Italy\\
$^{27}$ Dipartimento di Fisica e Astronomia dell'Universit\`{a} and Sezione INFN, Padova, Italy\\
$^{28}$ Dipartimento di Fisica `E.R.~Caianiello' dell'Universit\`{a} and Gruppo Collegato INFN, Salerno, Italy\\
$^{29}$ Dipartimento DISAT del Politecnico and Sezione INFN, Turin, Italy\\
$^{30}$ Dipartimento di Scienze MIFT, Universit\`{a} di Messina, Messina, Italy\\
$^{31}$ Dipartimento Interateneo di Fisica `M.~Merlin' and Sezione INFN, Bari, Italy\\
$^{32}$ European Organization for Nuclear Research (CERN), Geneva, Switzerland\\
$^{33}$ Faculty of Electrical Engineering, Mechanical Engineering and Naval Architecture, University of Split, Split, Croatia\\
$^{34}$ Faculty of Engineering and Science, Western Norway University of Applied Sciences, Bergen, Norway\\
$^{35}$ Faculty of Nuclear Sciences and Physical Engineering, Czech Technical University in Prague, Prague, Czech Republic\\
$^{36}$ Faculty of Physics, Sofia University, Sofia, Bulgaria\\
$^{37}$ Faculty of Science, P.J.~\v{S}af\'{a}rik University, Ko\v{s}ice, Slovak Republic\\
$^{38}$ Frankfurt Institute for Advanced Studies, Johann Wolfgang Goethe-Universit\"{a}t Frankfurt, Frankfurt, Germany\\
$^{39}$ Fudan University, Shanghai, China\\
$^{40}$ Gangneung-Wonju National University, Gangneung, Republic of Korea\\
$^{41}$ Gauhati University, Department of Physics, Guwahati, India\\
$^{42}$ Helmholtz-Institut f\"{u}r Strahlen- und Kernphysik, Rheinische Friedrich-Wilhelms-Universit\"{a}t Bonn, Bonn, Germany\\
$^{43}$ Helsinki Institute of Physics (HIP), Helsinki, Finland\\
$^{44}$ High Energy Physics Group,  Universidad Aut\'{o}noma de Puebla, Puebla, Mexico\\
$^{45}$ Horia Hulubei National Institute of Physics and Nuclear Engineering, Bucharest, Romania\\
$^{46}$ Indian Institute of Technology Bombay (IIT), Mumbai, India\\
$^{47}$ Indian Institute of Technology Indore, Indore, India\\
$^{48}$ INFN, Laboratori Nazionali di Frascati, Frascati, Italy\\
$^{49}$ INFN, Sezione di Bari, Bari, Italy\\
$^{50}$ INFN, Sezione di Bologna, Bologna, Italy\\
$^{51}$ INFN, Sezione di Cagliari, Cagliari, Italy\\
$^{52}$ INFN, Sezione di Catania, Catania, Italy\\
$^{53}$ INFN, Sezione di Padova, Padova, Italy\\
$^{54}$ INFN, Sezione di Pavia, Pavia, Italy\\
$^{55}$ INFN, Sezione di Torino, Turin, Italy\\
$^{56}$ INFN, Sezione di Trieste, Trieste, Italy\\
$^{57}$ Inha University, Incheon, Republic of Korea\\
$^{58}$ Institute for Gravitational and Subatomic Physics (GRASP), Utrecht University/Nikhef, Utrecht, Netherlands\\
$^{59}$ Institute of Experimental Physics, Slovak Academy of Sciences, Ko\v{s}ice, Slovak Republic\\
$^{60}$ Institute of Physics, Homi Bhabha National Institute, Bhubaneswar, India\\
$^{61}$ Institute of Physics of the Czech Academy of Sciences, Prague, Czech Republic\\
$^{62}$ Institute of Space Science (ISS), Bucharest, Romania\\
$^{63}$ Institut f\"{u}r Kernphysik, Johann Wolfgang Goethe-Universit\"{a}t Frankfurt, Frankfurt, Germany\\
$^{64}$ Instituto de Ciencias Nucleares, Universidad Nacional Aut\'{o}noma de M\'{e}xico, Mexico City, Mexico\\
$^{65}$ Instituto de F\'{i}sica, Universidade Federal do Rio Grande do Sul (UFRGS), Porto Alegre, Brazil\\
$^{66}$ Instituto de F\'{\i}sica, Universidad Nacional Aut\'{o}noma de M\'{e}xico, Mexico City, Mexico\\
$^{67}$ iThemba LABS, National Research Foundation, Somerset West, South Africa\\
$^{68}$ Jeonbuk National University, Jeonju, Republic of Korea\\
$^{69}$ Johann-Wolfgang-Goethe Universit\"{a}t Frankfurt Institut f\"{u}r Informatik, Fachbereich Informatik und Mathematik, Frankfurt, Germany\\
$^{70}$ Korea Institute of Science and Technology Information, Daejeon, Republic of Korea\\
$^{71}$ KTO Karatay University, Konya, Turkey\\
$^{72}$ Laboratoire de Physique des 2 Infinis, Ir\`{e}ne Joliot-Curie, Orsay, France\\
$^{73}$ Laboratoire de Physique Subatomique et de Cosmologie, Universit\'{e} Grenoble-Alpes, CNRS-IN2P3, Grenoble, France\\
$^{74}$ Lawrence Berkeley National Laboratory, Berkeley, California, United States\\
$^{75}$ Lund University Department of Physics, Division of Particle Physics, Lund, Sweden\\
$^{76}$ Nagasaki Institute of Applied Science, Nagasaki, Japan\\
$^{77}$ Nara Women{'}s University (NWU), Nara, Japan\\
$^{78}$ National and Kapodistrian University of Athens, School of Science, Department of Physics , Athens, Greece\\
$^{79}$ National Centre for Nuclear Research, Warsaw, Poland\\
$^{80}$ National Institute of Science Education and Research, Homi Bhabha National Institute, Jatni, India\\
$^{81}$ National Nuclear Research Center, Baku, Azerbaijan\\
$^{82}$ National Research and Innovation Agency - BRIN, Jakarta, Indonesia\\
$^{83}$ Niels Bohr Institute, University of Copenhagen, Copenhagen, Denmark\\
$^{84}$ Nikhef, National institute for subatomic physics, Amsterdam, Netherlands\\
$^{85}$ Nuclear Physics Group, STFC Daresbury Laboratory, Daresbury, United Kingdom\\
$^{86}$ Nuclear Physics Institute of the Czech Academy of Sciences, Husinec-\v{R}e\v{z}, Czech Republic\\
$^{87}$ Oak Ridge National Laboratory, Oak Ridge, Tennessee, United States\\
$^{88}$ Ohio State University, Columbus, Ohio, United States\\
$^{89}$ Physics department, Faculty of science, University of Zagreb, Zagreb, Croatia\\
$^{90}$ Physics Department, Panjab University, Chandigarh, India\\
$^{91}$ Physics Department, University of Jammu, Jammu, India\\
$^{92}$ Physics Program and International Institute for Sustainability with Knotted Chiral Meta Matter (SKCM2), Hiroshima University, Hiroshima, Japan\\
$^{93}$ Physikalisches Institut, Eberhard-Karls-Universit\"{a}t T\"{u}bingen, T\"{u}bingen, Germany\\
$^{94}$ Physikalisches Institut, Ruprecht-Karls-Universit\"{a}t Heidelberg, Heidelberg, Germany\\
$^{95}$ Physik Department, Technische Universit\"{a}t M\"{u}nchen, Munich, Germany\\
$^{96}$ Politecnico di Bari and Sezione INFN, Bari, Italy\\
$^{97}$ Research Division and ExtreMe Matter Institute EMMI, GSI Helmholtzzentrum f\"ur Schwerionenforschung GmbH, Darmstadt, Germany\\
$^{98}$ Saga University, Saga, Japan\\
$^{99}$ Saha Institute of Nuclear Physics, Homi Bhabha National Institute, Kolkata, India\\
$^{100}$ School of Physics and Astronomy, University of Birmingham, Birmingham, United Kingdom\\
$^{101}$ Secci\'{o}n F\'{\i}sica, Departamento de Ciencias, Pontificia Universidad Cat\'{o}lica del Per\'{u}, Lima, Peru\\
$^{102}$ Stefan Meyer Institut f\"{u}r Subatomare Physik (SMI), Vienna, Austria\\
$^{103}$ SUBATECH, IMT Atlantique, Nantes Universit\'{e}, CNRS-IN2P3, Nantes, France\\
$^{104}$ Sungkyunkwan University, Suwon City, Republic of Korea\\
$^{105}$ Suranaree University of Technology, Nakhon Ratchasima, Thailand\\
$^{106}$ Technical University of Ko\v{s}ice, Ko\v{s}ice, Slovak Republic\\
$^{107}$ The Henryk Niewodniczanski Institute of Nuclear Physics, Polish Academy of Sciences, Cracow, Poland\\
$^{108}$ The University of Texas at Austin, Austin, Texas, United States\\
$^{109}$ Universidad Aut\'{o}noma de Sinaloa, Culiac\'{a}n, Mexico\\
$^{110}$ Universidade de S\~{a}o Paulo (USP), S\~{a}o Paulo, Brazil\\
$^{111}$ Universidade Estadual de Campinas (UNICAMP), Campinas, Brazil\\
$^{112}$ Universidade Federal do ABC, Santo Andre, Brazil\\
$^{113}$ University of Cape Town, Cape Town, South Africa\\
$^{114}$ University of Houston, Houston, Texas, United States\\
$^{115}$ University of Jyv\"{a}skyl\"{a}, Jyv\"{a}skyl\"{a}, Finland\\
$^{116}$ University of Kansas, Lawrence, Kansas, United States\\
$^{117}$ University of Liverpool, Liverpool, United Kingdom\\
$^{118}$ University of Science and Technology of China, Hefei, China\\
$^{119}$ University of South-Eastern Norway, Kongsberg, Norway\\
$^{120}$ University of Tennessee, Knoxville, Tennessee, United States\\
$^{121}$ University of the Witwatersrand, Johannesburg, South Africa\\
$^{122}$ University of Tokyo, Tokyo, Japan\\
$^{123}$ University of Tsukuba, Tsukuba, Japan\\
$^{124}$ University Politehnica of Bucharest, Bucharest, Romania\\
$^{125}$ Universit\'{e} Clermont Auvergne, CNRS/IN2P3, LPC, Clermont-Ferrand, France\\
$^{126}$ Universit\'{e} de Lyon, CNRS/IN2P3, Institut de Physique des 2 Infinis de Lyon, Lyon, France\\
$^{127}$ Universit\'{e} de Strasbourg, CNRS, IPHC UMR 7178, F-67000 Strasbourg, France, Strasbourg, France\\
$^{128}$ Universit\'{e} Paris-Saclay Centre d'Etudes de Saclay (CEA), IRFU, D\'{e}partment de Physique Nucl\'{e}aire (DPhN), Saclay, France\\
$^{129}$ Universit\`{a} degli Studi di Foggia, Foggia, Italy\\
$^{130}$ Universit\`{a} del Piemonte Orientale, Vercelli, Italy\\
$^{131}$ Universit\`{a} di Brescia, Brescia, Italy\\
$^{132}$ Variable Energy Cyclotron Centre, Homi Bhabha National Institute, Kolkata, India\\
$^{133}$ Warsaw University of Technology, Warsaw, Poland\\
$^{134}$ Wayne State University, Detroit, Michigan, United States\\
$^{135}$ Westf\"{a}lische Wilhelms-Universit\"{a}t M\"{u}nster, Institut f\"{u}r Kernphysik, M\"{u}nster, Germany\\
$^{136}$ Wigner Research Centre for Physics, Budapest, Hungary\\
$^{137}$ Yale University, New Haven, Connecticut, United States\\
$^{138}$ Yonsei University, Seoul, Republic of Korea\\
$^{139}$  Zentrum  f\"{u}r Technologie und Transfer (ZTT), Worms, Germany\\
$^{140}$ Affiliated with an institute covered by a cooperation agreement with CERN\\
$^{141}$ Affiliated with an international laboratory covered by a cooperation agreement with CERN.\\

\end{flushleft} 
  
\end{document}